%% file: main.tex
\documentclass[11pt]{article}
\usepackage[toc,page]{appendix}
\usepackage[utf8x]{inputenc}
\usepackage{setspace}
\usepackage[colorlinks]{hyperref}
\usepackage{amsmath}
\usepackage{amssymb}
\usepackage{amsthm}
\usepackage[frak=esstix]{mathalpha}
\usepackage{mathtools}
\usepackage{physics}
\usepackage{booktabs}
\usepackage{color,soul}
\usepackage{tocloft}
\usepackage[dvipsnames]{xcolor}
\usepackage{geometry} 
\usepackage{fullpage}[cm] 
\usepackage[font=small,format=plain,labelfont=bf,up,textfont=normal,up]{caption}
\usepackage{cancel}

\usepackage{simpler-wick}
\usepackage[compat=1.1.0]{tikz-feynman}
\usepackage[compat=1.1.0]{tikz-feynhand}
\usepackage{braket}
\usepackage{parskip}

\swapnumbers

\selectcolormodel{cmy}
\hypersetup{
	colorlinks,
	linkcolor = {cyan},
	citecolor = {green},
	urlcolor  = {magenta}
}

\swapnumbers 

\selectcolormodel{cmy}
\hypersetup{
	colorlinks,
	linkcolor = {cyan},
	citecolor = {green},
	urlcolor  = {magenta}
}

\numberwithin{equation}{section}

\newcommand{\pr}[1]{\left(#1\right)}

\newcommand{\pd}{\ensuremath{\partial}}

\newcommand{\AAA}{\ensuremath{\mathcal{A}}}
\newcommand{\BB}{\ensuremath{\mathcal{B}}}
\newcommand{\DD}{\ensuremath{\mathcal{D}}}
\newcommand{\EE}{\ensuremath{\mathcal{E}}}

\newcommand{\LL}{\ensuremath{\mathcal{L}}}
\newcommand{\MM}{\ensuremath{\mathcal{M}}}

\newcommand{\pp}{\ensuremath{\varphi}}

\newcommand{\bb}{\ensuremath{\beta}}
\newcommand{\g}{\ensuremath{\gamma}}
\newcommand{\de}{\ensuremath{\delta}}

\newcommand{\la}{\ensuremath{\lambda}}

\newcommand{\sg}{\ensuremath{\sigma}}
\newcommand{\ep}{\ensuremath{\epsilon}}
\newcommand{\La}{\ensuremath{\Lambda}}

\newcommand{\pc}{\ensuremath{\phi_{\rm cl}}}
\newcommand{\Sc}{\ensuremath{S_{\rm cl}}}
\newcommand{\Tc}{\ensuremath{T_{\rm cl}}}
\newcommand{\Mg}{\ensuremath{M_{\rm gap}}}

\newcommand{\PolyG}{\ensuremath{\psi}}

\newcommand{\dph}{\ensuremath{\delta\phi}}

\newcommand{\vggb}{\ensuremath{V_{2NGB,1BP}}}
\newcommand{\vggg}{\ensuremath{V_{3NGB}}}
\newcommand{\vbbb}{\ensuremath{V_{3BP}}}
\newcommand{\vgggg}{\ensuremath{V_{4NGB}}}

\newcommand{\dphib}{\ensuremath{\de\phi_{\rm bulk}}}
\newcommand{\psib}{\ensuremath{\psi_{\rm bulk}}}

\newcommand{\vone}{\ensuremath{\mathcal{V}_{1}}}
\newcommand{\vtwo}{\ensuremath{\mathcal{V}_{2}}}
\newcommand{\vtwoA}{\ensuremath{\mathcal{V}_{2A}}}
\newcommand{\vtwoB}{\ensuremath{\mathcal{V}_{2B}}}
\newcommand{\vthree}{\ensuremath{\mathcal{V}_{3}}}
\newcommand{\vthreeA}{\ensuremath{\mathcal{V}_{3A}}}
\newcommand{\vthreeB}{\ensuremath{\mathcal{V}_{3B}}}

\newcommand{\Sb}{\sg_{\rm 1BP}}
\newcommand{\Sbb}{\sg_{\rm 2BP}}
\newcommand{\Sgg}{\sg_{\rm 2NGB}}

\newcommand{\beq}{\begin{equation}} \newcommand{\eeq}{\end{equation}}
\newcommand{\bea}{\begin{equation} \begin{aligned}} \newcommand{\eea}{\end{aligned} \end{equation}}

\begin{document}

\vspace{1.8cm}
\begin{center}
  \Huge Effective Strings in QED$_3$
\end{center}  

\vspace{0.7cm}
\begin{center}        
{\large Ofer Aharony, Netanel Barel, and Tal Sheaffer}
\end{center}

\vspace{0.15cm}
\begin{center}        
\emph{Department of Particle Physics and Astrophysics, Weizmann Institute of Science, \\ Rehovot 7610001, Israel}\\[.4cm]
             
e-mails: \tt ofer.aharony@weizmann.ac.il, netanel.barel@weizmann.ac.il, tal.sheaffer@weizmann.ac.il
\end{center}

\vspace{1.5cm}


\begin{abstract}

Effective string theory describes the physics of long confining strings in theories, like Yang-Mills theory, where the mass gap $M_{gap}^2$ is of the same order as the string tension $T$. In $2+1$ dimensions, there is a class of confining theories, including massive QED$_3$ as first analyzed by Polyakov, for which $M_{gap}^2\ll T$. These theories are weakly coupled at low energies of order $M_{gap}$, and may be analyzed perturbatively. In this paper, we analyze the physics of strings in such theories, focusing on QED$_3$, at energies of order $M_{gap}$ (but still well below $\sqrt{T}$). We argue that the width of the string in these theories should be of order $1/M_{gap}$ independently of its length, as long as the string is not exponentially long. We also compute at leading order in perturbation theory the ground state energy of a confining string on a circle, and the scattering of Nambu-Goldstone bosons on the string worldsheet. 
 
\end{abstract}
\thispagestyle{empty}


\clearpage

\setstretch{1.35}  
\tableofcontents
\thispagestyle{empty}

\setstretch{1.5}  


\section{Introduction and Summary}\label{sec_int}
\input{body/intoruction}


\section{Review of QED in \texorpdfstring{$2+1$}{2+1} Dimensions and Effective String Theory}\label{sec_rev}
\input{body/review}


\section{Preliminaries}\label{sec_pre}
\input{body/preliminaries}


\section{Flux-Tube Profile}\label{sec: flux-tube profile}
\input{body/profile}

\section{Ground-State Energy}\label{sec_cal}
\input{body/ground}


\section{Scattering}\label{sec: scattering}
\input{body/scattering}


\section{Conclusions}\label{sec: conc}
\input{body/conclusions}

\section*{Acknowledgments}

The authors would like to thank A. Athenodorou, M. Caselle, S. Dubovsky, and M. Teper for useful discussions.
This work was supported in part by an Israel Science Foundation (ISF) grant number 2159/22, by Simons Foundation grant 994296 (Simons
Collaboration on Confinement and QCD Strings), by the Minerva foundation with funding from the Federal German Ministry for Education and Research, by the German Research Foundation through a German-Israeli Project Cooperation (DIP) grant ``Holography and the Swampland'', and by a research grant from Martin Eisenstein. OA is the Samuel Sebba Professorial Chair of Pure and Applied Physics.

\appendix
\addtocontents{toc}{\protect\setcounter{tocdepth}{1}}


\section{Fluctuations Above the String}\label{app_basis}
\input{appendices/basis}


\section{Change of Variables}\label{app_change}
\input{appendices/change}


\section{Vertices in Pseudo-Momentum Space}\label{app_vertices}
\input{appendices/vertices}


\section{Computation of the Jacobian}\label{app_jacobian}
\input{appendices/jacobian}


\section{Scattering in the Direct Formalism}\label{app_scattering}
\input{appendices/scattering}


\section{IR divergences by Soft NGBs}\label{app_IR}
\input{appendices/IR}


\section{Integrals and Sums}\label{app_integrals}
\input{appendices/integrals}

\end{document}

%% file: body/intoruction.tex
Strings in confining gauge theories that have a mass gap exhibit many universal properties. Indeed, these properties are common to any theory that has a stable string-like excitation and a mass gap. In such theories, in the presence of a long string, the only light degrees of freedom are the massless transverse excitations of the string, which can be thought of as Nambu-Goldstone bosons (NGBs) associated with the translation symmetries spontaneously broken by the presence of the string. These modes are described by ``effective string theory'' (EST) (see \cite{Dubovsky:2012sh, Aharony:2013ipa} and references therein), and their effective action is universal\footnote{For $2+1$ dimensional theories, the effective action is equal to the Nambu-Goto string action.} up to terms going as a rather large power of the energy divided by the scale where extra degrees of freedom appear. This implies that properties of long strings, such as their excitation spectrum, width and low-energy scattering amplitudes of NGBs, are universal.

In many confining theories, such as Yang-Mills theories in $3+1$ or $2+1$ dimensions, there is a single mass scale governing the dynamics ($\Lambda_{QCD}$ in the $3+1$ dimensional case, $g_{YM}^2$ in the $2+1$ dimensional case). The mass gap is then of the same order as the square root of the string tension, and the EST breaks down at this scale (with corrections to the universal behavior involving a power series in the energy divided by this scale).  

However, in $2+1$ dimensional QED (with heavy electrons), the behavior is different. This theory has a $U(1)_T$ global symmetry generated by the current $J_{\mu} = \epsilon_{\mu \nu \rho} F^{\nu \rho}$ (a ``topological symmetry''), which is conserved to all orders in perturbation theory. The low-energy behavior then depends on the UV completion of the theory. For some UV completions, such as ``non-compact QED'', the symmetry is exact and is spontaneously broken at low energies by the expectation value of the scalar dual to the photon. In those cases, the low-energy behavior involves a free massless scalar field. However, in many UV completions, such as the Georgi-Glashow model, there are instantons carrying magnetic flux (``monopole-instantons''). As Polaykov showed in \cite{Polyakov:1975rs}, these lead to a non-perturbative potential for the scalar field, which explicitly breaks the global $U(1)_T$ symmetry and gives the scalar a mass $m$ (such that the mass gap $\Mg$ is $m$). When these effects occur at high energies compared to the gauge coupling squared $e^2$ (that scales as an energy), they are controllable, and the mass $m$ is exponentially small \cite{Polyakov:1976fu, Polyakov:1987ez, Banks:1977cc, Gopfert:1981er, Karliner:1983ab, Ito:1981tq}. As Polyakov showed (and as we review in the next section), the low-energy theory of the scalar field has solitonic string solutions, which can be identified as confining strings (in the sense that electric sources in the low-energy theory must be attached to such strings). These strings can break by forming electron-positron pairs, but when the electrons are heavy, they are approximately stable.

In the limit where the non-perturbative corrections are under control, the tension $T$ of these strings is much larger than the mass squared $m^2$ of the dual photon, since $T$ and $m$ are suppressed by the same exponentially small factor. In this case, the usual effective string description breaks down at the scale $m$, which is much smaller than $\sqrt{T}$. However, since the theory at the scale $m$ is still weakly coupled, one can still compute in a different effective theory that is valid up to the scale $\sqrt{T}$, and that includes both the NGBs and the bulk photon (BP)\footnote{In the case of QED$_3$ there are no additional degrees of freedom on the string worldsheet at the scale $m$, but other theories with solitonic strings, such as the $2+1$ dimensional $\phi^4$ theory, do have such additional degrees of freedom, which would then need to be included in the effective theory as well. A similar analysis to ours applies to general weakly coupled $2+1$ dimensional scalar field theories, where the strings we discuss are domain walls between different vacua of the scalar fields. The effective action on such strings/domain walls was also discussed in \cite{Gregory:1989gg, Silveira:1993am,Carter:1994ag, Blanco-Pillado:2024bev} in the Lorentzian context, and in \cite{Lin:1982ce, Zia:1985gt} in the Hamiltonian context.}. Thus, we can still compute properties of the string such as its width, energy levels, and NGB scattering amplitudes up to the scale $\sqrt{T}$, but, starting from scales of order $m$, they are different from the usual EST predictions. In this paper, we compute some of the simplest properties of strings in QED$_3$, at leading order in perturbation theory, highlighting their differences from the usual effective string description.

In the absence of experimental realizations of QED$_3$, it would be interesting to compare our theoretical predictions to lattice simulations of this theory. Earlier works \cite{Caselle:2014eka, Sterling:1983fs, Vadacchino:2014ota, Caselle:2016mqu, Caselle:2016iai} have looked at the properties of confining strings in QED$_3$ and noted that they behave differently from the usual EST. Strings and ``glueballs'' of QED$_3$ were also analyzed in \cite{Athenodorou:2018sab}.
Our predictions depend on having a large hierarchy of scales $m^2 \ll T$, and ideally, in a lattice simulation with lattice spacing $a$, we should also have $T \ll 1/a^2$. Achieving these hierarchies is difficult since it requires very large lattices, and in the existing simulations presented in the works mentioned, the ratios $m^2/T$ and $T a^2$ are not very small; so, it is not clear if our predictions can be directly applied.\footnote{Moreover, as discussed in \cite{Athenodorou:2018sab}, the standard U(1) lattice action does not have a continuum limit in which both the photon mass and the string tension remain finite as in the continuum theory we discuss here (following Polyakov); thus, obtaining this theory requires modifying the lattice theory (for instance, one could add the extra degrees of freedom of the Georgi-Glashow model).} Nevertheless, our results do seem to match some of the qualitative features observed in the lattice.
We hope that our work will motivate more precise lattice simulations, which will be able to test our predictions for the string width and ground state energy quantitatively.

The structure of the paper is the following. We begin in section \ref{sec_rev} by reviewing QED$_3$ and its confining strings, and by reviewing the usual EST (which is also valid in this theory at energies well below $m$). In section \ref{sec_pre}, we present the two formalisms we use to analyze the effective theory at the scale $m$, including both the NGB (a $1+1$ dimensional massless scalar field living on the string worldsheet) and the BP (a massive $2+1$ dimensional scalar field living in the bulk spacetime): the direct formalism and the stringy formalism.

The following sections present our results. In section \ref{sec: flux-tube profile} we analyze the width of the string (as measured, for instance, by the decay of the electric field as we go away from the center of the string/flux-tube). There are two contributions to the width. The first contribution comes from the classical width of the solitonic solution of the low-energy theory, which is of order $1/m$. The other contribution is the usual one appearing in EST, from the fluctuations of the zero modes of the transverse position of the string (the NGBs), and this is proportional to $1/\sqrt{T}$ but grows logarithmically with the length of the string \cite{Luscher:1980iy}. For standard confining strings, the latter contribution dominates, but in QED$_3$, it is actually the classical width that dominates, as long as the string is not exponentially long (compared to the scale $T$). Thus, unlike standard confining strings, here we expect the width of long (but not exponentially long) strings to be a constant of order $1/m$ rather than growing with the string length, and we expect the electric field profile to be exponential (in the distance from the string) rather than Gaussian.

In section \ref{sec_cal}, we analyze the ground state energy of the string. For strings of length $L \gg 1/m$, it is the same as in EST. However, the energy level has corrections proportional to powers of $e^{-mL}$. While being negligible for $L \gg 1/m$, they lead to a completely different (but still computable) energy once $L \lesssim 1/m$. 

Finally, in section \ref{sec: scattering} we compute (at leading order in perturbation theory) the scattering amplitudes of two NGBs, both to other NGBs and to BPs. The latter process can only occur at energies above $m$. 

We end in section \ref{sec: conc} with our conclusions. Several appendices contain technical results that are used in the main text. Appendix \ref{app_basis} deals with normalization for a continuum of states, related to their completeness and orthonormality relations. Appendix \ref{app_change} discusses the relation between the different variables in the two formalisms used in this work. In appendix \ref{app_vertices}, the vertices in the two formalisms are evaluated in momentum space. In appendix \ref{app_jacobian}, the Jacobian of the change of variables in the stringy formalism is calculated. Appendix \ref{app_scattering} compares scattering amplitudes calculated using the two formalisms. In appendix \ref{app_IR}, the IR divergences related to the emission of soft NGBs, and their cancellation when there is no recoil, are discussed. Finally, appendix \ref{app_integrals} elaborates on calculations of integrals and sums that appear in the main text.

%% file: body/review.tex
\subsection{Review of the Low-Energy Behavior of QED\texorpdfstring{$_3$}{QED in 2+1 d}}

The main example we focus on in this paper is QED in $2+1$ dimensions\footnote{If not mentioned explicitly, Lorentzian signature is used, and the metric is mostly negative. When Euclidean signature is used, it will be clear from the context or explicitly mentioned.} (with a heavy electron), with a UV completion that breaks the ``topological'' $U(1)_T$ symmetry (so that it allows for monopole-instantons). As mentioned in the introduction, this theory exhibits non-perturbatively confinement and a mass gap.

QED in $2+1$ dimensions, with the Lagrangian:
\begin{equation}
    {\cal L} = -\frac{1}{4e^2} F_{\mu \nu}^2 + \bar\psi (i \gamma^{\mu} D_{\mu} - m_e) \psi,
\end{equation}
is weakly coupled (at all scales) when the mass $m_e$ of the electron is much larger than the gauge coupling constant, $m_e \gg e^2$. Below the scale of the electron mass, the only degree of freedom is the photon. In $2+1$ dimensions, the gauge field $A_{\mu}$ can be replaced by its dual $\phi$, defined as:
\begin{equation}\label{dual_photon}
    F_{\mu\nu} = e \epsilon_{\mu \nu \rho} \partial^{\rho} \phi.
\end{equation}
$\phi$ is a periodic field, with periodicity $e$, i.e. $\phi \equiv \phi + e$, thanks to flux quantization. The gauge-kinetic term becomes a canonically-normalized free kinetic term for $\phi$. This term is invariant under $U(1)_T$, now acting locally on $\phi$, shifting it by a constant. In any UV completion of QED$_3$ that has Euclidean solutions carrying magnetic flux (``monopole-instantons'') with finite action, this symmetry is explicitly broken (non-perturbatively). Solutions carrying one unit of magnetic charge generate (at leading order in the dilute instanton gas approximation) a potential proportional to $\cos(2\pi \phi / e)$, with a non-perturbatively small coefficient. Terms of the form $\cos(2\pi n \phi / e)$ with integer $n > 1$ will also be generated, but with even smaller coefficients; therefore, they can be neglected for the purposes of this paper. Denoting $\beta \equiv 2\pi / e$, we will write the low-energy effective action as
\begin{equation}\label{eq_pre_action}
    S_{\rm sg} = \int dxdydt \pr{\frac{1}2(\pd\phi)^2 - \frac{m^2}{\bb^2}\pr{1 - \cos(\bb\phi)}},
\end{equation}
where $m^2$ is a non-perturbatively small coefficient, $m \ll e^2$ (this implies $m\bb^2\ll 1$, which is the perturbation expansion parameter in \eqref{eq_pre_action}). This action shows that the photon becomes massive with a mass $m$.

The definition \eqref{dual_photon} of $\phi$ implies that when we go around an electrically charged source with quantized charge $n$, $\phi$ shifts by $n\cdot e$. Using the effective action \eqref{eq_pre_action} we can find solutions for the electric flux between two distant charges, showing that it is confined to a string-like flux tube. To begin with, let us assume that the sources are infinitely far away from each other in the $y$ direction, and located at $x=x_0$.
The equation of motion (EOM) satisfied by $\phi$ is the sine-Gordon equation:
\begin{equation}\label{eq_pre_EOM}
    \pd^2\phi + \frac{m^2}{\bb}\sin(\bb\phi) = 0.
\end{equation}
This equation has a solitonic solution interpolating between two adjacent minima of the potential:
\begin{equation}\label{eq_pre_string_solution}
    \pc = \frac{4}{\bb}\arctan\pr{e^{m\pr{x-x_0}}}.
\end{equation}
$\pc$ describes a confining flux tube located at $x=x_0$, with the only non-zero component of the electro-magnetic field being an electric field
\begin{equation}
    E_y = F_{ty} = e \partial_x \pc = \frac{e^2m}{\pi}\frac{1}{\cosh(m(x-x_0))}.
\end{equation}
Substituting this solution into the action \eqref{eq_pre_action} gives the classical action:
\begin{equation}\label{eq_pre_clas_action}
    \Sc = \int dxdydt\pr{\frac{1}2(\pd\pc)^2 - \frac{m^2}{\bb^2} \pr{1 - \cos(\bb\pc)} }=-\frac{8m}{\bb^2}\int dydt,
\end{equation}
where we used:
\begin{equation}
    \int dx\frac{1}{2}(\pd_x\pc)^2 = \int dx\frac{m^2}{\bb^2}(1-\cos(\bb\pc)) = \frac{4m}{\bb^2}.
\end{equation}
The constant in front of the integrand in the right-hand side of \eqref{eq_pre_clas_action} can be interpreted as the confining string (classical) tension $T_{\rm cl} \equiv \frac{8m}{\bb^2}$.\footnote{The classical tension gets quantum corrections; see section \ref{sec_cal}. $ T$ denotes the true tension, and it is also much greater than $m^2$.} 
Note that $T_{\rm cl} \gg m^2$, such that there is a separation of scales between the photon's mass and the string tension. From $\Sc$ one can read off the classical ground state energy for a confining string wrapped around a circle of length $L_y$, $E_0 = L_y\frac{8m}{\bb^2}$. The background \eqref{eq_pre_string_solution} spontaneously breaks translation symmetry in the $x$-axis, as indicated by the free parameter $x_0$, and a NGB appears as a massless mode on the string worldsheet. It also spontaneously breaks reflection symmetry, and changing $x\to -x$ gives a solution with the opposite orientation of the string. 

One interesting scenario involves a string stretched between a charged particle (of charge $n$) and an anti-particle, sitting at positions $(x,y)=\pr{0,\pm\frac{R}2}$. The condition that $\phi$ needs to fulfill near each particle can be found from the electro-magnetic field for a free particle of charge $n$ sitting at $(x,y)=(0,0)$:
\begin{equation}
    F_{tx} = \frac{ne^2}{2\pi}\frac{x}{x^2+y^2}, \ F_{ty} = \frac{ne^2}{2\pi}\frac{y}{x^2+y^2}, \ 
    F_{xy} = 0.
\end{equation}
The field $\phi$ then satisfies:
\begin{equation}
    \pd_t\phi = 0, \ \pd_x\phi = \frac{ne}{2\pi}\frac{y}{x^2+y^2}, \ \pd_y\phi = -\frac{ne}{2\pi}\frac{x}{x^2+y^2},
\end{equation}
with the solution:
\begin{equation} \label{near_source}
    \phi_{\rm em} = -\frac{ne}{2\pi}\arctan\pr{\frac{y}{x}} + c,
\end{equation}
giving the behavior of $\phi$ for small $x^2+y^2$. It has a non-trivial holonomy around each charge; the winding number equals the charge.\footnote{The solution $\pc$ considered above, with the $y$-axis being infinite, is consistent with the electric field lines for the infinite string found above, where the positive charge is located at $y = -\infty$ and the negative at $y = \infty$. There is an electric flux of $e^2$, either circling the positive charge at $-\infty$ or the negative charge at $\infty$.}
When the distance between the sources is large and finite we expect a solution that interpolates between the behavior \eqref{near_source} near the particles, and (for $n=1$) the behavior \eqref{eq_pre_string_solution} in the region between them. However, it is not easy to find the  exact solution. Therefore, we will consider the case where the sources are infinitely far away, or without sources at all for the compact-$y$ case. When the distance of some observable that we will compute from the finite string is much smaller than the string's length, and it is not close to the charges, there should be no difference from the infinite / compact string. 

In this paper we will be interested in studying the flux-tube profile, the ground-state energy correction, and scattering / decay processes. The flux-tube profile is the same for the infinite and for the compact string. The ground-state energy quantum correction for the infinite string is actually a quantum correction to the classical string tension, and can be used as a renormalization condition. For the compact string, it is expected to have an extra ground-state correction coming from the quantum Casimir energy $-\frac{1}{6L}$ of the massless NGB, and additional 
exponentially small corrections going as $e^{-mL}$ from the massive BP. A similar behavior is expected also in higher orders in perturbation theory, where polynomials in $\beta$ (multiplying negative powers of $L$) are accompanied by exponentially small terms $\sim e^{-mL}$ in the long string limit $mL\gg 1$. Beyond the ground state, the infinite string allows the scattering of NGB states along the string, while for the compact string, there are excited states (which can decay to BPs if they have enough energy).

\subsection{Review of Effective String Theory}

In $(d-1)+1$ dimensional theories that have a mass gap $\Mg$ and that have stable strings with tension $T$, the low-energy theory (below the scale $\Mg$) in the presence of a long string includes only the NGBs associated with the $(d-2)$ translation symmetries broken by the string. For a string stretched in the $x^1$ direction, these scalar fields parameterize the position of the string in the transverse directions $x^2, \cdots, x^{d-1}$.

The effective action of the NGBs includes only terms with derivatives, and can be expanded at low energies in a derivative expansion. It is invariant under diffeomorphisms of the string worldsheet, since we have a freedom to choose with which coordinates to parameterize it. The leading terms in the derivative expansion are universal, and controlled by EST. For a string in $2+1$ dimensions, the effective action for a string located at $X^{\mu}(\sigma^a)$ (where $\sigma^a$ ($a=0,1$) is some parameterization of the $1+1$ dimensional string worldsheet) may be written in terms of the induced metric $g_{ab} \equiv \partial_a X^{\mu} \partial_b X_{\mu}$ as
\begin{equation} \label{ng_corrections}
    S_{\rm EST} = \int d^2\sigma [-T \sqrt{-\det(g)} + c \sqrt{-\det(g)} R[g]^2 + {\rm higher\ derivative\ terms}],
\end{equation}
where $R[g]$ is the Ricci curvature of the metric $g$, and $c$ is a non-universal constant. In theories like QCD$_3$ that have a single dynamical scale, one expects $c \sim 1/T$ and similarly for the coefficients of higher derivative terms, so that the derivative expansion is expected to break down at the scale $\sqrt{T}$. As discussed above, in QED$_3$, the BP has mass $m \ll \sqrt{T}$, so the effective string action is only valid below $m$, and above it the interactions with the BP must be taken into account.

In the way that the action is written above, it is not manifest that the only dynamical degree of freedom is the NGB. When expanding around a string stretched (without loss of generality) in the $y$ direction, we can make this manifest by gauge-fixing the diffeomorphisms to the static gauge, where we choose the two worldsheet coordinates to be $t$ and $y$. In this gauge, the leading term in the action \eqref{ng_corrections} takes the form
\begin{equation} \label{ng_action_static}
    S_{NG} = - T \int d^2\sigma \sqrt{-\det(\eta_{ab} + \partial_a X \partial_b X)} = T \int d^2\sigma \left(-1 - \frac{1}{2} \partial_a X \partial^a X + \cdots \right),
\end{equation}
where $X$ is the NGB describing the motion of the string in the transverse $x$ direction, the higher order terms in \eqref{ng_action_static} start at four-derivative order, and the leading correction to \eqref{ng_action_static}, proportional to $c$, starts at eight-derivative order.

Energy levels and scattering amplitudes well below the scale where the effective action breaks down are universal, with the leading non-universal corrections controlled by the constant $c$. At leading order, the energy levels of a string on a circle of circumference $L_y$ with no longitudinal momentum are given by
\begin{equation} \label{string_energy_levels}
    E_n = T L_y \sqrt{1 + \frac{8\pi}{T L_y^2}\left(n - \frac{d-2}{24}\right)},
\end{equation}
where $n$ is the excitation level, and $d=3$ for our case. The leading correction in the large $L_y$ limit is proportional to $c/L_y^7$, with higher-derivative terms leading to higher powers of $1/L_y$. The width $w(R)$ of a string of length $R$, coming from its quantum fluctuations (and defined, for instance, from the square root of the variance in the position in the transverse dimensions), can also be computed using EST, and at leading order it goes as:
\begin{equation}\label{eq_rev_EST_width}
    w(R)^2 = \frac{d-2}{2\pi T} \log\pr{\frac{R}{R_0}}
\end{equation}
for some constant $R_0$.

In this paper, we will be interested in how this low-energy effective action is modified by including the massive BP of mass $m$ in addition to the NGB. We expect that including this field will give universal modifications to the effective string predictions at energy scales between $m$ and $\sqrt{T}$.

%% file: body/preliminaries.tex
In this chapter, we present two formalisms for treating the quantum theory of fluctuations around a long string in QED$_3$. These formalisms are used in the following three chapters to derive the results of this work. In the first formalism, the ``direct formalism'', the NGB, like the other modes orthogonal to it, is included inside the fluctuations $\dph$ around the string $\pc$. In this formalism, the string is located in some fixed $x_0$, and there are no linear terms in the action (no ``tadpoles''). This formalism is simple and similar to what is usually done in QFT when expanding the field around some classical configuration. It is useful for calculating properties for which the physics of the string's location is unimportant, such as energy levels. However, when the position of the string can change, either by quantum fluctuations or by scattering, this formalism turns out to be less useful. Another disadvantage of this formalism is the treatment of IR divergences. Therefore, we also present a second formalism, the ``stringy formalism'', in which the NGB is ``exponentiated'' and stands as a separate field $x_0(y,t)$. The remaining fluctuations around the string contain only the bulk particle. The IR divergences now appear in the separate field $x_0$ and are easier to handle.

\subsection{Direct Formalism}\label{subsec_first_formalism}

In this formalism, we expand $\phi$ around $\pc$ as follows:
\begin{equation}\label{eq_pre_expansion_fst}
\begin{split}
        \phi(x,y,t) = \pc(x) + \dph(x,y,t)
        = \frac{4}{\bb}\arctan\pr{e^{mx}} + \dph(x,y,t).
\end{split}
\end{equation}
The spectrum of the theory is found by substituting inside the action \eqref{eq_pre_action}, giving the action for $\dph$:
\begin{equation}
    S = S_{\rm cl} + \int dxdydt\pr{\frac{1}2(\pd\dph)^2 - \frac{1}2m^2\cos\pr{\bb\pc}\dph^2} + O\pr{\dph^3}.
\end{equation}
The EOM for $\dph$ is:
\begin{equation}
    \pd^2\dph + m^2\cos(\bb\pc)\dph = 0.
\end{equation}
We can separate variables and choose the $y,t$-dependence to be a plane-wave:
\begin{equation}
    \frac{\pd^2\dph(y,t)}{\pd t^2} - \frac{\pd^2\dph(y,t)}{\pd y^2} + M^2\dph(y,t) = 0,
\end{equation}
giving the equation for the $x$-dependence:
\begin{equation}\label{eq_pre_eigenvalue}
    -\frac{d^2\dph(x)}{dx^2} + m^2\cos(\bb\pc(x))\dph(x) \equiv \hat D \dph(x) = M^2\dph(x).
\end{equation}
The general solution of this equation is:
\begin{equation}
    \dph(x) = c_1e^{-\sqrt{1-\frac{M^2}{m^2}}mx}\pr{\sqrt{1-\frac{M^2}{m^2}}+\tanh(mx)} + c_2e^{\sqrt{1-\frac{M^2}{m^2}}mx}\pr{-\sqrt{1-\frac{M^2}{m^2}}+\tanh(mx)}.
\end{equation}
States bound to the string should be normalizable, which happens only for $M=0$. Hence, there is a unique massless excitation, and there are no massive particles bound to the string.\footnote{Unlike for the potential $(\phi^2-a^2)^2$, where another bound state exists.} Its normalized wave function is:
\begin{equation}
    \phi_0(x) \equiv \sqrt{\frac{m}{2}}\sech(mx).   
\end{equation}
This mode is the NGB corresponding to the spontaneous breaking of the translation symmetry due to the string. Indeed, it satisfies $\phi_0 \propto \frac{d\pc}{dx}$ and generates translations of the string.

The other consistent solutions, which become oscillatory waves as $|x|\to\infty$, are obtained when $M^2=m^2+p^2$:
\begin{equation}
    c_1e^{-ipx}\pr{\frac{ip}{m}+\tanh(mx)} + c_2e^{ipx}\pr{-\frac{ip}{m}+\tanh(mx)}. 
\end{equation}
It is easy to see that the wave function in the second term, which is the complex conjugate of the one in the first term, is obtained from it by $p\to -p$; hence, we use only the second one (with any $p$) in our basis (and impose a reality condition on the field at the end). This solution, describing a bulk particle with mass $m$, corresponds to the BP, now interacting with the string when it passes through it. The interaction is described by the factor $-\frac{ip}{m} + \tanh(mx)$. This factor goes rapidly to a constant for $m|x|\gg 1$, where the wave-function turns into a simple plane wave, representing free propagation. However, the value of the constant changes as we cross the string, which implies that the wave changes by a phase. It will be conceptually beneficial to consider the transverse direction to the string to be compactified with periodicity $L_x$ ($mL_x\gg 1$) and to impose periodic boundary conditions. Then, when crossing the string from left to right, with $p>0$ one has:
\begin{equation}\label{eq_pre_pos_phase}
\begin{split}
    e^{i\de_i} &= -i\frac{p}{m} - 1, \ x = -\infty \ (= -L_x/2),
    \\
    e^{i\de_f} &= -i\frac{p}{m} + 1, \ x = +\infty \ (= +L_x/2),
    \\
    \de_{\rightarrow}(p) &= \de_f - \de_i = \pi - 2\arctan\pr{\frac{p}{m}}.
\end{split}    
\end{equation}
The condition for allowed momenta is:
\begin{equation}
    +\frac{pL_x}2 + \de_f = -\frac{pL_x}2 + \de_i + 2\pi n \Rightarrow p_nL_x + \de_{\rightarrow}(p_n) = 2\pi n   
\end{equation}
for any integer $n$. Similarly, the phase for a wave moving from right to left, with $p = -|p| < 0$ is:
\begin{equation}\label{eq_pre_neg_phase}
\begin{split}
    e^{i\de_i} &= i\frac{|p|}{m} + 1, \ x = +\infty \ (= +L_x/2), 
    \\
    e^{i\de_f} &= i\frac{|p|}{m} - 1, \ x = -\infty \ (= -L_x/2), 
    \\
    \de_{\leftarrow}(p) &= -(\pi - 2\arctan\pr{\frac{|p|}{m}}) = -\pi - 2\arctan\pr{\frac{p}{m}} = - \de_{\rightarrow}(-p).
\end{split}    
\end{equation}
The condition for allowed momenta is:
\begin{equation}
    -\frac{|p|L_x}2 + \de_f = +\frac{|p|L_x}2 + \de_i + 2\pi n \Rightarrow -|p_n|L_x + \de_\leftarrow(p_n) = p_nL_x + \de_\leftarrow(p_n) = 2\pi n.    
\end{equation}
Therefore, the same condition holds for all momenta, with the appropriate phase in the different domains. The phase is antisymmetric, as expected; hence, for every $n>0$ with $p_n>0$ there is a corresponding $p_{-n} = -p_n <0$ (reflecting the reflection symmetry of the action $x\rightarrow -x$ together with $\dph\to -\dph$). At high momenta, the phase vanishes, representing the fact that the photon does not ``feel'' the string. At $p = 0$, there is a discontinuity of $2\pi$ in the phase. This is a fake discontinuity because of the periodicity of the phase. Therefore, one can set $\de(p) = \de_{\rightarrow}(p)$ for all $p$, and as long as derivatives are considered, everything works well. However, when the phase itself is considered, the antisymmetric one should be used for the negative momentum.

Note that (in this linearized approximation) the bulk particles interact with the string only by a phase change, and their wave-functions do not split into returning and transversing waves. This is due to the integrability of the Sine-Gordon model in $1+1$d, where the interactions of particles with solitons are only by a phase shift.

From the condition for the phase, one can read off the density of states:
\begin{equation}\label{eq_pre_den_states}
\begin{split}
    & p_nL_x + \de(p_n) = 2\pi n \qquad \Rightarrow 
    \qquad \frac{\Delta n}{\Delta p} \to \frac{dn}{dp} = \frac{L_x}{2\pi} - \frac{1}{2\pi}\frac{d\de(p)}{dp} = \frac{L_x}{2\pi} - \frac{1}{2\pi}\frac{2m^2}{p^2 + m^2}.
\end{split}    
\end{equation}
Therefore, integrating over these states comes with the density factor\footnote{Note that the summation starts from $n=1$. There is no exact $p=0$ eigenvalue in $\dph_{\rm BP}$ for finite $L_x$; the photon has no standing wave. One can think of the massive $p=0$ eigenvalue in the vacuum as mapped into the massless NGB in the presence of the string. This ``matching'' of states is important for the canonical quantization of the theory; see subsection \ref{subsec_ground_can}.}:
\begin{equation}
    \sum_{n=1}^\infty \to \int_0^\infty dp\pr{\frac{L_x}{2\pi} - \frac{1}{2\pi}\frac{2m^2}{p^2 + m^2}}.
\end{equation}
The first term is the density of states in an empty space. The second term is the influence of the string, which reduces the density of states for $|p| \lesssim m$. For high momenta $|p|\gg m$, the density of states approaches its value in empty space, as the photon moves almost freely.

Regarding the normalization of these continuous states, we choose a $\de$-function normalization. The normalization factor $N(p)$ satisfies:
\begin{equation}
    |N(p_1)|^2\int dx e^{ip_1x}\left(-\frac{ip_1}{m}+\tanh(mx)\right)e^{-ip_2x}\left(+\frac{ip_2}{m}+\tanh(mx)\right) = 2\pi\de(p_1 - p_2).
\end{equation}
Integrating this equation over $p_2$ yields:
\begin{equation}
\begin{split}
    2\pi|N(p_1)|^{-2} &=  \int dxe^{ip_1x}\left(-\frac{ip_1}{m}+\tanh(mx)\right)\left(\frac{-\pd_x}{m}+\tanh(mx)\right)\int dp_2e^{-ip_2x}
    \\
    &= 2\pi\int dxe^{ip_1x}\left(-\frac{ip_1}{m}+\tanh(mx)\right)\left(\frac{-\pd_x}{m}+\tanh(mx)\right)\de\pr{x}
    \\
    &= 2\pi\pr{\frac{p_1^2}{m^2}+1}.
\end{split}
\end{equation}
Thus, the normalized wave-function is:
\begin{equation}
    \phi_p(x) \equiv \frac{ip-m\tanh(mx)}{ip-m}e^{ipx},
\end{equation}
and the full decomposition of the $x$-dependence of the fluctuation $\dph$ is:
\begin{equation} \label{dphi_expansion}
\begin{split}
    \dph(x) &\equiv \dph_{\rm NGB}(x) + \dph_{\rm BP}(x)
    \\
    &= \pp_0\phi_0(x) + \int \frac{dp}{2\pi}\pp_p\phi_p^*(x)
    \\
    &= \pp_0\sqrt{\frac{m}{2}}\sech(mx) + \int \frac{dp}{2\pi} \pp_p \frac{ip+m\tanh(mx)}{ip+m}e^{-ipx}.
\end{split}
\end{equation}
The orthogonality and completeness relations satisfied by $\dph_{\rm NGB}$ and $\dph_{\rm BP}$ are presented in appendix \ref{app_basis}.

$x$-reflection is expected to remain a symmetry of this action. Although the background $\pc$ is not invariant under $x\to -x$, the reflected solution represents the same string with opposite orientation. The fact that the background appears different is reflected by the transformation $\dph(x)\to -\dph(-x)$ (while in the vacuum $\phi$ can be invariant under reflection). This transformation dictates $\pp_0\to -\pp_0$ for the NGB, and $\pp_p\to \frac{-ip+m}{ip+m}\pp_p$ for the BP. This phase represents to fact that the $\pp_p$ mode with $\pc(-x)$ is shifted by the same phase that the $\pp_{-p}$ mode is shifted with $\pc(x)$. This symmetry prevents any odd power interactions $\dph_{\rm NGB}^{2k+1}$ for the NGB, as can be verified directly in the interactions presented below.

This formalism can be used either in the path integral (PI)  framework or in the canonical quantization framework. The ground-state energy will be calculated in both frameworks. We elaborate on both in the following two subsections. Then we consider the interactions.

\subsubsection{PI Framework}

For the PI framework, the offshell dependence of the fluctuations $\dph$ is:
\begin{equation}
\begin{split}
    \dph(x,\sg) = \int \frac{d^2q}{(2\pi)^2}\pp_0(q)\phi_0(x)e^{iq\sg} + \int \frac{dpd^2q}{(2\pi)^3} \pp(p,q)\phi_p^*(x)e^{iq\sg},
\end{split}
\end{equation}
where the two coordinates $y,t$ are denoted as $\sg^a$, and $q_a$ represents the corresponding momenta. The propagator $\hat G$ is the inverse of the kinetic term $\hat D$ (up to a factor of $-i$): $\hat G = i\hat{D}^{-1}$. In operatorial form, they can be written as:
\begin{align}
    \hat D &= \int \frac{d^2q}{(2\pi)^2}q^2\ket{0q}\bra{0q} + \int \frac{dpd^2q}{(2\pi)^3}(q^2-p^2-m^2)\ket{pq}\bra{pq}, 
    \\
    \hat G &= \int \frac{d^2q}{(2\pi)^2}\frac{i}{q^2}\ket{0q}\bra{0q} + \int \frac{dpd^2q}{(2\pi)^3}\frac{i}{q^2-p^2-m^2}\ket{pq}\bra{pq},
\end{align}
where $\ket{pq},\ket{0q}$ are the one-dimensional modes $\ket{p},\ket{0}$ extended to three dimensions with plane waves in the other two directions. Using wave-functions the propagator is:
\begin{equation}
    G(x,\sg;x',\sg') = \bra{x,\sg}\hat G\ket{x',\sg'} = \int \frac{d^2q}{(2\pi)^2}\frac{ie^{iq(\sg-\sg')}}{q^2}\phi_0(x)\phi_0(x') + \int \frac{dpd^2q}{(2\pi)^3}\frac{ie^{iq(\sg-\sg')}}{q^2-p^2-m^2}\phi_p(x)\phi_p^*(x').
\end{equation}
Since the functions $\phi_0(x)e^{-iq\sg},\phi_p(x)e^{-iq\sg}$ form an orthonormal basis, the measure of the path integral satisfies $\DD\dph(x,\sg) = \DD\pp_0(q)\DD\pp(p,q)$.

\subsubsection{Canonical Framework}

In this framework onshell fluctuations are considered. The full onshell dependence of the fluctuations $\dph$ is:
\begin{equation} \label{phi_expansion}
\begin{split}
    \dph(x,\sg) = \int \frac{dq}{2\pi}a_q\phi_0(x)e^{iqy-i|q|t} + \int \frac{dpdq}{(2\pi)^2} a_{pq} \phi_p(x)e^{iqy-i\sqrt{q^2+m^2}t} + c.c.
\end{split}
\end{equation}
The field $\dph$ and the classical coefficients $a_q,a_{pq}$ become operators in the quantum theory. We will not use separate notations for them, and it will be clear from the context whether the quantities are classical or quantum. 

The conjugate momentum $\pi$ to $\phi$ is $\pi = \frac{\pd\phi}{\pd t}$, and the Hamiltonian is:
\begin{equation}
    H = \int dx dy\pr{\frac{1}{2}\pi^2 + \frac{1}{2}(\nabla\phi)^2 + \frac{m^2}{\bb^2}(1 - \cos(\bb\phi))}.
\end{equation}
Since the string solution is static, $\pi$ is also the conjugate momentum of $\dph$. Expanding the Hamiltonian in $\dph$ gives:
\begin{equation}
\begin{split}
    H &= H_{\rm cl} + H_f + O\pr{\dph^3},
    \\
    H_{\rm cl} &= \int dxdy\pr{\frac{1}{2}(\nabla\pc)^2 + \frac{m^2}{\bb^2}(1 - \cos(\bb\pc))} = \frac{8m}{\bb^2}L_y, 
    \\
    H_f &= \int dxdy\pr{\frac{1}{2}\pi^2 + \frac{1}{2}(\nabla\dph)^2 + \frac{m^2}{2}\cos(\bb\pc)\dph^2} \equiv \int dxdy\pr{\frac{1}{2}\pi^2 + \frac{1}{2}\dph \hat D\dph}.
\end{split}
\end{equation}
Diagonalizing the quadratic part of the Hamiltonian follows the same steps as usual, with eigenvalues and eigenfunctions as those of \eqref{eq_pre_eigenvalue}, modified by a plane-wave in the $y$-direction, with eigenvalues $E_{pq} \equiv \sqrt{p^2 + q^2 + m^2},\ E_q \equiv |q|$ for the BP and the NGB, respectively. The creation and annihilation operators $a_p$, $a_{pq}$ in \eqref{phi_expansion} satisfy the usual canonical commutation relations.
The free part of the Hamiltonian reads:
\begin{align}
    H_f = \int \frac{dpdq}{(2\pi)^2} \frac{E_{pq}}{2}\pr{a_{pq}a^\dagger_{pq} + a^\dagger_{pq}a_{pq}} + \int \frac{dq}{2\pi} \frac{E_q}{2}\pr{a_q a^\dagger_q + a^\dagger_q a_q}.
\end{align}
This also holds true for the vacuum sector with its own modes.\footnote{There is a subtlety with the commutation relation $[a_{pq},a^\dagger_{pq}]$, which is not the same $\de(0)$ as in the vacuum sector, due to the different density of states. We will use the discrete version presented above, where this ambiguity will not appear.}

\subsubsection{Interactions}\label{subsubsec_interations}

The rest of the expansion of the action or the Hamiltonian in $\dph$ involves interactions. The higher terms in the potential contain the following terms:
\begin{align}
    \sin(\bb\pc) = -2\tanh(mx)\sech(mx), \ \qquad \cos(\bb\pc) = 1-2\sech^2(mx).
\end{align} 
Then, the potential can be expanded (the quadratic term was considered above):
\begin{equation}
\begin{split}
    V &\equiv \frac{m^2}{\bb^2}(1-\cos(\bb\pc + \bb\dph)) 
    \\
    &\supset m^2\sum_{k=1}^\infty \frac{(-1)^k}{(2k+1)!}\sin(\bb\pc)\bb^{2k-1}\dph^{2k+1} +  m^2\sum_{k=1}^\infty\frac{(-1)^k}{(2k+2)!}\cos(\bb\pc)\bb^{2k}\dph^{2k+2}
    \\
    &= \sum_{k=1}^\infty\frac{m^2(-1)^{k+1}}{(2k+1)!}2\tanh(mx)\sech(mx)\bb^{2k-1}\dph^{2k+1} + \sum_{k=1}^\infty\frac{m^2(-1)^k}{(2k+2)!}(1-2\sech^2(mx))\bb^{2k}\dph^{2k+2},
\end{split}
\end{equation} 
which will be denoted as $-\LL^{\rm int}_{\dph}$. The higher terms describe interactions between the NGB and the BP. They contain all possible vertices involving any number of NGBs / BPs. Only a few of them are needed for this work; the cubic vertex for three BPs is given in the next subsection, and the rest in appendix \ref{app_scattering}.
These terms contain both interactions that are confined to the string and interactions that are not confined to the string. The interactions of the NGBs are always confined, represented by their wave-function $\phi_0(x)$. As for the BPs, the odd terms $\dph^{2k+1}$ are confined (because of the $\sech(mx)$ factor), while the even terms $\dph^{2k+2}$ contain two parts -- one which is confined (the $\sech(mx)^2$ factor and part of the $1$), and another which is not confined (the other part of the $1$). This is because the BPs have interactions anywhere, as in the vacuum, where one expands around the trivial solution $\pc = 0$ and all of the interactions contain even powers of $\dph$. The interactions which are not confined to the string conserve the $x$-momentum, while the others do not.

\subsubsection{Counterterms}\label{subsubsec_counterterms}

Any periodic function of $\phi$ can be added as a counterterm to the action \eqref{eq_pre_action}, and will generally appear in the effective action. For the purposes of this work, only one of them is important, which is:
\begin{equation}
    S = \int dxdydt \pr{\frac{1}2(\pd\phi)^2 - \frac{m^2 + \de m^2}{\bb^2}\pr{1 - \cos(\bb\phi)}},
\end{equation}
where $\de m^2 = \de m^2_2\bb^2 + O(\bb^4)$. $\de m^2_2$ will be determined as usual by setting the mass of the photon in the vacuum to $m$. The cosmological constant counterterm $\de\La$ is important for the vacuum's energy but not for the string's energy (which is the energy added to the vacuum due to the presence of the string), nor for any correlation function, so it will not play any role in this paper.

An expansion in $\dph$ done for the potential holds for the counterterm as well, which also includes terms of lower order in $\dph$:
\begin{equation}
\begin{split}
    \frac{\de m^2}{\bb^2}(1-\cos(\bb\pc + \bb\dph)) 
    =& \sum_{k=0}^{\infty}\frac{\de m^2(-1)^{k+1}}{(2k+1)!}2\tanh(mx)\sech(mx)\bb^{2k-1}(\dph)^{2k+1}
    \\
    &+ \frac{\de m^2}{\bb^2} + \sum_{k=-1}^{\infty}\frac{\de m^2(-1)^k}{(2k+2)!}(1-2\sech^2(mx))\bb^{2k}(\dph)^{2k+2}
    \\
    \equiv & -\LL^{\rm cnt}_{\dph}.
\end{split}
\end{equation} 
The expansion of the counterterm includes $O(\bb^0,\bb^1)$ terms, which will correct the ground state energy and the one-point function, respectively. The higher terms will not be used here. 

\subsection{Stringy Formalism}

We will see in section \ref{sec: scattering} that IR divergences related to the soft NGB appear in various computations, indicating that some kind of resummation may be useful. In this subsection, we will reorganize the computation accordingly by rewriting the NGB as a field $x_0(\sg)$ describing the transverse position of the string. $\sg$ are now interpreted as the worldsheet coordinates in the static gauge. This will lead to a different expansion for the action, which naturally looks like a string theory coupled to a bulk field. We also define $X_0 \equiv \sqrt{\frac{8m}{\beta^2}}x_0$. $X_0$ is normalized to have a canonical kinetic term, and so our change of variables will be from $\phi$ to $X_0$ and $\dphib$, with the latter accounting for 3d ``bulk'' fluctuations of the field (the BP).

Since the string has a width, its location is not exactly well-defined. To find a reasonable way to define its location, we note the following identity, which is valid for any $x_0$ and $\sigma$ in the limit that locally $\phi$ is a small fluctuation around the (possibly shifted in the transverse space) classical solution:
\begin{equation} \label{xzerodef}
    \frac{1}{\bb}\int dx\phi(x,\sg)\sech(m(x-x_0)) \in \pr{0,\frac{2\pi^2}{m\bb^2}}.
\end{equation}
This is because, due to our boundary conditions, $\phi$ approaches a constant at large $|x|$ ($0$ or $\frac{2\pi}{\bb}$), and for large $|x_0|$ the function $\sech(m(x-x_0))$ is supported only at large values. We define $x_0(\sg)$ such that the integral \eqref{xzerodef} is midway between its two boundary values:\footnote{This is well-defined in the limit of small fluctuations around the classical solution \pc, and we will not need more than this for this work.}
\begin{equation}
    \frac{1}{\bb}\int dx\phi(x,\sg)\sech(m(x-x_0(\sg))) = \frac{\pi^2}{m\bb^2}.
\end{equation}
We then expand the field as follows:
\begin{equation}\label{eq_pre_expansion_snd}
\begin{split}
    \phi(x,\sg) & = \pc(x-x_0(\sg),\sg) + \dphib(x-x_0(\sg),\sg)
    \\
    & = \frac{4}{\bb}\arctan\pr{e^{m(x-x_0(\sg))}} + \dphib(x-x_0(\sg),\sg).
\end{split}
\end{equation}

Due to the choice of $x_0$, $\dphib$ is constrained to be orthogonal
to the NG mode:
\begin{equation}\label{eq: dphib orthogonality constraint}
\begin{split}
    \frac{1}{\bb}\int dx \pc(x-x_0)\sech(m(x-x_0)) &= \frac{1}{\bb}\int dx \pc(x-x_0) \pr{\frac{\bb}{2m}\frac{d\pc(x-x_0)}{dx}} = \frac{1}{4m}\pc^2|^\infty_{-\infty} = \frac{\pi^2}{m\bb^2}
    \\
    & \Rightarrow\int dx \dphib(x)\sech(mx) = 0,
\end{split}
\end{equation}
and so it has the form found in the direct formalism:
\begin{equation}
    \dphib(x,\sg) = \int \frac{dp}{2\pi}\phi_p^*(x)\dphib(p,\sg). 
\end{equation}
Effectively, we have ``exponentiated'' the NG mode, which is $x_0(\sg)$. Note that:
\begin{equation}\label{eq: parallel derivative identity}
    \pd_a\phi(x,\sg) = -\pd_a x_0(\sg)\pd_x\pc(x-x_0(\sg)) + (\pd_a-\pd_ax_0(\sg)\pd_x)\dphib(x-x_0(\sg),\sg).
\end{equation}

\subsubsection{Action}

Since the action includes an integral over $x$, all $ x_0$ dependencies can be shifted away once the various derivatives have been taken. Hence, $x_0$ enters the action only through its derivatives, as expected from a massless NGB, and it interacts with the BP close to $x=x_0(\sigma)$. This is the main effect of the exponentiation.

The variable $x_0(\sg)$ contains more information than the NGB in the direct formalism. As we will see momentarily, its kinetic term allows, in addition to the small fluctuations, a different orientation as well as a free motion of the whole string:
\begin{equation}
    x_0(y,t) = x_0 + ny + pt + \text{fluctuations}.   
\end{equation}
Ignoring these, one can relate the variables in the two formalisms by comparing \eqref{eq_pre_expansion_fst} and \eqref{dphi_expansion} to the expansion of \eqref{eq_pre_expansion_snd} around $x_0=0$. The results are presented in appendix \ref{app_change}.

It will be instructive to evaluate the action at $\dphib=0$:
\begin{equation}\label{eq_pre_kinetic_term_x0}
\begin{split}
    & \int dxd^2\sg\pr{\frac{1}{2}(\pd\pc(x-x_0))^2 - \frac{m^2}{\bb^2}(1-\cos(\bb\pc(x-x_0)))}
    \\
    x\to x+x_0(\sg) \Rightarrow =& \int dxd^2\sg \pr{-\frac{1}{2}(\pd_x\pc(x))^2 + \frac{1}{2}(\pd_x\pc(x))^2(\pd_ax_0)^2 - \frac{m^2}{\bb^2}(1-\cos(\bb\pc(x)))}
    \\
    =& \int d^2\sg\frac{8m}{\bb^2}\pr{-1 + \frac{1}{2}(\pd_ax_0)^2} \equiv \Sc + S^{\rm kin}_{x_0}.
\end{split}
\end{equation}
We obtain the classical tension $\frac{8m}{\bb^2}$ and a kinetic term for the NGB, with no interactions. Two comments are in order: 
\begin{enumerate}
    \item The $1$ in the parentheses can be interpreted naturally as $1 = \frac{1}{2}+\frac{1}{2} = \frac{1}{2}(\pd_a t)^2 - \frac{1}{2}(\pd_a y)^2$ evaluated in the static gauge. Therefore, the whole contribution can be viewed as $-\frac{1}{2}(\pd_a x^\mu)^2$ in a gauge-fixed form.
    \item If we use a canonical normalization for the NGB $x_0 \to \sqrt{\frac{\bb^2}{8m}}X_0$ then the kinetic term is of order $O(\bb^0)$:
    \begin{equation}
        \frac{8m}{\bb^2}\pr{-1+\frac{1}{2}(\pd_ax_0)^2} \to -\frac{8m}{\bb^2} + \frac{1}{2}(\pd_aX_0)^2.
    \end{equation}
    Since $X_0(\sg)$ is a function of two space-time coordinates only, it is dimensionless, as a 2d scalar field should be.
\end{enumerate}

We now switch to the canonical normalization and expand the action in orders of the coupling $\bb$\footnote{Expansion in $\dph$ in the direct formalism is also an expansion in the perturbation parameter $\bb$ (or $m\bb^2$, which is dimensionless). Here, this expansion is more general than an expansion in $\dphib$.}. Note that $\pc$ comes with a $\bb^{-1}$ factor. The expansion is performed after utilizing \eqref{eq: parallel derivative identity} and shifting away all dependence on $x_0$ in the action, as in the second line in \eqref{eq_pre_kinetic_term_x0}.
We will sometimes suppress the $\sg$ dependencies and integrals.

\paragraph{Order $\bb^{0}$:}
\begin{equation}\label{eq: action to order beta0}
\begin{split}
    \frac{1}{2}(\pd_aX_0)^2 + S^{\rm kin}_{\dphib},
\end{split}
\end{equation}
where
\begin{equation}
\begin{split}
    S^{\rm kin}_{\dphib} = \frac{1}{2}(\pd\dphib)^2 - \frac{1}{2}m^2\cos(\bb\pc)\dphib^2.
\end{split}
\end{equation}
We find that the quadratic in $\dphib$ terms are the same as in the direct formalism, so the same propagator for $\dphib$ can be used.

\paragraph{Order $\bb^1$ (Excluding the cosine potential):}
At this order we find that the original kinetic term contributes two cubic vertices. One comes from keeping the linear order in $\dphib$ and utilizing the equation of motion for $\pc$:
\begin{equation}
\begin{split}
    & = \int dx \pr{-\pd_ax_0\pd_x\pc(\pd_a-\pd_ax_0\pd_x)\dphib + \text{EOM}(\pc)\dphib  + \text{total derivative}}
    \\
    & = \frac{m\bb}{4}(\pd_a X_0)^2 \int dx \sech(mx)\pd_x\dphib,
\end{split}
\end{equation} 
where in the final line we used the orthogonality of $\dphib$ to $\pd_x \pc \propto \sech(mx)$.\footnote{It removes the quadratic term $\sim x_0 \dphib$, which would otherwise have appeared at order $O(\bb^0)$).} We will denote this vertex as $\vone$. By integration by parts, it can also be written as:
\begin{equation}\label{eq: vone}
\begin{split}
    \vone \equiv &  \frac{m\bb}{4}\int d^2\sg (\pd_aX_0)^2\int dx\sech(mx)\tanh(mx)\dphib.
\end{split}    
\end{equation}

The other vertex is quadratic in $\dphib$:
\begin{equation}
\begin{split} 
    \vthree \equiv & -\frac{\bb}{\sqrt{8m}}\int d^2\sg\pd_aX_0\int dx\pd_x\dphib\pd_a\dphib.
\end{split}
\end{equation} 

\paragraph{Order $\bb^2$ (excluding the cosine potential):}
Finally, we get one quartic vertex from the kinetic term:
\begin{equation}
\begin{split}
    \vtwo \equiv & \frac{\bb^2}{16m}\int d^2\sg(\pd_aX_0)^2\int dx\pr{\pd_x\dphib}^2.
\end{split}
\end{equation}

\paragraph{The cosine potential at order $\bb^{>0}$:}
All additional terms depend only on $\dphib$, and come from expanding the classical cosine potential and its counterterm. The expansions are the same as in subsection \ref{subsubsec_interations}, except for the spurious dependence on $x_0$, which is gone due to the $x$ integration. Therefore, the result is the same, replacing $\dph\to\dphib$.

\paragraph{Counterterms}

As for the potential, the counterterms are the same as in subsection \ref{subsubsec_counterterms}, except for the spurious dependence on $x_0$, which is gone due to the $x$ integration. Therefore, the result is the same, replacing $\dph\to\dphib$.

\subsubsection{Jacobian}

In the direct formalism, the transformation was between two orthogonal bases with respect to the metric on field-space, namely from the eigenbasis of $\partial^2$, to the eigenbasis of the kinetic operator in the string sector. In contrast, the Jacobian of the transformation from $\phi(x,\sg)$ to the $X_0(\sg),\dphib(x,\sg)$ variables gives a nontrivial contribution to the action. We here summarize the results and relegate the derivation to appendix \ref{app_jacobian}. 

The contribution $S_J$ is a sequence of 1-loop counterterms involving only $\dphib$:
\begin{equation}    
    S_J =  i\sum_{n=1}^{\infty}\frac{1}{n}\left(-\frac{\bb}{4}\right)^n\de^2(0)\int d^2\sg\left(\int dx\sech(mx)\pd_x\dphib(x,\sg)\right)^n.
\end{equation}
These are one-loop in terms of their suppression by the perturbative parameter $\bb$ relative to the classical action. The UV-divergence $\de^2(0)$ is the formal expression $\int\frac{d^2q}{(2\pi)^2}$. It is roughly $\Lambda^2$, with $\Lambda$ the UV cutoff, or $a^{-2}$, with $a$ the lattice spacing. This will help cancel divergences that come from the loop:
\begin{equation}
    \langle \pd_aX_0\pd^aX_0\rangle_{\rm free} = \int\frac{d^2q}{(2\pi)^2}\frac{i}{q^2}iq_a(-iq^a) = i\de^2(0).    
\end{equation}

Altogether, the PI in this formalism is:
\begin{equation}
    \int\DD X_0(\sg)\DD\dphib(p,\sg) e^{i\pr{S^{\rm kin}_{X_0} + S^{\rm kin}_{\dphib} + \vone + \vtwo + \vthree + S^{\rm int}_{\dphib} + S^{\rm cnt}_{\dphib} + S_J }},
\end{equation}
where $S^{\rm int}_{\dphib}$ and $S^{\rm cnt}_{\dphib}$ denote the interaction terms coming from the cosine potential and its counterterm, respectively.

\subsubsection{Pseudo-Momentum Space}\label{subsubsec_pre_psu_mom}

Perturbation theory typically simplifies in momentum space due to momentum conservation. Furthermore, the kinetic term of $\dphib$ is diagonalized in pseudo-momentum space and becomes the standard scalar propagator. I.e., with the expansions:
\begin{equation}
    \dphib(x,\sg) = \int \frac{dpd^2q}{(2\pi)^3}\phi_p^*(x)e^{iq\sg}\dphib(p,q),\ \qquad\ x_0(\sg) = \int \frac{d^2q}{(2\pi)^2}e^{iq\sg}x_0(q),
\end{equation}
the kinetic terms become:
\begin{equation}
\begin{split}
    S^{\rm kin} =& \int d^2\sg\frac{1}{2}(\pd_a X_0)^2 + \int dxd^2\sg \pr{\frac{1}{2}(\pd\dphib)^2 - \frac{1}{2}m^2\cos(\bb\pc)\dphib^2}
    \\ 
    =& \int \frac{d^2q}{(2\pi)^2} q^2 X_0(q)X_0(-q) + \int \frac{dpd^2q}{(2\pi)^3}(q^2 - p^2 - m^2)\dphib(p,q)\dphib(-p,-q),
\end{split}
\end{equation}
from which the propagators for $\dphib(p,q)$ and $X_0(q)$ can be deduced:
\begin{equation}
    \frac{i}{q^2 - p^2 - m^2},\ \frac{i}{q^2}.    
\end{equation}
Note that the true integration variables in the PI are $\dphib(p,\sg)$ rather than $\dphib(x,\sg)$, since the decomposition makes manifest the constraint \eqref{eq: dphib orthogonality constraint}.

The pseudo-momentum $p$ is not conserved, due to the presence of the string, which breaks the $x$-translation symmetry. There is, however, a certain simplification; integrals naturally decompose into conserving parts and compactly-supported parts. Of course, the conserving part does not contribute to NGB interactions since they are confined to the string and cannot happen at any point in space-time. Another way to see that is from the direct formalism, where conserving parts exist only within BP interactions. The results (derived in appendix \ref{app_vertices}) are:
\begin{equation}\label{eq: v1 px-space}
\begin{split}
    \vone = \frac{\pi\bb}{8m} & \int \pr{\prod_{i=1}^3 \frac{d^2q_i}{(2\pi)^2}}(q_1\cdot q_2)(2\pi)^2\de(q_1+q_2+q_3)\cdot
    \\
    & \int \frac{dp}{2\pi}(m-ip)\sech(\frac{\pi p}{2m})X_0(q_1)X_0(q_2)\dphib(p,q_3).
\end{split}
\end{equation}
$\vtwo,\vthree$ can be decomposed as $\vtwo = \vtwoA + \vtwoB$, with:
\begin{equation}\label{eq: v2a px space}
\begin{split}
    \vtwoA = & \frac{\bb^2}{16m} \int \pr{\prod_{i=1}^4 \frac{d^2q_i}{(2\pi)^2}} (-q_1\cdot q_2)(2\pi)^2\de(q_1+q_2+q_3+q_4)\cdot
    \\
    & \int\frac{dp}{2\pi}p^2 X_0(q_1)X_0(q_2)\dphib(p,q_3)\dphib(-p,q_4),
\end{split}
\end{equation} 
\begin{equation}\label{eq: v2b px space}
\begin{split}
    \vtwoB = \frac{\bb^2}{16m} & \int \pr{\prod_{i=1}^4 \frac{d^2q_i}{(2\pi)^2}}(-q_1\cdot q_2)(2\pi)^2\de(q_1+q_2+q_3+q_4) \int\frac{dp_1}{2\pi}\frac{dp_2}{2\pi} 
    \\
    & \frac{\frac{2\pi}{3}(p_1+p_2)(p_1^2 - p_1p_2 + p_2^2 + m^2)\csch\pr{\frac{\pi(p_1+p_2)}{2m}}}{(ip_1+m)(ip_2+m)}X_0(q_1)X_0(q_2)\dphib(p_1,q_3)\dphib(p_2,q_4),
\end{split}
\end{equation} 
and $\vthree = \vthreeA + \vthreeB$, with:
\begin{equation}\label{eq: v3a px space}
    \vthreeA = \frac{1}{2}\frac{\bb}{\sqrt{8m}}\int \pr{\prod_{i=1}^3 \frac{d^2q_i}{(2\pi)^2}} (q_1\cdot(q_2-q_3))(2\pi)^2\de(q_1+q_2+q_3) \intop\frac{dp}{2\pi}\pr{ip}x_0(q_1)\dphib\pr{p,q_2}\dphib\pr{-p,q_3},
\end{equation}
\begin{equation}\label{eq: v3b px space}
\begin{split}
     \vthreeB = & \frac{i\pi}{4}\frac{\bb}{\sqrt{8m}}\int \pr{\prod_{i=1}^3 \frac{d^2q_i}{(2\pi)^2}} (q_1\cdot (q_2-q_3))(2\pi)^2\de(q_1+q_2+q_3) \intop\frac{dp_1}{2\pi}\frac{dp_2}{2\pi} \cdot 
    \\
    & \frac{\pr{p_1^{2}-p_2^{2}}}{\pr{ip_1+m}\pr{ip_2+m}}\csch\left(\frac{\pi\pr{p_1 + p_2}}{2m}\right) x_0(q_1)\dphib\pr{p_1,q_2}\dphib\pr{p_2,q_3}.
\end{split}
\end{equation} 
We can see that the vertices $\vtwoA$ and $\vthreeA$ conserve pseudo-momentum, whereas $\vone,\vtwoB$ and $\vthreeB$ are compactly supported. These three vertices come from the kinetic term, and therefore, in perturbation theory, they appear with $i$, not with $(-i)$ as usual for vertices.

We will also need the cubic vertex $\vbbb$ from $\LL^{\rm int}_{\dphib}$:
\begin{equation}
\begin{split}
    \vbbb = & -\frac{\pi m\bb}{24} \int \pr{\prod_{i=1}^3 \frac{d^2q_i}{(2\pi)^2}}(2\pi)^2\de(q_1+q_2+q_3) \int \frac{dp_1}{2\pi}\frac{dp_2}{2\pi}\frac{dp_3}{2\pi}\cdot
    \\
    & \frac{3m^4 + (-p_1+p_2+p_3)(p_1-p_2+p_3)(p_1+p_2-p_3)(p_1+p_2+p_3) + 2m^2(p_1^2+p_2^2+p_3^2)}{(ip_1+m)(ip_2+m)(ip_3+m)}\cdot
    \\
    & \sech\pr{\frac{\pi(p_1+p_2+p_3)}{2m}} \dphib(p_1,q_1)\dphib(p_2,q_2)\dphib(p_3,q_3).
\end{split}
\end{equation}
This vertex comes from the potential term, therefore, in perturbation theory it appears with a $(-i)$.

%% file: body/profile.tex
\subsection{String Width in Effective String Theory}

An important characteristic of flux tubes, say of length $R$ stretched between two electric sources, is their width. From the field theory perspective, this is usually taken to involve the profile of the electric field or some other observable (e.g., energy density) and, in particular, its decay rate as one moves away from the would-be center of the string. In EST \cite{Luscher:1980iy}, the profile is predicted to be approximately Gaussian, with a squared-width that is inversely proportional to the string tension $T$, and that grows logarithmically with the length of the string $R$, see \eqref{eq_rev_EST_width}. This is simply due to the tendency of massless scalars in $1+1$ dimensions to spread out. On the other hand, from the classical solution \eqref{eq_pre_string_solution}, one expects an exponential decay of the profile, with a width of order $1/m$ independent of the length.

A simple interpretation is that the exponential profile represents an intrinsic width, whereas the Gaussian profile represents the string's wave-function (in particular, the wave-function of the NGBs). By analogy, an electron's Compton wavelength is like its intrinsic size or radius, whereas the spread of its wave-function may depend on its circumstances (e.g. propagating, bound to an atom, etc.). The observed electron density will depend on both.

In this section, we show that -- as predicted by the above intuition, and as suggested by the form \eqref{eq_pre_expansion_snd} of the field $\phi$ -- the observed electric field profile is given by the convolution of the classical fixed-width profile with an $R$-dependent Gaussian wave-function. Since in our regime ($m^2 \ll T$) the classical width is much larger (unless the string is exponentially long), this implies that the classical profile should dominate measurements of the electric field.
As already said, instead of considering an open flux tube, for which the classical solution is difficult to obtain, a closed flux tube on a circle of length $L_y$ is considered. We freeze the zero mode of its transverse position by hand.\footnote{It does not exist for the open string due to the sources attached to its boundary.} This procedure is expected to give similar results for the width as the open flux tube case.

We will study the profile of the $y$ component of the electric field $E_y = e\pd_x\phi$ as a function of $x$. For lengths that are not exponentially long, the general behavior of this profile follows simply from the properties mentioned above. In our theory, since we control the effective action also at scales of order $m$, we can be more precise, and in the next subsection we discuss the full computation in our effective action. The computation can also be used for exponentially long strings. Euclidean signature will be used all along this section and the corresponding part of Appendix \ref{app_integrals}. 

The electric field operator is:
\bea \label{electric_field}
\pd_x\phi\pr{x,\sg}=\frac{2m}{\bb}\sech\left(m\left(x-x_0\pr{\sg}\right)\right)+\pd_x\dphib\left(x-x_0(\sg),\sg\right).
\eea
Consider the expectation value of the electric field, or of some general operator $O\pr{x-x_0(\sg)}$ constructed with $x_0$, inserted without loss of generality
at $\sg=0$. At leading order, ignoring the fluctuations of the bulk modes and keeping just the kinetic term of $x_0$, we can write\footnote{The corrections $O\pr{\bb^2}$ come from the BP and are beyond EST.}:
\bea\label{eq: convolution with Gaussian wave-function}
\left\langle O\left(x-x_0(0)\right)\right\rangle  & =\intop d\tilde{x}O(x-\tilde{x})\left\langle \de\pr{\tilde{x}-x_0(0)}\right\rangle \\
& =\intop d\tilde{x} O(x-\tilde{x})\intop\frac{dp}{2\pi}\left\langle e^{ip\pr{\tilde{x}-x_0(0)}}\right\rangle 
\\
& =\intop d\tilde{x} O(x-\tilde{x}) \intop\frac{dp}{2\pi}e^{ip\tilde{x}}\left\langle e^{-ipx_0(0)}\right\rangle 
\\
& =\intop d\tilde{x} O(x-\tilde{x}) \intop\frac{dp}{2\pi}e^{ip\tilde{x}}e^{-\frac{1}{2}p^2\pr{\frac{\bb^2}{8m}G_{x_0}\pr{0;0}}} + O\pr{\bb^2}
\\
& =\intop d\tilde{x}\sqrt{\frac{1}{2\pi}\frac{\Tc}{G_{x_0}\pr{0;0}}}e^{-\frac{1}{2}\frac{\Tc}{G_{x_0}\pr{0;0}}\tilde{x}^2}O(x-\tilde{x}) + O\pr{\bb^2},
\eea
where $G_{x_0}\pr{0;0}$ is the propagater of $x_0(0)$.\footnote{The expectation value $\left\langle e^{-ipx_0(0)}\right\rangle$ comes from the Gaussian PI with a source, in Euclidean signature, given by:
\begin{equation}
    Z[J] \equiv \int \DD\phi e^{-\frac{1}{2}\int d\vec x\phi(\vec x)\pr{-\nabla^2 + m^2}\phi(\vec x)  + i\int d\vec x J(\vec x)\phi(\vec x)} = Z[0]e^{-\frac{1}{2}\int d\vec x d\vec y J(\vec x)G(\vec x;\vec y) J(\vec y)},
\end{equation}
with $G(\vec x;\vec y) = \int \frac{d\vec p}{(2\pi)^d}\frac{e^{i\vec p(\vec x - \vec y)}}{\vec p^2 + m^2}$. In our case $m=0, \ \vec x \to \sg$, and $J(\sg) = -\frac{p\bb}{\sqrt{8m}}\de(\sg)$.} 
At leading order in perturbation theory, we get a normalized Gaussian wave-function, like in \cite{Luscher:1980iy}. However, we must deal with both IR and UV divergences in $G_{x_0}\pr{0;0}$. The IR divergences are dealt with, as we already indicated above, by fixing the state $x_0(n_y=0,t)=0$. The UV divergences
should be canceled by a more precise analysis that goes beyond EST, which we will perform in the next subsection; for now we will just illustrate the general form of the result.

If we use dimensional regularization, the divergence has the form (in the compact case $q_y = \frac{2\pi n_y}{L_y}$; the remaining momenta in the $1+\ep$ non-compact directions paralleling the worldsheet are denoted by $q$):
\bea\label{eq: x0 self energy}
G_{x_0}\pr{0;0} & =\frac{1}{L_y}\sum_{n_y\neq0}\mu^{-\ep}\intop\frac{d^{1+\ep}q}{\left(2\pi\right)^{1+\ep}}\frac{1}{q^2+\left(\frac{2\pi n_y}{L_y}\right)^2}\\
 & =\frac{2\pi^{\frac{1+\ep}{2}}}{\left(2\pi\right)^{1+\ep}\Gamma\pr{\frac{1+\ep}{2}}}2^{\ep-1}\pi^{\ep}\zeta(1-\ep)\sec\pr{\frac{\pi\ep}{2}}\left(\mu L_y\right)^{-\ep}\\
 & =-\frac{1}{2\pi\ep}+\frac{2\log\left(\mu L_y\right)+\gamma-\log(4\pi)}{4\pi}+O\pr{\ep}.
\eea
In any cutoff regularization, we would thus get a Gaussian width squared growing with $\log\pr{L_y}$,
\bea\label{eq: Luscher log Lambda L}
\frac{\log\left(M L_y\right)}{2\pi \Tc}+\dots,
\eea
like in \cite{Luscher:1980iy}, where $M$ is some cutoff above the scale where the EST breaks down. A priori, we do not know in our theory whether this will be of order $m$ or of order $\sqrt{T}$, but this will not affect the leading order behavior (and will be determined by a more precise computation in the next subsection). 

At leading order in $\bb$, the profile is given by convolving the
classical value with the wavefunction:
\bea
\left\langle \pd_x\phi(x)\right\rangle =\intop d\tilde{x}\sqrt{\la_M}e^{-\pi\la_M\tilde{x}^2}\left(\frac{2m}{\bb}\sech(m(x-\tilde{x}))\right) + O\pr{\bb},
\eea
where $\lambda_M \equiv \Tc / \log(M L_y)$. We see that when $L_y$ is not exponentially long, $\la_M$ is large compared to the inverse classical width squared $m^2$, and the Gaussian wave-function is narrow compared to the classical wave-function. The integral can be approximated by a saddle point analysis near $\tilde{x}=0$, so the $\sech$ is just expanded at small $\tilde{x}$, with successive terms being indeed suppressed by the Gaussian measure against which they are integrated. This coincides with the naive small $\bb$ expansion, where there is no need to define the coupling $\la_M$ in the first place because there are no problematic soft modes. Thus, in this case, we find the profile is the classical exponentially decaying $\sech(mx)+\text{small corrections}$. 

\begin{figure}
\captionsetup{singlelinecheck = false, justification=justified}
\includegraphics[scale=0.9]{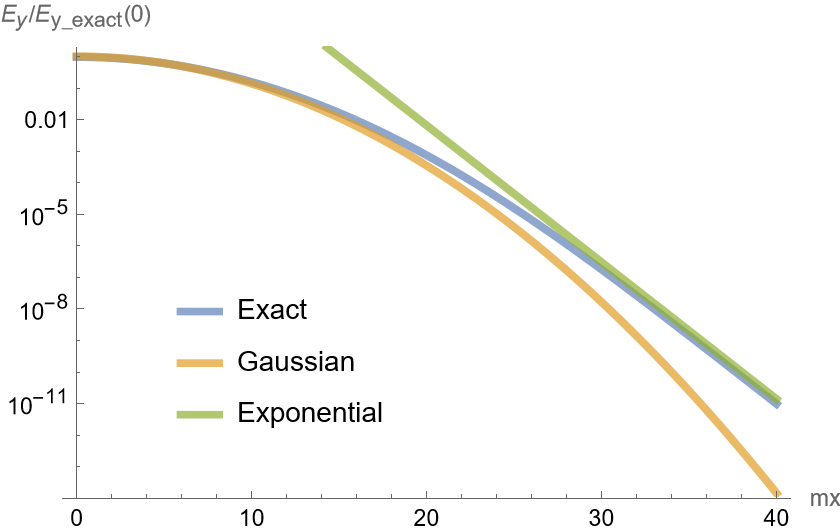}
\caption{The electric field as a function of the distance from the string, for $m=1$ and $\la_M = 6.35\cdot 10^{-3}$ (blue). One can see the transition from Gaussian (orange) to exponential (green) profile.}
\label{fig: e-field profile}
\end{figure}

When $L_y$ is exponentially long, such that $\la_M$ is small compared to $m^2$, but $x$ is not ``too large'', the saddle is $\tilde{x} \approx x$. We can expand the Gaussian around that point, with successive terms suppressed by $\la_M$. So, the profile is now a Gaussian, with variance $\sim\log (L_y)$, plus small corrections. In this case, the observed width of the profile reflects the quantum uncertainty in the string's position. However, this is invalidated when $x\gtrsim\frac{m}{2\pi\la_M}$, since then the Gaussian's fast decay is outlasted by the fast -- but still slower -- exponential decay. This is the signature of massive bulk
propagation. The saddle point equation is:
\begin{equation}
\begin{split}
    &0 = \pd_{\tilde{x}}\log\pr{\sech(m(x-\tilde{x}))e^{-\pi\la_M\tilde{x}^2}}
    \\
    \Rightarrow & \tanh\left(m\pr{\tilde{x}-x}\right) = -\frac{2\pi\la_M}{m}\tilde{x}.
\end{split}
\end{equation}
Hence, at large $x \gtrsim \frac{m}{2\pi\la_M}$, the Gaussian term on the right-hand side becomes important if one assumes as before $\tilde{x}\approx x$. Instead, the saddle is independent of $x$ and located at $\tilde{x} \sim \frac{m}{2\pi\la_M} \lesssim x$, and the profile is exponential again:
\begin{equation}
    \sim e^{-m\pr{x-\frac{m}{2\pi\la_M}} - \la_M\pr{\frac{m}{2\pi\la_M}}^2} = e^{-m\pr{x -\frac{m}{4\pi\la_M}}}.    
\end{equation}
The bottom line is that the Gaussian profile is only apparent when $\frac{m}{2\pi\la_M}\gg \frac{1}{m}$, hence when $L_y\gg\frac{1}{M} e^{\frac{2\pi \Tc}{m^2}}$, and otherwise we have an exponential profile with a width independent of $L_y$. The resulting leading contribution to the electric field profile, summarizing this discussion, is plotted in figure \ref{fig: e-field profile}.

In \cite{Caselle:2016mqu}, it was observed that an interpolation is needed between the exponential and Gaussian regimes, the latter of which is supposed to appear at a sufficiently long string length. The above computation provides an analytic solution to this problem in the context of QED$_3$. We note that the profile computed on the lattice in \cite{Caselle:2016mqu} cannot be directly compared with the computation here because it is observed at a distance from the string where it becomes sensitive to the edges of the string. In contrast, here we considered a closed string winding around a compact direction. Furthermore, the analysis in \cite{Caselle:2016mqu} was conducted in the regime $m\sim 1/a$ (with $a$ the lattice spacing), where the effective theory we employ here is invalid. 

It is sometimes claimed that the Gaussian profile should be observed at lengths $L_y>1/m$. This is true in theories with a single length scale, but we see that it is incorrect when $m^2 \ll T$. It was observed in \cite{Caselle:2016mqu} that the profile is not Gaussian. This fact was ascribed there to the presence, in the framework of EST, of higher-derivative terms, specifically the extrinsic curvature, which gives the string a rigidity. We showed here that even without taking into account this subleading correction, the Gaussian profile can be observed only in the \textbf{exponentially long} string regime, so lattice simulations (in theories with $m^2 \ll T$) are expected to see an exponential profile with a width independent of $L_y$. Note that this exponential hierarchy is possible because of the separation between the string tension scale $T$ and the UV cutoff scale $m^2$ for the validity of the EST (which governs the classical width of the string), a separation that is present in QED$_3$ but not, say, in Yang-Mills theory.

\subsection{Divergence Cancellation}

To cancel more carefully the UV divergence in the Gaussian width, we must utilize the fact that it arises from high energy ``hard'' modes that should be amenable to perturbation theory in $\bb$, even though the full range of energies contributing to \eqref{eq: Luscher log Lambda L} is not. Thus, we must introduce some scale $M$, which probably should be around $M\sim m$, that splits the modes of $x_0$ into ``soft'' and ``hard'', and expand-out only the ``hard'' modes perturbatively. Before we continue, it is convenient to Fourier transform the field \footnote{Note that this Fourier decomposition is distinct from the mode decomposition we used in the previous section for the bulk field. The latter was parameterized by the pseudo-momentum $p$ and the wave-functions $\phi_p(x)$, while here it is the usual transform with plane waves.}:
\bea
\phi(p,\sg) & =\intop dxe^{-ipx}\left(\frac{4}{\bb}\arctan\pr{e^{m(x-x_0(\sg))}}+\dphib\left(x-x_0(\sg),\sg\right)\right)\\
 & =e^{-ipx_0(\sg)}\intop dxe^{-ipx}\pr{\frac{4}{\bb}\arctan\pr{e^{mx}}+\dphib\pr{x,\sg}},
\eea
so that all the $x_0$ dependence is in the form of an exponential
``vertex operator''. The Fourier transform of the electric field \eqref{electric_field} is then:
\begin{equation}\label{electric_field_Fourier}
(\pd_x\phi)(p,\sg)=e^{-ipx_0(\sg)}\left(\frac{2\pi}{\bb}\sech\left(\frac{\pi p}{2m}\right)+ip\intop dxe^{-ipx}\dphib\pr{x,\sg}\right).
\end{equation}

We now propose a particular way to split the modes into ``soft'' and ``hard'' modes. We defined above:
\bea
\la_M\equiv\frac{8m}{\bb^2\log\pr{ML_y}}
\Rightarrow L_y=\frac{1}{M}\exp\pr{\frac{8m}{\la_M\bb^2}},
\eea
and we furthermore define:
\bea
\frac{1}{\ep_M}\equiv\frac{1}{\ep}-\left(2\log\pr{\mu/M}+\gamma-\log(4\pi)\right).
\eea
We will expand only in $\bb$, and not in $\la_M$, although
in the exponentially long string regime, $\la_M$ is also small.
We can reexpress \eqref{eq: x0 self energy} as:
\bea
G_{x_0}\pr{0;0} & =-\frac{1}{2\pi\ep}+\frac{2\log\left(\mu L_y\right)+\gamma-\log(4\pi)}{4\pi}
\\
& = -\frac{1}{2\pi\ep_M}+\frac{\log\pr{ML_y}}{2\pi}\\
& =-\frac{1}{2\pi\ep_M}+\frac{8m}{\bb^2}\frac{1}{2\pi\lambda_M},
\eea
separating it into a finite IR contribution and a diverging but $\bb^2$-suppressed UV contribution.
For vertex operators, we define a kind of normal-ordering:
\bea\label{eq: normal ordered vertex op times divergence}
e^{-ipx_0(\sg)} & =\underbrace{e^{-ipx_0(\sg)-\frac{1}{2}\frac{\bb^2}{8m}\frac{1}{2\pi\ep_M}p^2}}_{\equiv:e^{-ipx_0(\sg)}:_M}e^{\frac{1}{2}\frac{\bb^2}{8m}\frac{1}{2\pi\ep_M}p^2},
\eea
and expand the last factor to orders in $\bb^2$. For each order, we expect the UV divergence to be canceled. Turning now to the electric field \eqref{electric_field_Fourier}, at leading order $O\pr{\bb^{-1}}$ one finds:
\beq
\left\langle (\pd_x\phi)(p,\sg)\right\rangle _{\bb^{-1}}=\left\langle :e^{-ipx_0(\sg)}:_M\right\rangle _{\bb^{0}}\frac{2\pi}{\bb}\sech\left(\frac{\pi p}{2m}\right),
\eeq
where:
\beq
\left\langle :e^{-ipx_0(\sg)}:_M\right\rangle _{\bb^{0}}=e^{-\frac{1}{2}\frac{p^2}{2\pi\lambda_M}}
\eeq
is essentially the Gaussian wave-function of the string in momentum space, like the fourth line in \eqref{eq: convolution with Gaussian wave-function}, but with the divergence removed.

We demonstrate the cancellation of the divergence in \eqref{eq: normal ordered vertex op times divergence} at first nontrivial order in appendix \ref{app_integrals}. We find there (see \eqref{eq: NLO e-field profile}):
\bea
\left\langle (\pd_x\phi)(p,\sg)\right\rangle _{\bb} & =\left\langle :e^{-ipx_0(\sg)}:_M\right\rangle _{\bb^{0}}\frac{\pi\bb}{8m}\sech\left(\frac{\pi p}{2m}\right)p^2\pr{\frac{1}{2\pi}\log\pr{\frac{m}{2M}}+O\pr{L_y^{-1}}}+\cdots.
\eea
This suggests that a natural choice is $M=\frac{m}{2}$. We see that choices that far exceed the scale of $m$ will cause an enhancement of sub-leading terms in the expansion, thereby invalidating it. This means that $M\sim m$ gives the correct width of the Gaussian wave-function, with small perturbative corrections.

%% file: body/ground.tex
In this section, we will find the leading (1-loop, $O(\bb^0)$) quantum correction to the ground state energy $\frac{8m}{\bb^2}L_y$ of the string. For the infinite string, this is really a correction to the string tension, as the energy is proportional to $L_y$. For the finite string, we will find the Casimir energy $-\frac{\pi}{6L_y}$ of a massless scalar, and then additional exponentially-decaying contributions due to the massive BP.

We need the first correction $\de m_2^2$ in the coefficient $\de m^2=\de m_2^2\bb^2 + O\pr{\bb^4}$ of the counterterm. It will be fixed as usual by implementing the physical mass condition for the BP in the vacuum. The first correction to the mass of order $\bb^2$ comes from the quartic vertex $-\frac{m^2\bb^2}{4!}\dph^4$ and the quadratic counterterm $\frac{\de m_2^2\bb^2}{2!}\dph^2$. The 1-loop 1PI propagator is:
\begin{equation}
    -i\Sigma(p^2) = -i\frac{-m^2\bb^2}{2}\int \frac{d^3k}{(2\pi)^3}\frac{i}{k^2 - m^2 + i\ep} + i(p^2\bb^2\de z_2^2 -\bb^2\de m_2^2).
\end{equation}
Demanding:
\begin{equation}
    \Sigma(m^2) = 0, \ \frac{d\Sigma(p^2)}{dp^2}\bigg|_{p^2 = m^2} = 0,
\end{equation}
results in:
\begin{equation}
    \de z_2^2 = 0, \ \de m_2^2 = \frac{m^2}{2}\int \frac{d^3k}{(2\pi)^3}\frac{i}{k^2 - m^2 + i\ep}.
\end{equation}
For the integral we Wick-rotate $k_0 \to ik_0$, and therefore:
\begin{equation}\label{eq_ground_cntr_gen}
    \de m_2^2 = \frac{m^2}{2}\int \frac{d^3k}{(2\pi)^3}\frac{i}{k^2 - m^2 + i\ep} = 
    \frac{m^2}{2}\int \frac{d^3k}{(2\pi)^3}\frac{1}{k^2 + m^2}.
\end{equation}
The result depends on the regularization implemented in each subsection and will be presented there.

\subsection{PI Framework}

In the PI framework, the ground-state energy $E_{\rm gs}$ is extracted as the limit $\lim_{L_t\to\infty} \frac{F}{L_t}$, where $F$ is the (Euclidean) free energy and $L_t$ is the length of the time direction. The contribution to the ground state energy $\de E_{\rm gs}$ at order $\bb^0$ comes from the 1-loop determinants of both $X_0$ and $\dphib$ (in the stringy formalism), or $\dph_{\rm NGB}$ and $\dph_{\rm BP}$ (in the direct formalism), and from counterterms; the cosmological constant counterterm $\de\La = \de\La_0 + O\pr{\bb^2}$
and the potential counterterm $\frac{\de m^2}{\bb^{2}}\left(1-\cos\pr{\bb\pc}\right)$. This calculation is identical in the two formalisms. Collecting all of the contributions:
\begin{equation}\label{eq_ground_Egs}
\begin{split}
    \de E_{\rm gs} = & \underbrace{\frac{1}{2}L_y\intop\frac{d^{2}q}{\left(2\pi\right)^{2}}\log\pr{q^{2}}}_{X_0}+\underbrace{\frac{1}{2}L_y\intop dx\intop\frac{d^2q dp}{\left(2\pi\right)^3}\log\pr{q^{2}+p^{2}+m^{2}}\frac{p^{2}+m^{2}\tanh^{2}(mx)}{p^{2}+m^{2}}}_{\dphib}
    \\
    & \qquad + \underbrace{L_yL_x\de\La_0}_{\text{cosmological counterterm}}+\underbrace{\de m_{2}^{2}L_y\intop dx\left(1-\cos\left(\phi_{{\rm cl}}\right)\right)}_{\text{potential counterterm}}
    \\
    = & \underbrace{L_y\left(\frac{1}{2}\intop\frac{d^{2}q}{\left(2\pi\right)^{2}}\log\pr{q^{2}}-\intop\frac{d^2q dp}{\left(2\pi\right)^3}\log\pr{q^{2}+p^2+m^{2}}\frac{m}{p^{2}+m^{2}}+\frac{4\de m_{2}^{2}}{m}\right)}_{\de E_{\rm string}}
    \\
    & \qquad + \underbrace{\frac{1}{2}L_yL_x\intop\frac{d^3q}{\left(2\pi\right)^3}\log\pr{q^{2}+m^{2}}+L_yL_x\de\La_0}_{E_{{\rm vac}}}.
\end{split}    
\end{equation}
The energy is separated into a part proportional to $L_x$, coming from the bulk and containing the vacuum energy (which is fixed to some value by the choice of $\de\Lambda_0$ in the noncompact limit $L_y\to\infty$), and a part depending only on $L_y$, which is identified as the string's energy. We will be interested in the latter (or, equivalently, in the difference between the energy of the string ground state and the vacuum).

One possible regularization scheme in this framework is dimensional regularization (``dim-reg''), which can be consistently used to regulate both the vacuum and the string states. But we see that specifically for integrals like 1-loop determinants, with logarithms, dim-reg is not helpful (there is no ``gap'' of values for $\ep$ that is free of both IR and UV divergences). We can add another ad-hoc UV cutoff $\La$, such as an exponential decay, on top of dim-reg. It is ad-hoc because it applies covariantly only to the BP, while for the NGB it suppresses only the parallel momenta. We first take $\La\to\infty$ and then analytically continue the dimensional regulator $\ep$. This way, whenever dim-reg is enough by itself (which probably will be most of the time), it alone will be used. In the current calculation, dim-reg alone is enough for the string energy, at least after carrying out the $p$ integral and combining the two integrands. The counterterm $\de m_2^2$ in this regularization, using \eqref{eq_ground_cntr_gen}, is independent of the regulator:
\begin{equation}
    \de m_2^2 = \frac{m^2}{2}\frac{1}{(4\pi)^{\frac{3}{2}}}\frac{\Gamma\pr{1-\frac{3}{2}}}{\Gamma(1)}\pr{\frac{1}{m}}^{1-\frac{3}{2}} = -\frac{m^3}{8\pi}.
\end{equation}

We now compactify by taking $q_{y}\to\frac{2\pi n}{L_y}$ and $L_y\intop\frac{dq_{y}}{2\pi}\to\sum_{n}$. For the 1-loop string energy, we find in appendix \ref{app_integrals} (see \eqref{eq: appendix E string final result}):
\begin{equation}\label{eq_ground_final}
    \de E_{\rm string}=-\frac{m^{2}L_y}{4\pi}-\frac{\pi}{6L_y}+\frac{1}{\pi L_y}\text{Li}_{2}\pr{e^{-mL_y}}.
\end{equation}


Three things to note about this result:
\begin{enumerate}
    \item The tension correction is $-\frac{m^{2}}{4\pi}$. The tension is expanded with the perturbation parameter $m\bb^2$:
    \begin{equation}
        T = \frac{8m}{\bb^2} - \frac{m^{2}}{4\pi} + O(\bb^2) = \frac{8m}{\bb^2}\pr{1 - \frac{m\bb^2}{32\pi} + O((m\bb^2)^2)}.
    \end{equation}
    A physical tension can be used to set a renormalization condition for $\bb$ and determine $\de\bb$. This counterterm will not be used in this work. 

    \item The Casimir part is the result expected from one ($3-2=1$) massless scalar on a circle:
    \bea
    -\frac{\pi}{6L_y} = \frac{2\pi}{L_y}\left(-\frac{\pr{3-2}}{12}\right).
    \eea

    \item At large $L_y$:
    \bea
    \frac{1}{\pi L_y}\text{Li}_{2}\pr{e^{-mL_y}} = \frac{1}{\pi L_y}e^{-mL_y} + O\pr{e^{-2mL_y}},
    \eea
    which decays exponentially, as expected from the contribution of a massive particle. It is non-perturbative in $1/mL_y$. 

\end{enumerate}
    
$E_{\rm gs}$ can be expanded perturbatively in $m\bb^2$, or, equivalently, expressing $\bb$ as a series in $\frac{T}{m^2}$, in $1/T L_y^2$. It is expected that every energy level, to any order in perturbation theory, when expressed as a function of $TL_y^2$ and $m L_y$, should agree with the EST expression at large $m L_y$ as a function of $TL_y^2$ (given by \eqref{string_energy_levels} plus higher order corrections starting at order $1/L_y^7$), and contain also additional contributions non-perturbative in $1/mL_y$, as in \eqref{eq_ground_final}.

\begin{figure}
    \captionsetup{singlelinecheck = false, justification=justified}
    \includegraphics[scale=0.9]{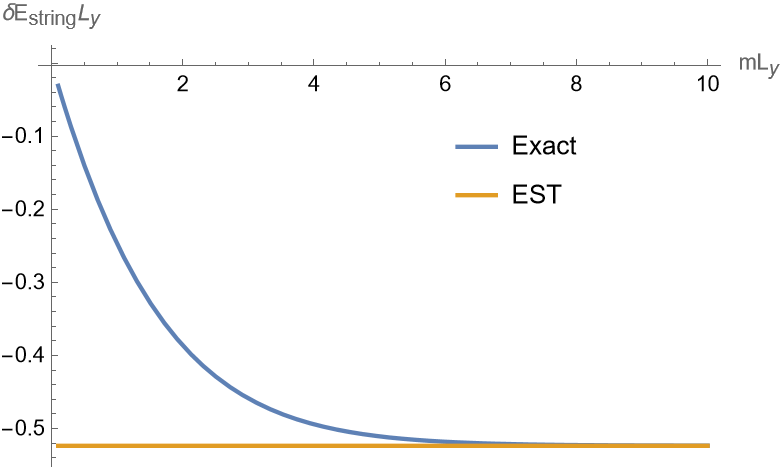}
    \caption{Ground-state energy of the string (in units of the string's length) as a function of the string's length (in units of the BP mass), compared to the EST result.}
    \label{fig_ground_energy}
\end{figure}

For $mL_y\gg 1$ one can ignore $\frac{1}{\pi L_y}\text{Li}_{2}\pr{e^{-mL_y}}$, and the EST result is recovered. However, for $mL_y \sim 1$, this term becomes important. Since $m\bb^2 \sim \frac{m^2}{T} \ll 1$ for perturbation theory to be valid, and also $TL_y^2\gg 1$ for the string description to be relevant, the region $mL_y \sim 1$ can be in the domain of validity only if the separation of scales between $m^2$ and $T$ is large enough. For example, for $\frac{m^2}{T} = 10^{-4}$, our approximations are valid for $0.1 \lesssim mL_y$. The ground state energy at one-loop order (without the contribution from the tension) is plotted in figure \ref{fig_ground_energy}, and compared to the EST result that arises at large $m$.

\subsection{Canonical Framework}\label{subsec_ground_can}

Canonical quantization suggests different regularization and renormalization schemes than dim-reg used in the PI framework. It will be convenient to use a sharp cutoff $\La$ for the $x$-momentum. For the $y$-axis, since we also consider finite $L_y$, using a sharp cutoff does not eliminate all UV divergences. Therefore, we use a smooth version with a weight factor of $e^{-s|q_y|}$. There is no integral of $q_t$, and the one needed for the counter-term \eqref{eq_ground_cntr_gen} is convergent. In this regularization, it reads:
\begin{equation}\label{eq_ground_cntr_can}
    \de m_2^2 = \frac{m^2}{2}\int_{-\infty}^\infty \frac{dq_y}{2\pi} e^{-s|q_y|}\int_{-\La}^\La \frac{dp}{2\pi}\int_{-\infty}^\infty \frac{dq_t}{2\pi}\frac{1}{p^2 + q_y^2 + q_t^2 + m^2} = \frac{m^2}{4}\pr{\frac{1}{\pi^2 s}(\g + \log(2\La s)) - \frac{m}{2\pi}}.
\end{equation}
This regularization is not usual, and it might not be suitable when comparing results from the vacuum and the string states. However, we will see that it indeed gives the same result as \eqref{eq_ground_final}. The calculation follows similar lines to the correction of the soliton mass in the $1+1$d Sine-Gordon model \cite{Rajaraman:1982is}. The calculation first appeared in \cite{Dashen:1975hd}, based on the same methods used for the kink \cite{Dashen:1974cj}. We explain the subtleties by first recovering the $1+1$d result $m_{\rm soliton} = \frac{8m}{\bb^2} - \frac{m}{\pi} + O(\bb^2)$\footnote{$\bb$ is dimensionless in $1+1$d.}, and then we will continue to the $2+1$d case. In $1+1$d the mass counterterm takes the form $\de m_{2,1d}^2 = -\frac{m^2\bb^2}{8\pi}\log\pr{\frac{m^2}{4\La^2}}$. 

The naive expression for the $1+1$d energy correction is given by the zero-point energy, with the two counterterms:
\begin{equation}
    E_{{\rm gs},1d} = \frac{8m}{\bb^2} + \frac{1}{2}\cdot 0 + \frac{1}{2} \int dp\pr{\frac{L_x}{2\pi} + \frac{1}{2\pi}\frac{d\de(p)}{dp}}\sqrt{p^2+m^2} + 
    \frac{4}{m\bb^2}\de m_{2,1d}^2 + 
    L_x\de\La_{0,1d} + O(\bb^2).
\end{equation}
The vacuum energy should be subtracted from this equation:
\begin{align}
    E_{{\rm vac},1d} &= \frac{1}{2}\int dp\frac{L_x}{2\pi}\sqrt{p^2+m^2} + L_x\de\La_{0,1d} + O(\bb^2),
    \\
    m_{\rm soliton} &= E_{{\rm gs},1d} - E_{{\rm vac},1d} \equiv \frac{8m}{\bb^2} + \de m_{\rm soliton},
    \\
    \de m_{\rm soliton} &= \int \frac{dp}{2\pi}\frac{d\de(p)}{dp}\sqrt{p^2+m^2}
    + \frac{4}{m\bb^2}\de m_{2,1d}^2. 
    \label{eq_ground_dem_naive}
\end{align}
However, this method does not yield the right result, nor does it eliminate all of the divergences appearing in $2+1$d. In \cite{Dashen:1974cj}, they used a finite but very large $L_x$, such that the modes are discrete. Since $L_x$ is very large, the modes are approximately the same as in the continuum, with the phase shifts \eqref{eq_pre_pos_phase},\eqref{eq_pre_neg_phase} dictating the allowed momenta ${p'}_n$ by \eqref{eq_pre_den_states}. Then, the integrals turn into discrete sums:
\begin{equation}\label{eq_ground_dem_dis}
    \de m_{\rm soliton} = \frac{1}{2}\sum_{n\in \mathbb{Z}\setminus\{0\}}^\infty \pr{\sqrt{{p'}_n^2+m^2} - \sqrt{p_n^2+m^2}} + \frac{1}{2}(0 - m) 
    + \frac{4}{m\bb^2}\de m_{2,1d}^2,
\end{equation}
where $p_n = \frac{2\pi n}{L_x}$ are the allowed momenta in the vacuum. For all $n\neq 0$, the momenta in the soliton sector are related to the momenta in the vacuum sector by ${p'}_n \approx p_n - \frac{\de(p_n)}{L_x}$. There is no $p=0$ eigenvalue in the soliton sector; the bulk scalar has no standing wave. The massive $p=0$ eigenvalue in the vacuum is mapped into the massless NGB in the presence of the soliton. Hence, the $n=0$ state is the bulk scalar at rest in the vacuum with energy $m$ and the NGB in the one-soliton sector with zero energy, and we treat it separately. For any function $f$:
\begin{equation}
    \sum_{n=1}^\infty (f({p'}_n) - f(p_n)) \approx \sum_{n=1}^\infty \frac{df(p)}{dp}\Bigg|_{p=p_n}\pr{-\frac{\de(p_n)}{L_x}} \to -\int\frac{dp}{2\pi}\frac{df(p)}{dp}\de(p),
\end{equation}
where we change to a simple integral since the density of states is constant in the vacuum. By integration by parts, this relation can be modified:
\begin{equation}\label{eq_ground_sum_int}
    \sum_{n=1}^\infty (f({p'}_n) - f(p_n)) \to \int_0^\infty \frac{dp}{2\pi}\frac{d\de(p)}{dp}f(p) + \frac{1}{2\pi}f(p)\de(p)|^0_\infty.
\end{equation}
A couple of things to note:
\begin{enumerate}
    \item The derivative $\frac{d\de(p)}{dp}$ is not sensitive to the discontinuity of $\de(p)$ at $p=0$. However, here, we need the boundary value of $\de(p)$. Therefore, we considered half of the sum. A similar formula holds for the sum from $-\infty$ to $-1$. 
    \item The summation is from $n=1$ and does not include $n=0$.
    \item \eqref{eq_ground_sum_int} is almost the same as the naive subtraction in \eqref{eq_ground_dem_naive}, the only difference is the boundary terms.
\end{enumerate}
Using \eqref{eq_ground_sum_int} for the sum inside \eqref{eq_ground_dem_dis} gives:
\begin{equation}
\begin{split}
    \frac{1}{2}\sum_{n\in\mathbb{Z}\setminus\{0\}} \pr{\sqrt{{p'}_n^2+m^2} - \sqrt{p_n^2+m^2}} 
    &\to \frac{1}{2}\pr{2\int_0^\infty \frac{dp}{2\pi} \pr{\frac{d\de(p)}{dp}\sqrt{p_n^2+m^2}} + 2\frac{1}{2\pi}\sqrt{p_n^2+m^2}\de(p)|^0_\infty} 
    \\
    &\to \frac{m}{2\pi}\log\pr{\frac{m^2}{4\La^2}} -\frac{m}{\pi} + \frac{m}{2},
\end{split}    
\end{equation}
where we implemented the sharp cutoff. Putting all parts together we obtain:
\begin{equation}
    \de m_{\rm soliton} = \pr{\frac{m}{2\pi}\log\pr{\frac{m^2}{4\La^2}} -\frac{m}{\pi} + \frac{m}{2}} -\frac{m}{2} + \frac{4}{m\bb^2}\pr{-\frac{m^2\bb^2}{8\pi}\log\pr{\frac{m^2}{4\La^2}}} = -\frac{m}{\pi},
\end{equation}
which is the desired result.

Now, we move to $2+1$d. The analog of \eqref{eq_ground_dem_dis} is:
\begin{equation}
\begin{split}
    \de E_{\rm string} = &\frac{1}{2}\sum_{l=-\infty}^\infty \sum_{n\in \mathbb{Z}\setminus\{0\}}^\infty  \pr{\sqrt{q_l^2 + {p'}_n^2 + m^2} - \sqrt{q_l^2 + p_n^2 + m^2}} 
    \\
    + &\frac{1}{2}\sum_{l=-\infty}^\infty \pr{|q_l| - \sqrt{q_l^2 + m^2}} + \frac{4}{m\bb^2}\de m_2^2.
\end{split}    
\end{equation}
Repeating the same steps as before for each $y$-momentum $q_l = \frac{2\pi l}{L_y}$ yields:
\begin{equation}\label{eq_ground_deE_dis}
\begin{split}
    \de E_{\rm string} = \sum_{l = -\infty}^\infty e^{-s|q_l|} \Bigg(& \frac{q_l\arctan\left(\frac{m^2-q_l^2}{2mq_l}\right) - q_l\arctan\left(\frac{m}{q_l}\right)}{\pi} + \frac{m}{2\pi}\log \left(\frac{m^2+q_l^2}{4\La^2}\right) 
    \\
    &+ \frac{\pi\sqrt{m^2+q_l^2}-2m}{2\pi} + \frac{1}{2}\pr{|q_l| - \sqrt{m^2+q_l^2}} \Bigg) + \frac{4}{m\bb^2}\de m_2^2L_y,
\end{split}    
\end{equation}
where we implemented the $y$-momentum regulator. The first two terms in the sum come from the integral, the third term from the boundary terms, and the fourth term is the $n=0$ contribution. The last term is the counterterm contribution. Using the Euler-Maclaurin formula, it can be seen that the counterterm is canceled by the integral of the $\log$, and the string tension correction and the Casimir energy of the massless NGB are obtained. These are the perturbative terms in $mL_y$, and indeed, the Euler-Maclaurin series is a series in $mL_y$. To obtain the non-perturbative part one needs to evaluate the sums exactly. This is done in appendix \ref{app_integrals}. The final result is: 
\begin{equation}
    \de E_{\rm string} = -\frac{m^2}{4\pi}L_y - \frac{\pi}{6L_y} + \frac{1}{\pi L_y}\text{Li}_2\pr{e^{-mL_y}},
\end{equation}
as in \eqref{eq_ground_final}.



%% file: body/scattering.tex
In this section, the simplest scattering amplitudes along the string are calculated, those of two NGBs (a single NGB cannot decay by momentum conservation). For simplicity we discuss the case of an infinite string, avoiding the complications of recoil of the string when BPs are emitted.

The calculation can be done in each one of the formalisms used so far. However, while the BP interactions appear the same in both formalisms, the NGB interaction vertices are fewer in the stringy formalism. There are only three of them, rather than an infinite number in the direct formalism. Moreover, they lead to simpler integrals. Hence, we will use the stringy formalism. In appendix \ref{app_scattering}, we calculate the two simple processes of 2NGB $\to$ 1BP and 2NGB $\to$ 2NGB in the direct formalism and verify that the results are the same as presented here.

The three vertices in the stringy formalism correspond to three processes that can happen at tree level, involving NGBs and BPs:
\begin{enumerate}
    \item Two NGBs to one bulk-photon.
    \item Two NGBs to two bulk-photons.
    \item One NGB to two bulk-photons. This process is prohibited by kinematics since the BP is massive and the NGBs are massless.
\end{enumerate}
The second process is suppressed with respect to the first one by $\bb$. Still, it is kinematically favored, because the measure of the outgoing particles' phase space is non-zero, unlike the first process. Two additional processes of order $\bb^2$ are:
\begin{enumerate}
    \item Two NGBs to two NGBs.
    \item Two NGBs to one NGB and one BP.
\end{enumerate}

To be careful with this theory, which violates $x$-momentum conservation and involves fields ``living'' in different dimensions, the scattering amplitudes will be derived from the correlators (with the canonical variable $X_0$) using the LSZ reduction formula. In the correlation function, we use general offshell momenta, and for the residue at the poles, onshell momenta are inserted. For the process $\{k_1,k_2\}$ into some out state, $\{-k_1,-k_2\}$ are used in the correlator. In this section, we will use the center of mass frame. Putting the momenta of the two NGBs onshell gives $k_1^2 = k_2^2 = 0,\ k_1\cdot k_2 = \frac{s}{2}$ where, as usual, $s \equiv (k_1 + k_2)^2$ is the total energy squared. The other two Mandelstam variables are also used with two-momentum only: $t \equiv (k_1 - k_3)^2,\ u \equiv (k_1 - k_4)^2$. We will denote the spatial part of $k$ as usual by $\vec k$, although it has only one component.

Reducing the usual derivation of the total cross section from the scattering amplitude from $3+1$d to $1+1$d yields the same formula (see, e.g., \cite{Peskin:1995ev} pp.102-106, and the final result (4.79)). In $1+1$d, the total cross section is just a probability. The integration over the final state is different for a NGB, with $\frac{d\vec k}{2\pi}\frac{1}{2|\vec k|}$, and a BP, with $\frac{d\vec kdp}{(2\pi)^2}\frac{1}{2\sqrt{\vec k^2 + p^2 + m^2}}$. The prefactor $\frac{1}{2E_12E_2}\frac{1}{|v_1-v_2|}$ equals $\frac{1}{2s}$ for scattering of two NGBs.

In the following, we draw the Feynman diagrams for the momentum-space correlation functions calculated. Dashed lines denote NGBs, while solid lines are BPs. The momentum direction depends on a convention; we choose the following convention:
\begin{equation*}
    \wick{\c X_0(q_1) \c X_0(q_2)} =\quad 
    \begin{tikzpicture}[baseline=(b.base)]
    \begin{feynhand}
    \vertex (a) [dot,label=$X_0(q_1)$] at (-2,0) {}; 
    \vertex (b) [dot,label=$X_0(q_2)$] at (2,0) {}; 
    \propag [scalar, mom={$q_2$}] (a) to (b); 
    \end{feynhand}
    \end{tikzpicture}
    =\quad 
    \begin{tikzpicture}[baseline=(b.base)]
    \begin{feynhand}
    \vertex (a) [dot,label=$X_0(q_1)$] at (-2,0) {}; 
    \vertex (b) [dot,label=$X_0(q_2)$] at (2,0) {}; 
    \propag [scalar, revmom={$q_1=-q_2$}] (a) to (b); 
    \end{feynhand}
    \end{tikzpicture}
\end{equation*}
and similarly for BPs.

\subsection{2 NGB to 1 BP}

The three-point function $\langle X_0(-k_1)X_0(-k_2)\dphib(p,k_3)\rangle$ gets a tree-level contribution only from one diagram with one insertion of $\vone$:
\begin{equation}\label{eq_sctr_2NGB1BP_cor}
\begin{split}
    &\langle X_0(-k_1)X_0(-k_2)\dphib(p,k_3)\rangle_{\rm tree} =
    \vcenter{\hbox{
    \begin{tikzpicture}[baseline=(c.base)]
    \begin{feynhand}
    \vertex (a) at (-1.4,-1.4) {}; 
    \vertex (b) at (1.4,-1.4) {}; 
    \vertex (c) [dot,label=right:{$\vone$}] at (0,0) {};
    \vertex (d) at (0,1.8) {};
    \propag [scalar, mom={$k_1$}] (a) to (c);
    \propag [scalar, mom={$k_2$}] (b) to (c); 
    \propag [plain, mom={$(p,k_3)$}] (c) to (d);
    \end{feynhand}
    \end{tikzpicture}
    }}
    \\
    &=\frac{i}{k_1^2}\frac{i}{k_2^2}\frac{i}{k_3^2 - p^2 - m^2}(2\pi)^2\de(-k_1-k_2+k_3) \frac{i\pi\bb}{4m}(k_1\cdot k_2)(m+ip){\rm sech}\left(\frac{\pi p}{2m}\right).
\end{split}    
\end{equation}
Putting the momenta onshell and implementing the scattering kinematics yield $(k_3,p)=(\sqrt{p^2 + m^2},0,p)$ with $p^2+m^2 = s$. Then:
\begin{equation}\label{eq_sctr_2NGB1BP_sctr}
    i\MM(\vec k_1,\vec k_2\to (p,\vec k_3)) = \frac{i\pi\bb s}{8m}(m\pm i\sqrt{s-m^2}){\rm sech}\left(\frac{\pi \sqrt{s-m^2}}{2m}\right),
\end{equation}
depending on the sign of $p$. The total cross section is given by:
\begin{equation}\label{eq_sctr_tot_1BP}
\begin{split}
    \sg_{\rm 1BP} &= \frac{1}{2s}\int\frac{d\vec k_3dp}{(2\pi)^2 2\sqrt{\vec k_3^2+m^2+p^2}} \left|\MM(\vec k_1,\vec k_2\to (p,\vec k_3))\right|^2 (2\pi)^2 \de(\vec k_3)\de\left(\sqrt{\vec k_3^2+m^2+p^2}-\sqrt{s}\right) 
    \\
    &= \frac{\pi^2\bb^2 s^2}{128m^2\sqrt{s-m^2}}\sech\left(\frac{\pi\sqrt{s-m^2}}{2m}\right)^2.
\end{split}    
\end{equation}

\subsection{2 NGB to 2 NGB}

The scattering amplitude of two NGBs to two NGBs can be deduced from the four-point function $\langle X_0(-k_1)X_0(-k_2)X_0(k_3)X_0(k_4)\rangle$. It gets contributions from diagrams with two $\vone$ vertices, 
and splits into $s,t,u$ channels:
\begin{equation}\label{eq_sctr_2GB2GB_cor}
\begin{split}
    &\langle X_0(-k_1)X_0(-k_2)  X_0(k_3)X_0(k_4)\rangle_{\rm tree} = 
    \\
    & \vcenter{\hbox{
    \begin{tikzpicture}[baseline=(c.base)]
    \begin{feynhand}
    \vertex (a) at (-1.4,-1.4) {}; 
    \vertex (b) at (1.4,-1.4) {}; 
    \vertex (c) [dot,label=right:{$\vone$}] at (0,0) {};
    \vertex (d) [dot,label=right:{$\vone$}] at (0,1.8) {};
    \vertex (e) at (-1.4,3.2) {};
    \vertex (f) at (1.4,3.2) {};
    \propag [scalar, mom={$k_1$}] (a) to (c);
    \propag [scalar, mom={$k_2$}] (b) to (c); 
    \propag [plain, mom={$(p,k_1+k_2)$}] (c) to (d);
    \propag [scalar, mom={$k_3$}] (d) to (e);
    \propag [scalar, mom={$k_4$}] (d) to (f); 
    \end{feynhand}
    \end{tikzpicture}
    }}
    + 
    \vcenter{\hbox{
    \begin{tikzpicture}[baseline=(c.base)]
    \begin{feynhand}
    \vertex (a) at (-1.4,-1.4) {}; 
    \vertex (b) at (-1.4,1.4) {}; 
    \vertex (c) [dot,label=below:{$\vone$}] at (0,0) {};
    \vertex (d) [dot,label=below:{$\vone$}] at (1.8,0) {};
    \vertex (e) at (3.2,-1.4) {};
    \vertex (f) at (3.2,1.4) {};
    \propag [scalar, mom={$k_1$}] (a) to (c);
    \propag [scalar, mom={$k_3$}] (c) to (b); 
    \propag [plain, mom={$(p,k_1-k_3)$}] (c) to (d);
    \propag [scalar, mom'={$k_2$}] (e) to (d);
    \propag [scalar, mom'={$k_4$}] (d) to (f); 
    \end{feynhand}
    \end{tikzpicture}
    }}
    + 
    \vcenter{\hbox{
    \begin{tikzpicture}[baseline=(c.base)]
    \begin{feynhand}
    \vertex (a) at (-1.4,-1.4) {}; 
    \vertex (b) at (-1.4,1.4) {}; 
    \vertex (c) [dot,label=below:{$\vone$}] at (0,0) {};
    \vertex (d) [dot,label=below:{$\vone$}] at (1.8,0) {};
    \vertex (e) at (3.2,-1.4) {};
    \vertex (f) at (3.2,1.4) {};
    \propag [scalar, mom={$k_1$}] (a) to (c);
    \propag [scalar, mom={$k_4$}] (c) to (b); 
    \propag [plain, mom={$(p,k_1-k_4)$}] (c) to (d);
    \propag [scalar, mom'={$k_2$}] (e) to (d);
    \propag [scalar, mom'={$k_3$}] (d) to (f); 
    \end{feynhand}
    \end{tikzpicture}
    }}
    \\
    & = \frac{i}{k_1^2}\frac{i}{k_2^2}\frac{i}{k_3^2}\frac{i}{k_4^2}(2\pi)^2\de(k_1 + k_2 - k_3 - k_4)\pr{-\frac{i\pi^2\bb^2}{16m^2}}(k_1\cdot k_2)(k_3\cdot k_4)\cdot
    \\
    & \quad\quad\quad \int\frac{dp}{2\pi}\frac{m^2 + p^2}{(k_1 + k_2)^2 - p^2 - m^2 + i\epsilon}{\rm sech}\pr{\frac{\pi p}{2m}}^2
    + (k_1\leftrightarrow -k_3) + (k_2\leftrightarrow -k_3),
\end{split}    
\end{equation}
where the $t$-channel is related to the $s$-channel by $-k_2\leftrightarrow k_3$, and the $u$-channel by $-k_1\leftrightarrow k_3$. The integral is evaluated in appendix \ref{app_integrals}:
\begin{equation}
    I_{\rm 2NGB}(s,m) \equiv \int\frac{dp}{2\pi}\frac{m^2 + p^2}{s - p^2 - m^2 +i\epsilon}{\rm sech}\pr{\frac{\pi p}{2m}}^2 = -\frac{m}{\pi^2}\pr{2 + \frac{s}{m\sqrt{-s+m^2}}\PolyG_1\pr{\frac{1}{2}+\frac{\sqrt{-s+m^2}}{2m}} },
\end{equation}
where $\PolyG_1$ is the Polygamma function. The square root $\sqrt{-s+m^2}$ takes the negative imaginary root for $s>m^2$ (see appendix \ref{app_integrals}). Using onshell momenta gives 
$k_1\cdot k_2 = k_3\cdot k_4 = \frac{s}{2}$. Without loss of generality we can take $t=0$ and $u=-s$, and the final result is:
\begin{equation}\label{eq_sctr_2GB2GB_sctr}
\begin{split}
    i\MM(\vec k_1,&\vec k_2 \to \vec k_3,\vec k_4) = i\bb^2\pr{\frac{s^2}{16m} + \frac{s^3}{64m^2}\frac{\PolyG_1\pr{\frac{1}{2}+\frac{\sqrt{-s+m^2}}{2m}}}{\sqrt{-s+m^2}} - \frac{s^3}{64m^2}\frac{\PolyG_1\pr{\frac{1}{2}+\frac{\sqrt{s+m^2}}{2m}}}{\sqrt{s+m^2}} }. 
\end{split}
\end{equation}

$\MM(\vec k_1,\vec k_2 \to \vec k_3,\vec k_4)$ has an imaginary part for $s > m^2$ as expected. This comes from the creation of a BP, and thus should be related to the amplitude of 2NGB $\to$ 1BP by the optical theorem. The integral for the optical theorem and the result are given in \eqref{eq_sctr_tot_1BP} (up to the $1/(2s)$ factor). Note that unlike the standard situation in QFTs where there are integrals over $d$-dimensional momenta and a $d+1$-dimensional $\de$-function, here the $x$-momentum does not have a $\de$-function conservation. This also enables the appearance of a branch-cut in $2\to 2$ scattering. The imaginary part of \eqref{eq_sctr_2GB2GB_sctr} comes from the $s$-channel solely for $s>m^2$:
\begin{equation}\label{eq_sctr_opt_3}
\begin{split}
    2\Im(\MM(\vec k_1,\vec k_2 \to \vec k_1,\vec k_2)) &= 2\Im\pr{\frac{\bb^2s^3}{64m^2}\frac{\PolyG_1\pr{\frac{1}{2}-i\frac{\sqrt{s-m^2}}{2m}}}{-i\sqrt{s-m^2}}} 
    \\
    &= \frac{\bb^2s^3}{64m^2\sqrt{s-m^2}}2\Re\pr{\PolyG_1\pr{\frac{1}{2}-i\frac{\sqrt{s-m^2}}{2m}}}.
\end{split}    
\end{equation}
For evaluating the real part of $\PolyG_1$ we use the Polygamma reflection formula:
\begin{equation}
    \PolyG_n(z) - (-1)^n\PolyG_n(1-z) = -\pi\frac{d^n}{dz^n}\cot(\pi z)
\end{equation}
with $n=1$ and $z=\frac{1}{2} - i\frac{\sqrt{s-m^2}}{2m}$, which gives:
\begin{equation}\label{eq_sctr_opt_4}
    2\Re\pr{\PolyG_1\pr{\frac{1}{2}-i\frac{\sqrt{s-m^2}}{2m}}} = \frac{\pi^2}{\cosh^2\pr{\frac{\pi\sqrt{s-m^2}}{2m}}}.
\end{equation}
Inserting \eqref{eq_sctr_opt_4} into \eqref{eq_sctr_opt_3} reproduces \eqref{eq_sctr_tot_1BP}, up to a factor $2s$ present in the total cross section and not in the optical theorem.

Another point that should be addressed is the comparison to EST. Among all of the processes considered in this section, this is the only one existing in EST, as it involves only NGBs. Therefore, it can be compared to the EST result and should agree with it for $\sqrt{s}\ll m$. Indeed, the scattering amplitude for 2NGB $\to$ 2NGB is (see (24) in \cite{Dubovsky:2012sh}, where $i=j=k=l$ and $t+u=-s$):
\begin{equation}
    \MM^{\rm EST}_{\rm tree}(\vec k_1,\vec k_2\to \vec k_3,\vec k_4) = \frac{1}{2}\ell_s^2s^2,
\end{equation}
which is the leading term of \eqref{eq_sctr_2GB2GB_sctr}, with $\frac{\bb^2}{16m} = \frac{1}{2\Tc} = \frac{\ell_s^2}{2}$. The other terms are suppressed by the factor $\frac{\sqrt{s}}{m}$ for $\sqrt{s}\ll m$, as expected.

The total cross section is given by:
\begin{equation}\label{eq_sctr_tot_2GB}
\begin{split}
    \sg_{\rm 2GB} &= \frac{1}{2s}\int\frac{d\vec k_3}{2\pi(2|\vec k_3|)}\frac{d\vec k_4}{2\pi(2|\vec k_4|)} \left|\MM(\vec k_1,\vec k_2\to \vec k_3,\vec k_4)\right|^2 (2\pi)^2 \de(\vec k_3 + \vec k_4)\de\left(|\vec k_3|+|\vec k_4|-\sqrt{s}\right) 
    \\
    &= \frac{1}{4s^2}\left|\MM(\vec k_1,\vec k_2\to \vec k_1,\vec k_2)\right|^2.
\end{split}    
\end{equation}

\subsection{2 NGB to 1 NGB and 1 BP}\label{subsec_sctr_1NGB1BP}

The scattering amplitude of two NGBs to one NGB and one BP can be computed from the correlator $\langle X_0(-k_1)X_0(-k_2)X_0(k_3)\dphib(p,k_4)\rangle$. The leading contribution comes from the insertion of $\vone$ and $\vthree$ vertices in three different channels; each diagram splits to $x$-momentum conserving / violating parts due to $\vthreeA,\vthreeB$. The correlation function is:
\begin{equation}
\begin{split}
    & \langle X_0(-k_1)X_0(-k_2)X_0(k_3)\dphib(p,k_4)\rangle_{\rm tree} = 
    \\
    &
    \vcenter{\hbox{
    \begin{tikzpicture}[baseline=(c.base)]
    \begin{feynhand}
    \vertex (a) at (-1.4,-1.4) {}; 
    \vertex (b) at (1.4,-1.4) {}; 
    \vertex (c) [dot,label=right:{$\vone$}] at (0,0) {};
    \vertex (d) [dot,label=right:{$\vthree$}] at (0,1.8) {};
    \vertex (e) at (-1.4,3.2) {};
    \vertex (f) at (1.4,3.2) {};
    \propag [scalar, mom={$k_1$}] (a) to (c);
    \propag [scalar, mom={$k_2$}] (b) to (c); 
    \propag [plain, mom={$(p',k_1+k_2)$}] (c) to (d);
    \propag [scalar, mom={$k_3$}] (d) to (e);
    \propag [plain, mom={$(p,k_4)$}] (d) to (f); 
    \end{feynhand}
    \end{tikzpicture}
    }}
    + 
    \vcenter{\hbox{
    \begin{tikzpicture}[baseline=(c.base)]
    \begin{feynhand}
    \vertex (a) at (-1.4,-1.4) {}; 
    \vertex (b) at (-1.4,1.4) {}; 
    \vertex (c) [dot,label=below:{$\vone$}] at (0,0) {};
    \vertex (d) [dot,label=below:{$\vthree$}] at (1.8,0) {};
    \vertex (e) at (3.2,-1.4) {};
    \vertex (f) at (3.2,1.4) {};
    \propag [scalar, mom={$k_1$}] (a) to (c);
    \propag [scalar, mom={$k_3$}] (c) to (b); 
    \propag [plain, mom={$(p',k_1-k_3)$}] (c) to (d);
    \propag [scalar, mom'={$k_2$}] (e) to (d);
    \propag [plain, mom'={$(p,k_4)$}] (d) to (f); 
    \end{feynhand}
    \end{tikzpicture}
    }}
    + 
    \vcenter{\hbox{
    \begin{tikzpicture}[baseline=(c.base)]
    \begin{feynhand}
    \vertex (a) at (-1.4,-1.4) {}; 
    \vertex (b) at (-1.4,1.4) {}; 
    \vertex (c) [dot,label=below:{$\vthree$}] at (0,0) {};
    \vertex (d) [dot,label=below:{$\vone$}] at (1.8,0) {};
    \vertex (e) at (3.2,-1.4) {};
    \vertex (f) at (3.2,1.4) {};
    \propag [scalar, mom={$k_1$}] (a) to (c);
    \propag [plain, mom={$(p,k_4)$}] (c) to (b); 
    \propag [plain, mom={$(p',k_1-k_4)$}] (c) to (d);
    \propag [scalar, mom'={$k_2$}] (e) to (d);
    \propag [scalar, mom'={$k_3$}] (d) to (f); 
    \end{feynhand}
    \end{tikzpicture}
    }}
    \\
    &=\frac{i}{k_1^2}\frac{i}{k_2^2}\frac{i}{k_3^2}\frac{i}{k_4^2 - p^2 - m^2}(2\pi)^2\de(-k_1-k_2+k_3+k_4)\cdot
    \\
    &\quad \Bigg(\frac{i\pi\bb^2}{(8m)^{\frac{3}{2}}}(k_1\cdot k_2)(k_3^2+2(k_3\cdot k_4))p(m+ip){\rm sech}\left(\frac{\pi p}{2m}\right)\frac{i}{(k_1+k_2)^2 - p^2 - m^2}
    \\
    &\quad\quad\quad -\frac{\pi^2\bb^2}{2(8m)^\frac{3}{2}}(k_1\cdot k_2)\frac{k_3^2 + 2(k_3\cdot k_4)}{ip-m}\int\frac{dp'}{2\pi}{\csch}\pr{\frac{\pi(p+p')}{2m}}{\sech}\pr{\frac{\pi p'}{2m}}\frac{p'^2-p^2}{(k_1+k_2)^2-p'^2-m^2+i\ep}
    \Bigg)
    \\
    + & (k_1\leftrightarrow -k_3) + (k_2\leftrightarrow -k_3),
\end{split}
\end{equation}
where the second row comes from $\vthreeA$ and the third row from $\vthreeB$. The integral is evaluated in appendix \ref{app_integrals}:
\begin{equation}
\begin{split}
    I_{\rm 1NGB1BP}(s,m,p)  \equiv & \int\frac{dp'}{2\pi}{\csch}\pr{\frac{\pi(p+p')}{2m}}{\sech}\pr{\frac{\pi p'}{2m}}\frac{p'^2-p^2}{s-p'^2-m^2+i\ep} = 
    \\
    & \frac{-2p\left(\sqrt{-s+m^2}-m\right)+i \left(-s+m^2+p^2\right) \left(H_{-\frac{ip}{2m} + \frac{\sqrt{-s+m^2}}{2m}} - H_{\frac{ip}{2m}+\frac{\sqrt{-s+m^2}}{2m}}\right)}{2\pi\sqrt{-s+m^2}},
\end{split}
\end{equation}
where $H_n$ is the (analytic continuation of the) harmonic number $n$. We put the momenta onshell, and use the kinematics $k_1\cdot k_2 = \frac{s}{2}$, $k_3\cdot k_4 = \frac{s-p^2-m^2}{2}$, and similar expressions for the other multiplications. Energy-Momentum conservation gives $|\vec k_3| = |\vec k_4| = \frac{s-p^2-m^2}{2\sqrt{s}}$ and $\sqrt{\vec k_4^2 + p^2 + m^2} = \frac{s+p^2+m^2}{2\sqrt{s}}$, while $t=p^2+m^2-s$ or $0$ and $u=p^2+m^2-s-t$.\footnote{The value of $t$ is easy to see using $(k_1-k_3)^2$, since both are lightlike. However, it is also equal to $(k_2-k_4)^2$ as can be easily checked, although $k_4$ is timelike.}

The scattering amplitude is given by:
\begin{equation}\label{eq_sctr_amp_1NP1NGB}
\begin{split}
    i \MM(\vec k_1,\vec k_2 \to \vec k_3, (p,\vec k_4)) = & -\frac{\pi\bb^2 (m+ip)}{64\sqrt{2}(m)^{\frac{3}{2}}}\Bigg(2p\left(m^2+p^2\right) \sech\left(\frac{\pi p}{2 m}\right)
    \\
    & + \pi s\left(1-\frac{s}{m^2+p^2}\right) \left(I_{\rm 1NGB1BP}\left(m^2+p^2-s,m,p\right) + I_{\rm 1NGB1BP}(s,m,p)\right)\Bigg).
\end{split}
\end{equation}
The total cross section is given by:
\begin{equation}\label{eq_sctr_tot_21GB1BP}
\begin{split}
    \sg_{\rm 1NGB + 1BP} =& \frac{1}{2s}\int\frac{d\vec k_3}{2\pi(2|\vec k_3|)}\frac{d\vec k_4dp}{2\pi\pr{2\sqrt{\vec k_4^2 + p^2 + m^2}}}\cdot 
    \\
    &\left|\MM(\vec k_1,\vec k_2\to \vec k_3,(p,\vec k_4))\right|^2 (2\pi)^2 \de(\vec k_3 + \vec k_4)\de\left(|\vec k_3|+\sqrt{\vec k_4^2 + p^2 + m^2} -\sqrt{s}\right) 
    \\
    = &\frac{1}{2\pi s}\int_0^{\sqrt{s-m^2}}dp\frac{\left|\MM(\vec k_1,\vec k_2 \to \vec k_3, (p,\vec k_4))\right|^2}{s-p^2-m^2}.
\end{split}    
\end{equation}
This integral cannot be calculated analytically; but in any case such a computation does not make sense because the integral suffers from an IR divergence, when the $x$-momentum of the BP $p$ is close to $\sqrt{s-m^2}$, and the momentum $k_3$ of the NGB approaches zero. Naively one does not expect such a divergence to appear, since the NGB interactions come with positive momentum powers. These powers should cancel the measure factor $\frac{d\vec k_3}{2|\vec k_3|}$. However, the BP internal propagator in the $s$-channel of the insertion of the momentum-conserving part $\vthreeA$ becomes onshell when $k_3\to 0$, which makes $\MM(\vec k_1,\vec k_2\to \vec 0,(p,\vec 0))$ finite and $\sg_{\rm 1NGB + 1BP}$ divergent. 

Usually, an IR divergence in theories of massless particles such as massless QED indicates that the external states were not identified correctly, and that external states of massive particles need to be dressed with `clouds' of soft massless particles. In our case, we believe that this is not the correct interpretation. One reason is that there are no IR divergences in one-loop diagrams involving the same process without the external soft NGB, that can cancel the divergence in the tree-level cross section. Another reason is that the one-loop correction to the BP propagator does not suffer from an IR divergence, so there is no indication that the propagation of this state needs to be corrected.

Instead, we believe that the correct interpretation is that our formalism does not correctly take into account the recoil of the string when it absorbs or emits particles carrying transverse momentum. Such a recoil is guaranteed by momentum conservation; naively one would think that it can be ignored for an infinite string (which can absorb finite momentum while remaining at zero velocity) but it seems that this is not the case. Since the recoil involves turning on low-momentum NGBs, it apparently leads to an IR divergence when we expand around the wrong final state which does not take this into account.

As a consistency check on our interpretation, we can show that IR divergences of this type appear only when the total transverse momentum of the emitted BPs is non-zero, such that the string indeed recoils. Consider a general diagram in which some number of bulk particles are emitted, and they in turn emit (from their external lines) NGBs. In the corresponding cross section computation, we have, as in \eqref{eq_sctr_tot_21GB1BP}, an integral $\frac{d\vec{k}_i}{2|\vec{k}_i|}$ over the momentum of each such NGB. By itself this would give an IR divergence. The coupling in the $\vthree$ vertex contains an extra power of $k_i$ that cancels this divergence. However, we can still get a divergence, if in the $k_i \to 0$ limit the extra internal BP propagator attached to this vertex becomes onshell. Since the external state is onshell, this happens only when the momentum $p$ of this internal propagator is the same as that of the external particle, namely, it arises (as in the computation above) when we use the \vthreeA\ vertex and not the \vthreeB\ vertex. Using the form of the \vthreeA\ vertex, one can show that for each NGB that is emitted from a leg corresponding to a BP with transverse momentum $p_j$, the logarithmic IR divergence in the $\vec{k_i} \to 0$ limit is proportional to $p_j$ (with a coefficient that does not depend on which external leg the NGB comes from). Thus, the IR divergences cancel if and only if the external BPs obey $\sum p_j = 0$ (counting incoming and outgoing particles with opposite sign). This is precisely the condition for the amplitude to conserve the transverse momentum, namely to have no recoil of the string. Details of this argument are given in appendix \ref{app_IR}.

It would be interesting to properly take into account the recoil, by expanding around incoming and outgoing states with a general transverse momentum of the string, and to confirm that this indeed cancels the IR divergences we find.



\subsection{2 NGB to 2 BP}

Finally, the scattering amplitude of two NGBs to two BPs can be computed from the four-point function $\langle X_0(-k_1)X_0(-k_2)\dphib(p_1,k_3)\dphib(p_2,k_4)\rangle$. It gets contributions from four diagrams (two of which have two different channels).

The first diagram involves the four-point vertex $\vtwo$, and it splits into two parts due to $\vtwoA,\vtwoB$, giving:
\begin{equation}
\begin{split}
    & \langle X_0(k_1)X_0(k_2)\dphib(p_1,k_3)\dphib(p_2,k_4)\rangle_{\rm tree,1} = 
    \\
    &
    \vcenter{\hbox{
    \begin{tikzpicture}[baseline=(c.base)]
    \begin{feynhand}
    \vertex (a) at (-1.4,-1.4) {}; 
    \vertex (b) at (1.4,-1.4) {}; 
    \vertex (c) [dot,label=right:{$\vtwo$}] at (0,0) {};
    \vertex (d) at (-1.4,1.4) {};
    \vertex (e) at (1.4,1.4) {};
    \propag [scalar, mom={$k_1$}] (a) to (c);
    \propag [scalar, mom={$k_2$}] (b) to (c); 
    \propag [plain, mom={$(p_1,k_3)$}] (c) to (d);
    \propag [plain, mom={$(p_2,k_4)$}] (c) to (e); 
    \end{feynhand}
    \end{tikzpicture}
    }}
    \\
    & =\frac{i}{k_1^2}\frac{i}{k_2^2}\frac{i}{k_3^2 - p_1^2 - m^2}\frac{i}{k_4^2 - p_2^2 - m^2}(2\pi)^3\de(k_1 + k_2 - k_3 - k_4)\de(p_1 + p_2)\frac{i\bb^2}{4m}(k_1\cdot k_2)p_1p_2
    \\
    & + \frac{i}{k_1^2}\frac{i}{k_2^2}\frac{i}{k_3^2 - p_1^2 - m^2}\frac{i}{k_4^2 - p_2^2 - m^2}(2\pi)^2\de(k_1 + k_2 - k_3 - k_4)\cdot
    \\
    & \quad\quad \pr{-\frac{i\pi\bb^2}{6m}}(k_1\cdot k_2)\frac{(p_1+p_2)(p_1^2 -p_1p_2 + p_2^2 + m^2)}{(ip_1 - m)(ip_2 - m)}{\rm csch}\pr{\frac{\pi(p_1 + p_2)}{2m}}.
\end{split}
\end{equation}

The second diagram, with an intermediate NGB, comes from two $\vone$'s 
in the $t,u$-channels:
\begin{equation}
\begin{split}
    & \langle X_0(k_1) X_0(k_2)\dphib(p_1,k_3)\dphib(p_2,k_4)\rangle_{\rm tree,2} = 
    \\
    &
    \vcenter{\hbox{
    \begin{tikzpicture}[baseline=(c.base)]
    \begin{feynhand}
    \vertex (a) at (-1.4,-1.4) {}; 
    \vertex (b) at (-1.4,1.4) {}; 
    \vertex (c) [dot,label=below:{$\vone$}] at (0,0) {};
    \vertex (d) [dot,label=below:{$\vone$}] at (1.8,0) {};
    \vertex (e) at (3.2,-1.4) {};
    \vertex (f) at (3.2,1.4) {};
    \propag [scalar, mom={$k_1$}] (a) to (c);
    \propag [plain, mom={$(p_1,k_3)$}] (c) to (b); 
    \propag [scalar, mom={$k_1-k_3$}] (c) to (d);
    \propag [scalar, mom'={$k_2$}] (e) to (d);
    \propag [plain, mom'={$(p_2,k_4)$}] (d) to (f); 
    \end{feynhand}
    \end{tikzpicture}
    }}
    + 
    \vcenter{\hbox{
    \begin{tikzpicture}[baseline=(c.base)]
    \begin{feynhand}
    \vertex (a) at (-1.4,-1.4) {}; 
    \vertex (b) at (-1.4,1.4) {}; 
    \vertex (c) [dot,label=below:{$\vone$}] at (0,0) {};
    \vertex (d) [dot,label=below:{$\vone$}] at (1.8,0) {};
    \vertex (e) at (3.2,-1.4) {};
    \vertex (f) at (3.2,1.4) {};
    \propag [scalar, mom={$k_1$}] (a) to (c);
    \propag [plain, mom={$(p_2,k_4)$}] (c) to (b); 
    \propag [scalar, mom={$k_1-k_4$}] (c) to (d);
    \propag [scalar, mom'={$k_2$}] (e) to (d);
    \propag [plain, mom'={$(p_1,k_3)$}] (d) to (f); 
    \end{feynhand}
    \end{tikzpicture}
    }}
    \\
    & = \frac{i}{k_1^2}\frac{i}{k_2^2}\frac{i}{k_3^2 - p_1^2 - m^2}\frac{i}{k_4^2 - p_2^2 - m^2}(2\pi)^2\de(k_1 + k_2 - k_3 - k_4)\cdot
    \\
    & \quad\quad \pr{-\frac{i\pi^2\bb^2}{16m^2}}\frac{(k_1\cdot(k_3-k_1))(k_2\cdot(k_4-k_2))}{(k_3-k_1)^2}(ip_1 + m)(ip_2 + m){\rm sech}\pr{\frac{\pi p_1}{2m}}{\rm sech}\pr{\frac{\pi p_2}{2m}}
    \\
    &+ ((k_3,p_1) \leftrightarrow (k_4,p_2)).
\end{split}
\end{equation}

The third diagram, with an intermediate BP, comes from two $\vthree$'s 
in the $t,u$-channels, which splits into three due to $\vthreeA,\vthreeB$:
\begin{equation}
\begin{split}
    & \langle X_0(k_1)  X_0(k_2)\dphib(p_1,k_3)\dphib(p_2,k_4)\rangle_{\rm tree,3} = 
    \\
    &
    \vcenter{\hbox{
    \begin{tikzpicture}[baseline=(c.base)]
    \begin{feynhand}
    \vertex (a) at (-1.4,-1.4) {}; 
    \vertex (b) at (-1.4,1.4) {}; 
    \vertex (c) [dot,label=below:{$\vthree$}] at (0,0) {};
    \vertex (d) [dot,label=below:{$\vthree$}] at (1.8,0) {};
    \vertex (e) at (3.2,-1.4) {};
    \vertex (f) at (3.2,1.4) {};
    \propag [scalar, mom={$k_1$}] (a) to (c);
    \propag [plain, mom={$(p_1,k_3)$}] (c) to (b); 
    \propag [plain, mom={$(p',k_1-k_3)$}] (c) to (d);
    \propag [scalar, mom'={$k_2$}] (e) to (d);
    \propag [plain, mom'={$(p_2,k_4)$}] (d) to (f); 
    \end{feynhand}
    \end{tikzpicture}
    }}
    + 
    \vcenter{\hbox{
    \begin{tikzpicture}[baseline=(c.base)]
    \begin{feynhand}
    \vertex (a) at (-1.4,-1.4) {}; 
    \vertex (b) at (-1.4,1.4) {}; 
    \vertex (c) [dot,label=below:{$\vthree$}] at (0,0) {};
    \vertex (d) [dot,label=below:{$\vthree$}] at (1.8,0) {};
    \vertex (e) at (3.2,-1.4) {};
    \vertex (f) at (3.2,1.4) {};
    \propag [scalar, mom={$k_1$}] (a) to (c);
    \propag [plain, mom={$(p_2,k_4)$}] (c) to (b); 
    \propag [plain, mom={$(p',k_1-k_4)$}] (c) to (d);
    \propag [scalar, mom'={$k_2$}] (e) to (d);
    \propag [plain, mom'={$(p_1,k_3)$}] (d) to (f); 
    \end{feynhand}
    \end{tikzpicture}
    }}
    \\
    & = \frac{i}{k_1^2}\frac{i}{k_2^2}\frac{i}{k_3^2 - p_1^2 - m^2}\frac{i}{k_4^2 - p_2^2 - m^2}(2\pi)^3\de(k_1 + k_2 - k_3 - k_4)\de(p_1 + p_2)\cdot
    \\
    & \quad\quad\frac{i\bb^2}{8m}\frac{(k_1\cdot(2k_3-k_1))(k_2\cdot(2k_4-k_2))p_1p_2}{(k_3-k_1)^2 - p_1^2 - m^2 }
    \\
    & + ((k_3,p_1) \leftrightarrow (k_4,p_2))
    \\
    & + \frac{i}{k_1^2}\frac{i}{k_2^2}\frac{i}{k_3^2 - p_1^2 - m^2}\frac{i}{k_4^2 - p_2^2 - m^2}(2\pi)^2\de(k_1 + k_2 - k_3 - k_4)\cdot
    \\
    & \quad\quad \frac{i\pi^2\bb^2}{32m}\frac{(k_1\cdot(2k_3-k_1))(k_2\cdot(2k_4-k_2))}{(ip_1-m)(ip_2-m)}\cdot
    \\
    & \quad\quad \int\frac{dp}{2\pi}\frac{1}{(k_3-k_1)^2 - p^2 - m^2 + i\ep}\frac{(p_1^2-p^2)(p_2^2-p^2)}{p^2 + m^2}{\rm csch}\pr{\frac{\pi(p_1-p)}{2m}}{\rm csch}\pr{\frac{\pi(p_2+p)}{2m}}
    \\
    & + ((k_3,p_1) \leftrightarrow (k_4,p_2))
    \\
    & + \frac{i}{k_1^2}\frac{i}{k_2^2}\frac{i}{k_3^2 - p_1^2 - m^2}\frac{i}{k_4^2 - p_2^2 - m^2}(2\pi)^2\de(k_1 + k_2 - k_3 - k_4)\cdot
    \\
    & \quad\quad \frac{i\pi\bb^2}{16m}\frac{(k_1\cdot(2k_3-k_1))(k_2\cdot(2k_4-k_2))}{(k_3-k_1)^2 - p_1^2 - m^2}\frac{p_1(p_2^2-p_1^2)}{(ip_1-m)(ip_2-m)}\cdot{\rm csch}\pr{\frac{\pi(p_1+p_2)}{2m}}
    \\
    & + ((k_3,p_1) \leftrightarrow (k_4,p_2)) + (k_1 \leftrightarrow k_2) + (k_1 \leftrightarrow k_2,(k_3,p_1) \leftrightarrow (k_4,p_2)).
\end{split}
\end{equation}
The first two summands come from the same vertex ($\vthreeA,\vthreeB$); therefore, there are two different options when changing the momenta. The third summand comes from two different vertices (one $\vthreeA$ and one $\vthreeB$); therefore, there are four different options. The integral is evaluated in appendix \ref{app_integrals}:
\begin{equation}
\begin{split}
    & I^2_{\rm 2BP}(t,m,p_1,p_2) \equiv \int\frac{dp}{2\pi}\frac{1}{t - p^2 - m^2 + i\ep}\frac{(p_1^2-p^2)(p_2^2-p^2)}{p^2 + m^2}{\rm csch}\pr{\frac{\pi(p_1-p)}{2m}}{\rm csch}\pr{\frac{\pi(p_2+p)}{2m}} = 
    \\
    & \csch\left(\frac{\pi(p_1+p_2)}{2 m}\right) \Biggl(
    \frac{i\left(m^2+p_1^2\right) \left(m^2+p_2^2\right) \left(-H_{\frac{m+i p_1}{2 m}}+H_{\frac{m-i p_1}{2 m}}-H_{\frac{m+i p_2}{2 m}}+H_{\frac{m-i p_2}{2 m}}\right)}{2 \pi m t} 
    \\
    &- 
    \frac{i\left(m^2+p_1^2-t\right) \left(m^2+p_2^2-t\right)\left(-H_{\frac{ip_1}{2m}+\frac{\sqrt{-t+m^2}}{2m}}+H_{\frac{\sqrt{-t+m^2}}{2m}-\frac{i p_1}{2 m}}-H_{\frac{i p_2}{2 m}+\frac{\sqrt{-t+m^2}}{2m}}+H_{\frac{\sqrt{-t+m^2}}{2m}-\frac{i p_2}{2 m}}\right)}{2\pi t \sqrt{-t+m^2}}
    \\
    &
    +\frac{(p_1+p_2) \left(m^2+p_1 p_2-t \right)}{\pi t}\Biggl(1 -
    \frac{m}{\sqrt{-t+m^2}}\Biggr)
    \Biggr).
\end{split}
\end{equation}

The fourth diagram comes from insertions of $\vone$ and $\vbbb$:
\begin{equation}
\begin{split}
    & \langle X_0(k_1)X_0(k_2)\dphib(p_1,k_3)\dphib(p_2,k_4)\rangle_{\rm tree,4} = 
    \\
    &
    \vcenter{\hbox{
    \begin{tikzpicture}[baseline=(c.base)]
    \begin{feynhand}
    \vertex (a) at (-1.4,-1.4) {}; 
    \vertex (b) at (1.4,-1.4) {}; 
    \vertex (c) [dot,label=right:{$\vone$}] at (0,0) {};
    \vertex (d) [dot,label=right:{$\vbbb$}] at (0,1.8) {};
    \vertex (e) at (-1.4,3.2) {};
    \vertex (f) at (1.4,3.2) {};
    \propag [scalar, mom={$k_1$}] (a) to (c);
    \propag [scalar, mom={$k_2$}] (b) to (c); 
    \propag [plain, mom={$(p',k_1+k_2)$}] (c) to (d);
    \propag [plain, mom={$(p_1,k_3)$}] (d) to (e);
    \propag [plain, mom={$(p_2,k_4)$}] (d) to (f); 
    \end{feynhand}
    \end{tikzpicture}
    }}
    \\
    & = \frac{i}{k_1^2}\frac{i}{k_2^2}\frac{i}{k_3^2 - p_1^2 - m^2}\frac{i}{k_4^2 - p_2^2 - m^2}(2\pi)^2\de(k_1 + k_2 - k_3 - k_4) \frac{i\pi^2\bb^2}{16m}\frac{(k_1\cdot k_2)}{(ip_1-m)(ip_2-m)}\cdot
    \\
    & \quad\quad \int \frac{dp}{2\pi}\Biggl(\frac{{\rm sech}\pr{\frac{\pi(p_1+p_2+p)}{2m}}{\rm sech}\pr{\frac{\pi p}{2m}}}{(k_1+k_2)^2 - p^2 - m^2 + i\ep}\cdot
    \\
    & \quad\quad \pr{3m^4 + (-p_1+p_2+p)(p_1-p_2+p)(p_1+p_2-p)(p_1+p_2+p) + 2m^2(p_1^2+p_2^2+p^2)}\Biggr).
\end{split}
\end{equation}
The integral in the last two lines, denoted as $I^1_{\rm 2BP}(s,m,p)$, is evaluated in appendix \ref{app_integrals}:
\begin{equation}
\begin{split}
    I^1_{\rm 2BP}&(s,m,p_1,p_2) 
    \\
    &\frac{1}{6\pi m}\csch\left(\frac{\pi(p_1+p_2)}{2 m}\right) \Biggl(
    -2m(p_1+p_2) \bigl(8m^2 + 5p_1^2 -2p_1p_2 + 5p_2^2 -3s\bigr)
    \\
    &-\frac{3im}{\sqrt{-s+m^2}}\left(\left((p_1-p_2)^2-s\right) \left((p_1+p_2)^2-s\right) - 4m^2s\right)\cdot  
    \\
    &\biggl(2 H_{\frac{i (p_1+p_2)}{m}+\frac{\sqrt{-s+m^2}}{m}} - H_{\frac{i(p_1+p_2)}{2m}+\frac{\sqrt{-s+m^2}}{2m}}-2 H_{\frac{\sqrt{-s+m^2}}{m}-\frac{i (p_1+p_2)}{m}}+H_{\frac{\sqrt{-s+m^2}}{2m}-\frac{i (p_1+p_2)}{2m}}\biggr) \Biggr).
\end{split}
\end{equation}

The first thing to note is that the $x$-momentum conserving part must vanish. Indeed, onshell, from the first contribution, one has:
\begin{equation}
    \frac{i\bb^2}{4m}(k_1\cdot k_2)p_1p_2 = \frac{i\bb^2}{8m}sp_1p_2,
\end{equation} 
while from the third contribution, combining the $t,u$-channels, and using $t = (k_3-k_1)^2 = p_1^2 + m^2 - 2k_1\cdot k_3,\ u = (k_4-k_1)^2 = p_2^2 + m^2 - 2k_1\cdot k_4$:
\begin{equation}
\begin{split}
    & \frac{i\bb^2}{8m}\frac{(k_1\cdot(2k_3-k_1))(k_2\cdot(2k_4-k_2))p_1p_2}{(k_3-k_1)^2 - p_1^2 - m^2 } + \frac{i\bb^2}{8m}\frac{(k_1\cdot(2k_4-k_1))(k_2\cdot(2k_3-k_2))p_1p_2}{(k_4-k_1)^2 - p_1^2 - m^2 } =
    \\
    &\frac{i\bb^2}{8m}p_1p_2(t-p_1^2-m^2 + u-p_2^2-m^2).
\end{split}
\end{equation}
Adding the two vanishes, due to $s+t+u = \sum_{i=1}^4k_i^2 = p_1^2 + p_2^2 + 2m^2$.

\begin{figure}
    \captionsetup{singlelinecheck = false, justification=justified}
    \includegraphics[scale=0.9]{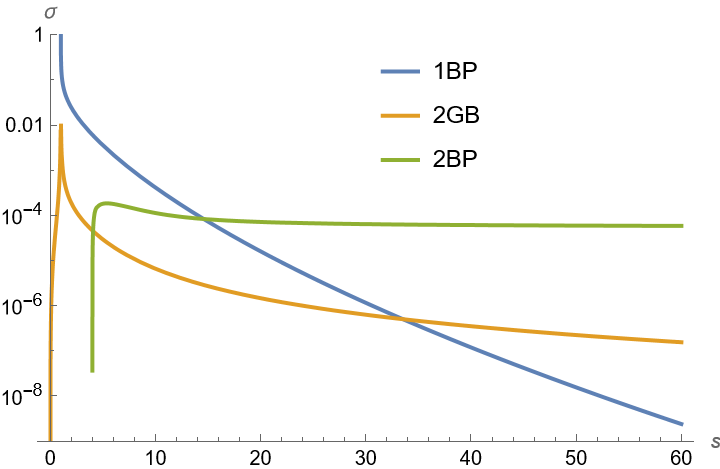}
    \caption{Different total cross sections for scattering of two NGBs as a function of the COM energy $s$ (in units of $m^2$); to one BP, two NGBs, and two BPs, for $m=1,\bb=0.5$.}
    \label{fig_sctr_csc}
\end{figure}

Going to the scattering amplitude, we put the momenta onshell and use the kinematics $k_1\cdot k_2 = \frac{s}{2}$, $k_3\cdot k_4 = \frac{s-p^2-m^2}{2}$, and similar expressions for the other multiplications. Energy-Momentum conservation gives:
\begin{equation*}
    |\vec k_3| = |\vec k_4| = \sqrt{\frac{\left(2 m^2+p_1^2+p_2^2-s\right)^2-4 \left(m^2+p_1^2\right) \left(m^2+p_2^2\right)}{4s}}.   
\end{equation*}
Inserting inside $t,u$ yields two symmetric values: 
\begin{equation*}
\begin{split}
    t = \frac{1}{2}\left(2m^2 + p_1^2 + p_2^2 - s \pm \sqrt{\left(2 m^2+p_1^2+p_2^2-s\right)^2-4 \left(m^2+p_1^2\right) \left(m^2+p_2^2\right)}\right), 
    \\
    u = \frac{1}{2}\left(2m^2 + p_1^2 + p_2^2 - s \mp \sqrt{\left(2 m^2+p_1^2+p_2^2-s\right)^2-4 \left(m^2+p_1^2\right) \left(m^2+p_2^2\right)}\right),   
\end{split}
\end{equation*}
satisfying $s+t+u = 2m^2 + p_1^2 + p_2^2$. Summing the four contributions gives:
\begin{equation}
\begin{split}
     i\MM(\vec k_1,\vec k_2 \to (p_1,\vec k_3), (p_2,\vec k_4)) =& \frac{i\pi^2\beta^2s\left(I^1_{\rm 2BP}(s,m,p_1,p_2) -tI^2_{\rm 2BP}(t,m,p_1,p_2) -uI^2_{\rm 2BP}(u,m,p_2,p_1)\right)}{32m(ip_1-m)(ip_2-m)} 
     \\
     -&\frac{i\pi\beta^2(p_1+p_2) \left(s \left(4 m^2+(p_1+p_2)^2\right)+3 \left(p_1^2-p_2^2\right)^2\right) \csch\left(\frac{\pi  (p_1+p_2)}{2 m}\right)}{48m(ip_1-m)(i p_2-m)}
     \\
     +&\frac{i\pi^2\beta^2s(m+ip_1) (m+ip_2) \sech\left(\frac{\pi  p_1}{2 m}\right) \sech\left(\frac{\pi  p_2}{2 m}\right)}{32 m^2}.
\end{split}
\end{equation}
The total cross section comes from an integration of the scattering amplitude on the allowed phase space:
\begin{equation}\label{eq_sctr_tot_1BP1GB}
\begin{split}
    \sg_{\rm 2BP} =& \frac{1}{2s}\int\frac{d\vec k_3dp_1}{(2\pi)^2 \pr{2\sqrt{\vec k_3^2+p_1^2+m^2}}}\frac{d\vec k_4dp_2}{(2\pi)^2 \pr{2\sqrt{\vec k_4^2+p_2^2+m^2}}}\cdot 
    \\
    &\left|\MM(\vec k_1,\vec k_2 \to (p_1,\vec k_3), (p_2,\vec k_4))\right|^2 (2\pi)^2 \de(\vec k_3+\vec k_4)\de\left(\sqrt{\vec k_3^2+p_1^2+m^2}+\sqrt{\vec k_4^2+p_2^2+m^2}-\sqrt{s}\right) 
    \\
    =& \frac{1}{4\pi^2s}\int_0^{\sqrt{s-2\sqrt{s}m}} dp_1\int_0^{\sqrt{s+p_1^2-2\sqrt{s}\sqrt{p_1^2+m^2}}}dp_2 \frac{\left|\MM(\vec k_1,\vec k_2 \to (p_1,\vec k_3), (p_2,\vec k_4))\right|^2}{\sqrt{\left(2 m^2+p_1^2+p_2^2-s\right)^2-4 \left(m^2+p_1^2\right) \left(m^2+p_2^2\right)}}.
\end{split}    
\end{equation}
This cross section can not be evaluated analytically, so numerical integration was used. The three finite cross sections are compared in figure \ref{fig_sctr_csc}, for $m=1,\bb=0.5$ ($m\bb^2$ is small). $\Sgg$ vanishes at $s=0$, and $\Sbb$ at $s=4m^2$; the latter is due to zero measure of the phase space. This does not happen for $\Sb$ at $s=m^2$ because the phase space is a point for all $s$. 

$\Sgg$ decays at large $s$ polynomially as $s^{-2}$, while $\Sb$ decays exponentially. Therefore, although $\Sb$ is leading in $\bb$, it must go below $\Sgg$ at some $s(\bb)$. This is expected to hold for the full result, not only for the tree-level approximation.

%% file: body/conclusions.tex
In this paper we made first steps towards analyzing the effective action on confining strings in theories like QED$_3$, for which the mass gap $m$ in the bulk is much smaller than the scale set by the string tension. Above the scale $m$, the effective action must include both the NGB on the string and the massive bulk particle, and we found that this leads to several changes in the properties of the strings.

In section \ref{sec: flux-tube profile} we found that the width of the string changes from the usual effective string behavior (where the width is dominated by quantum fluctuations) for exponentially long strings, to the classical width (of order $1/m$, independent of the length) when the strings are long compared to the string tension scale, but not exponentially long. It would be interesting to observe this transition in lattice simulations of QED-like theories. However, as we discussed in the introduction, it is challenging to significantly separate the scales of the mass, of the tension, and of the lattice spacing on a finite lattice.

In section \ref{sec_cal} we computed the ground state energy of the string, finding large corrections to the usual EST results when the length $L$ of the string is of order $1/m$ or smaller. It would be interesting to compute the corrections also to other energy levels, and to measure the ground state energy and other energies in lattice simulations. We expect qualitatively similar results for other energy levels below $m$, while energy levels above $m$ can mix with the bulk particles and decay into them.

In section \ref{sec: scattering} we computed (at leading order in perturbation theory) the scattering amplitudes of two NGBs. Somewhat surprisingly, we found that the scattering amplitudes to final states including BPs and NGBs exhibit IR divergences that need to be regularized, related to the massless NGBs. Usually NGBs do not give rise to IR divergences because their couplings are proportional to their momenta, but it turns out that when they scatter into bulk fields, this is not enough to eliminate the IR divergences. We argued that, unlike in other theories of massless particles, the IR divergences here are not related to a dressing of the massive particles by soft massless particles. Instead, we believe that these IR divergences are related to the recoil of the string when it emits or absorbs bulk particles carrying transverse momentum, forced by the momentum conservation of the full theory, which we did not take into account in our computations. It would be interesting to understand better how to take this recoil into account, and to confirm that doing this cancels the IR divergences that we find, and leads to finite cross sections.

There are various possible generalizations of our work. Our analysis essentially uses just the fact that the low-energy action of QED$_3$ (with any matter content, as long as monopole-instantons are allowed) is the sine-Gordon theory. It would be interesting to generalize the analysis to other weakly coupled low-energy theories, such as the massive $\phi^4$ theory, where we expect to get extra modes living on the string (since the $\phi^4$ soliton in $1+1$ dimensions has an extra bound state below the continuum). It would also be interesting to study examples in $3+1$ dimensions in which the bulk mass gap is well below the string tension scale.

A final remark concerns the Lorentz symmetry, which is not manifest in our action (in both formalisms). In general, effective actions on string-like objects are constrained by Lorentz symmetry, which rotates/boosts the string, and acts non-linearly on the NGBs describing its transverse position. In our case, when we also include the bulk mode $\dphib$ in the effective action, Lorentz transformations mix the NGB with $\dphib$ in a complicated non-linear way. We expect that order by order in perturbation theory Lorentz symmetry will give some constraints on the effective action, but we leave their analysis to future work. The specific effective action that we compute is definitely Lorentz-invariant, since we derive it from an underlying Lorentz-invariant theory.

%% file: appendices/basis.tex
This appendix summarizes the orthonormality and completeness relations among the wave-functions presented in subsection \ref{subsec_first_formalism}. We consider the $x$-axis only; the $y,t$-axes are trivial. Using bra-ket notation, the orthonormality and completeness relation of the $x$-basis are written as:
\begin{equation}
\begin{split}
    \braket{x_1|x_2} &= \de(x_1-x_2), \qquad
    \int dx\ket{x}\bra{x} = \hat{I}.
\end{split}
\end{equation}
For the other continuous basis $\ket{p}$ (ignoring for a moment the discrete NGB mode), there are three coefficients among these relations and the matrix-elements between the $\ket{x},\ket{p}$ bases:
\begin{equation}
\begin{split}
    \braket{p_1|p_2} &= C_1(p_1)\de(p_1-p_2),
    \\
    \int dp C_2(p)\ket{p}\bra{p} &= \hat{I}, 
    \\
    \braket{x|p} &= C_3(p)f_p(x), 
    \\
    \int dx f_{p_1}(x)f_{p_2}^*(x) &= N(p_1)\de(p_1-p_2).
\end{split}
\end{equation}
Consistency between these relations gives:
\begin{equation}
    C_1(p)C_2(p) = 1,\qquad N(p)|C_3(p)|^2 = C_1(p).
\end{equation}
For $f_p = \phi_p$, we have $N(p) = 2\pi$. If we choose $C_3(p) = 1$ then $C_1(p) =2\pi, C_2(p) = \frac{1}{2\pi}$, and this was used all everywhere in this paper.

Adding the NGB mode $\ket{0}$, the relations read:
\begin{equation}
\begin{split}
    &\braket{p_1|p_2} = 2\pi\de(p_1-p_2), \ \braket{p|0} = 0, \ \braket{0|0} = 1, 
    \\
    &\int \frac{dp}{2\pi}\ket{p}\bra{p} + \ket{0}\bra{0} = \hat{I}, 
    \\
    &\braket{x|p} = \phi_p(x), \  \braket{x|0} = \phi_0(x).
\end{split}
\end{equation}
Using the wave-functions, these relations reads:
\begin{equation}
\begin{split}
    & \int dx \phi_{p_1}(x)\phi_q^*(x)  = 2\pi\de(p-q), \ \int dx \phi_p(x)\phi_0(x) = 0, \ \int dx \phi_0^2(x) = 1, 
    \\
    & \int \frac{dp}{2\pi}\phi_p(x_1)\phi_p^*(x_2) + \phi_0(x_1)\phi_0(x_2) = \de(x_1-x_2). 
\end{split}
\end{equation}

%% file: appendices/change.tex
By comparing \eqref{eq_pre_expansion_fst} and \eqref{dphi_expansion} to the expansion of \eqref{eq_pre_expansion_snd} around $x_0=0$, one can write the change of variables between the NGB $\pp_0(\sg)$ and the BP $\pp(p,\sg)$ in the direct formalism and the NGB $X_0(\sg)$ and the BP $\dphib(p,\sg)$ in the stringy formalism: 
\begin{equation}
\begin{split}
    \pc(x) + \dph_{\rm GB}(x,\sg) + \dph_{\rm BP}(x,\sg) &= \pc(x) + \phi_0(x)\pp_0(\sg) + \int\frac{dp}{2\pi}\phi_p^*(x)\pp(p,\sg) 
    \\
    = \pc(x-x_0(\sg)) + \dphib(x-x_0(\sg),\sg) &= \sum_{n=0}^\infty \frac{(-1)^n}{n!}\frac{d^n\pc(x)}{dx^n}x_0^n(\sg) + \sum_{n=0}^\infty \frac{(-1)^n}{n!}\frac{d^n\dphib(x)}{dx^n}x_0^n(\sg) 
    \\
    &= \sum_{n=0}^\infty \frac{(-1)^n}{n!}\frac{d^n\pc(x)}{dx^n}x_0^n(\sg) + \sum_{n=0}^\infty \frac{(-1)^n}{n!}\int \frac{dp}{2\pi}\frac{d^n\phi_p^*(x)}{dx^n}\dphib(p,\sg)x_0^n(\sg).
\end{split}    
\end{equation}
The $n=0$ term of the first series is just $\pc(x)$, which cancels on both sides. Using the orthonormality relations of $\phi_0(x),\phi_p(x)$ one finds for $\pp_0(\sg)$:
\begin{equation}\label{eq_change_NGB}
\begin{split}
    \pp_0(\sg) =& \sum_{n=1}^\infty \frac{(-1)^n}{n!}\pr{\int dx\phi_0(x)\frac{d^n\pc(x)}{dx^n}}x_0^n(\sg) + \sum_{n=0}^\infty \frac{(-1)^n}{n!}\int \frac{dp}{2\pi}\pr{\int dx \phi_0(x)\frac{d^n\phi_p^*(x)}{dx^n}}\dphib(p,\sg)x_0^n(\sg) 
    \\
    =& -X_0(\sg) + \sum_{n=2}^\infty c_{1n} \bb^{n-1}X_0^n(\sg) + \sum_{n=1}^\infty \int \frac{dp}{2\pi} c_{2n}(p)\bb^n \dphib(p,\sg)X_0^n(\sg),
    \\
    c_{1n} \equiv& \frac{1}{n!}\pr{-\frac{1}{\sqrt{8m}}}^n\pr{\int dx\phi_0(x)\frac{d^n(\bb\pc(x))}{dx^n}},\ 
    c_{2n}(p) \equiv \frac{1}{n!}\pr{-\frac{1}{\sqrt{8m}}}^n\int dx \phi_0(x)\frac{d^n\phi_p^*(x)}{dx^n}.
\end{split}
\end{equation}
For $\pp(p,\sg)$ one finds:
\begin{equation}
\begin{split}
    \pp(p,\sg) =& \sum_{n=1}^\infty \frac{(-1)^n}{n!}\pr{\int dx\phi_p(x)\frac{d^n\pc(x)}{dx^n}}x_0^n(\sg) + \sum_{n=0}^\infty \frac{(-1)^n}{n!}\int \frac{dp'}{2\pi}\pr{\int dx \phi_p(x)\frac{d^n\phi_{p'}^*(x)}{dx^n}}\dphib(p',\sg)x_0^n(\sg) 
    \\
    =& \dphib(p,\sg) + \sum_{n=2}^\infty c_{3n} \bb^{n-1}X_0^n(\sg) + \sum_{n=1}^\infty\int\frac{dp'}{2\pi} c_{4n}(p')\bb^nX_0^n(\sg)\dphib(p',\sg),
    \\
    c_{3n} \equiv& \frac{1}{n!}\pr{-\frac{1}{\sqrt{8m}}}^n\pr{\int dx\phi_p(x)\frac{d^n(\bb\pc(x))}{dx^n}},\ 
    c_{4n}(p') \equiv \frac{1}{n!}\pr{-\frac{1}{\sqrt{8m}}}^n\int dx \phi_p(x)\frac{d^n\phi_{p'}^*(x)}{dx^n}.
\end{split}
\end{equation}
Hence, both variables are the same to order $\bb^0$, and deviate by terms of order $\bb^{>0}$. Using states and operators language, one can write the relations in a compact form (note that $\ket{p}$ is not an eigenvector of the $x$-momentum operator $\hat{P}$):
\begin{align}
    \pp_0(\sg) &= -X_0(\sg) + \bra{\pc}\pr{e^{-i\hat{P}x_0(\sg)} + i\hat{P}x_0(\sg) - \hat{I}}\ket{0} + \int\frac{dp}{2\pi}\dphib(p,\sg)\bra{p}e^{-i\hat{P}x_0(\sg)}\ket{0},
    \\
    \pp(p,\sg) &= \dphib(p,\sg) + \bra{\pc}\pr{e^{-i\hat{P}x_0(\sg)} + i\hat{P}x_0(\sg) - \hat{I}}\ket{p} + \int\frac{dp'}{2\pi}\dphib(p',\sg)\bra{p'}\pr{e^{-i\hat{P}x_0(\sg)} - \hat{I}}\ket{p}.
\end{align}

%% file: appendices/vertices.tex
In this appendix, we evaluate the vertices needed in this work in pseudo-momentum space, for the two formalisms. For each vertex, $\sg$-integrals and $\pp_0 / x_0$-dependencies will be suppressed along the calculations, and will be added after performing the $x$-integration.

\subsection{Direct Formalism}


\paragraph{\vggb:}
\begin{equation}
\begin{split}
    \vggb \subset & -\frac{m^2\bb}{3!} \int dx (-2\sech(mx)\tanh(mx))\dph^3 
    \\
    \propto & \frac{2m^2\bb}{3!}3 \int dx \sech(mx)\tanh(mx)\phi_0^2(x)\phi_p^*(x)
    \\
    = & \frac{\pi\bb}{16m}(m-ip)(m^2+p^2)\sech\pr{\frac{\pi p}{2m}}.
\end{split}
\end{equation}
The vertex is then:
\begin{equation}
\begin{split}
    \vggb =& \int \pr{\prod_{i=1}^3 \frac{d^2q_i}{(2\pi)^2}} (2\pi)^2\de(q_1+q_2+q_3) \int \frac{dp}{2\pi} \cdot
    \\
    & \frac{\pi\bb}{16m}(m-ip)(m^2+p^2)\sech\pr{\frac{\pi p}{2m}}
    \pp(p,q_1)\pp_0(q_2)\pp_0(q_3).
\end{split}
\end{equation}

\paragraph{\vgggg:}
\begin{equation}
\begin{split}
    \vgggg \subset & -\frac{m^2\bb^2}{4!} \int dx (1-2\sech^2(mx))\dph^4
    \\
    \propto & -\frac{m^2\bb^2}{4!} \int dx (1-2\sech^2(mx))\phi_0^4(x) = \frac{m^3\bb^2}{4!\cdot 5}.
\end{split}
\end{equation}
The vertex is then:
\begin{equation}
\begin{split}
    \vgggg = \frac{m^3\bb^2}{4!\cdot 5} \int \pr{\prod_{i=1}^4 \frac{d^2q_i}{(2\pi)^2}}(2\pi)^2\de(q_1+q_2+q_3+q_4).
\end{split}
\end{equation}

\subsection{Stringy Formalism}
\paragraph{\vone:}
\begin{equation}
\begin{split}
    \vone \propto & -\frac{2m^2}{\bb}\intop dx\sech(mx){\rm tanh}(mx)\dphib(x)
     \\
    = & -\frac{2m^2}{\bb}\intop dx\sech(mx){\rm tanh}(mx)\intop\frac{dp}{2\pi}e^{-ipx}\frac{ip+m\tanh(mx)}{ip+m}\dphib(p)
    \\
    = & -\frac{\pi}{\bb}\intop\frac{dp}{2\pi}\pr{m-ip}\sech\left(\frac{\pi p}{2m}\right)\dphib(p),
\end{split}
\end{equation}
so\footnote{One can also evaluate it by using the vertex with the derivative on $\dphib$.}:
\begin{equation}
    \vone = -\frac{\pi}{\bb}\frac{\bb^2}{8m}\int d^2\sg(\pd_aX_0(\sg))^2\intop\frac{dp}{2\pi}\pr{m-ip}\sech\left(\frac{\pi p}{2m}\right)\dphib(p,\sg).
\end{equation}
Finally:
\begin{equation}
\begin{split}
    \vone = \frac{\pi\bb}{8m} & \int \pr{\prod_{i=1}^3 \frac{d^2q_i}{(2\pi)^2}}(q_1\cdot q_2)(2\pi)^2\de(q_1+q_2+q_3)\cdot
    \\
    & \int \frac{dp}{2\pi}\pr{m-ip}\sech\left(\frac{\pi p}{2m}\right)x_0(q_1)x_0(q_2)\dphib(p,q_3).
\end{split}    
\end{equation}

\paragraph{\vtwo:}
\begin{equation}
\begin{split}
    \vtwo \propto & \frac{1}{2}\intop dx\left(\pd_{x}\dphib(x)\right)^2
    \\
    = & \frac{1}{2}\intop\frac{dp_1}{2\pi}\frac{dp_2}{2\pi}\dphib\pr{p_1}\dphib\pr{p_2}\cdot
    \\
    & \intop dxe^{-ix\pr{p_2+p_1}} \cdot \pr{\left(-ip_1+\pd_{x}\right)\left(\frac{ip_1+m\tanh(mx)}{ip_1+m}\right)} \cdot \left(1\to2\right).
\end{split}
\end{equation}
When we distribute the expression, written schematically as $\left(-ip_1+\pd_{x}\right)\left(-ip_2+\pd_{x}\right)$,
we get four terms. The term proportional to $\pr{-ip_1}\pr{-ip_2}$
can be simplified using eigenfunction orthogonality, giving the momentum
conserving part:
\begin{equation}
\begin{split}
    \vtwoA = & \frac{1}{2}\frac{\bb^2}{8m} \int d^2\sg(\pd_aX_0)^2\intop\frac{dp}{2\pi}p^2\pr{\dphib}^2 \Rightarrow
    \\
    \vtwoA =& \frac{\bb^2}{16m}\int \pr{\prod_{i=1}^4 \frac{d^2q_i}{(2\pi)^2}} (-q_1\cdot q_2)(2\pi)^2\de(q_1+q_2+q_3+q_4)\cdot
    \\
    & \intop\frac{dp}{2\pi}p^2x_0(q_1)x_0(q_2)\dphib(p,q_3)\dphib(-p,q_4).
\end{split}
\end{equation}
The remaining three terms are: 
\begin{equation}
\begin{split}
    & m^2\sech^2(mx)\left(p^2_{2}-ip_2m\tanh(mx)\right)
    +m^2\sech^2(mx)\left(p^2_{1}-ip_1m\tanh(mx)\right)
    +m^4\sech^4(mx) =
    \\
    & m^2\sech^2(mx)\pr{p_1^2+p_2^2} 
    -im^3\pr{p_1+p_2}\sech^2(mx)\tanh(mx)
    +m^4\sech^4(mx).
\end{split}
\end{equation}
The integral over $x$ is then:
\begin{equation}
\begin{split}
    & \intop dx e^{-ix\pr{p_2+p_1}}\frac{m^2\sech^2(mx)\pr{p_1^2+p_2^2} 
    -im^3\pr{p_1+p_2}\sech^2(mx)\tanh(mx)
    +m^4\sech^4(mx)}{\pr{ip_1+m}\pr{ip_2+m}} 
    \\
    &\qquad \qquad = \frac{\frac{2\pi}{3}(p_1+p_2)(p_1^2 - p_1p_2 + p_2^2 + m^2)\csch\pr{\frac{\pi\pr{p_1+p_2}}{2m}}}{(ip_1+m)(ip_2+m)}.
\end{split}
\end{equation}
The vertex is then:
\begin{equation}
\begin{split}
    \vtwoB = \frac{1}{2}\frac{\bb^2}{8m} & \int d^2\sg(\pd_aX_0(\sg))^2\cdot
    \\
    & \intop\frac{dp_1}{2\pi}\frac{dp_2}{2\pi}\frac{\frac{2\pi}{3}(p_1+p_2)(p_1^2 - p_1p_2 + p_2^2 + m^2)\csch\pr{\frac{\pi\pr{p_1+p_2}}{2m}}}{(ip_1+m)(ip_2+m)}\dphib\left(p_1,\sg\right)\dphib\left(p_2,\sg\right) \Rightarrow
    \\
    \vtwoB = \frac{\bb^2}{16m} & \int \pr{\prod_{i=1}^4 \frac{d^2q_i}{(2\pi)^2}} (-q_1\cdot q_2)(2\pi)^2\de(q_1+q_2+q_3+q_4) \intop\frac{dp_1}{2\pi}\frac{dp_2}{2\pi} \cdot
    \\
    & \frac{\frac{2\pi}{3}(p_1+p_2)(p_1^2 - p_1p_2 + p_2^2 + m^2)\csch\pr{\frac{\pi\pr{p_1+p_2}}{2m}}}{(ip_1+m)(ip_2+m)}x_0(q_1)x_0(q_2)\dphib\pr{p_1,q_3}\dphib\pr{p_2,q_4}.
\end{split}
\end{equation}

\paragraph{\vthree:}
\begin{equation}
\begin{split}
    \vthree \propto  & -\intop dx\pd_{x}\dphib\pd_a\dphib\\
    = & -\intop\frac{dp_1}{2\pi}\frac{dp_2}{2\pi} \dphib\pr{p_1}\pd_a\dphib\pr{p_2}\cdot \\
    & \intop dxe^{-ix\pr{p_2+p_1}} \pr{\left(-ip_1+\pd_{x}\right)\left(\frac{ip_1+m\tanh(mx)}{ip_1+m}\right)}\left(\frac{ip_2+m\tanh(mx)}{ip_2+m}\right).
\end{split}
\end{equation}
From the $-ip_1$ factor, using orthogonality, we get the momentum-conserving vertex:
\begin{equation}
\begin{split}
    \vthreeA = & -\frac{\bb}{\sqrt{8m}}\int d^2\sg \pd_aX_0\intop\frac{dp}{2\pi}\pr{-ip}\dphib\pr{p}\pd_a\dphib\pr{-p} \Rightarrow
    \\
    \vthreeA = & \frac{\bb}{\sqrt{8m}}\int \pr{\prod_{i=1}^3 \frac{d^2q_i}{(2\pi)^2}} (-q_1\cdot q_3)(2\pi)^2\de(q_1+q_2+q_3) \intop\frac{dp}{2\pi}\pr{ip}x_0(q_1)\dphib\pr{p,q_2}\dphib\pr{-p,q_3}.
\end{split}
\end{equation}
This form is not symmetric in the two BP fields. Using momentum conservation, one can modify it to:
\begin{equation}
    \vthreeA = \frac{\bb}{2\sqrt{8m}}\int \pr{\prod_{i=1}^3 \frac{d^2q_i}{(2\pi)^2}} (q_1\cdot(q_2-q_3))(2\pi)^2\de(q_1+q_2+q_3) \intop\frac{dp}{2\pi}\pr{ip}x_0(q_1)\dphib\pr{p,q_2}\dphib\pr{-p,q_3}.
\end{equation}
The $q_1^2$ term vanishes because it is anti-symmetric with respect to the change of integration variables $q_2\leftrightarrow q_3$ together with $p\to -p$ due to the additional $p$ factor.

From the other term, we get:
\begin{equation}
\begin{split}
    -\intop dx e^{-ix\pr{p_2+p_1}} \frac{m^2\sech^2(mx)}{ip_1+m}\frac{ip_2+m\tanh(mx)}{ip_2+m} 
    = \frac{\frac{i\pi}{2}\pr{p_1^2-p_2^2} \csch\left(\frac{\pi(p_1+p_2)}{2 m}\right)}{(ip_1+m)(ip_2+m)},
\end{split}
\end{equation}
so we find:
\begin{equation}
\begin{split}
    \vthreeB = & \frac{\bb}{\sqrt{8m}}\int d^2\sg \pd_aX_0\intop\frac{dp_1}{2\pi}\frac{dp_2}{2\pi} \cdot
    \\
    & \frac{\frac{i\pi}{2}\pr{p_1^2-p_2^2}}{\pr{ip_1+m}\pr{ip_2+m}}\csch\left(\frac{\pi\pr{p_1 + p_2}}{2m}\right) \dphib\pr{p_1}\pd_a\dphib\pr{p_2} \Rightarrow
    \\
    \vthreeB = & \frac{\bb}{\sqrt{8m}}\int \pr{\prod_{i=1}^3 \frac{d^2q_i}{(2\pi)^2}} (-q_1\cdot q_3)(2\pi)^2\de(q_1+q_2+q_3) \intop\frac{dp_1}{2\pi}\frac{dp_2}{2\pi} \cdot 
    \\
    & \frac{\frac{i\pi}{2}\pr{p_1^2-p_2^2}}{\pr{ip_1+m}\pr{ip_2+m}}\csch\left(\frac{\pi\pr{p_1 + p_2}}{2m}\right) x_0(q_1)\dphib\pr{p_1,q_2}\dphib\pr{p_2,q_3}.
\end{split}
\end{equation}
Again, this form is not symmetric in the two BP fields. Using momentum conservation, one can modify it to:
\begin{equation}
\begin{split}
    \vthreeB = & \frac{i\pi\bb}{4\sqrt{8m}}\int \pr{\prod_{i=1}^3 \frac{d^2q_i}{(2\pi)^2}} (q_1\cdot (q_2-q_3))(2\pi)^2\de(q_1+q_2+q_3) \intop\frac{dp_1}{2\pi}\frac{dp_2}{2\pi} \cdot 
    \\
    & \frac{\pr{p_1^2-p_2^2}}{\pr{ip_1+m}\pr{ip_2+m}}\csch\left(\frac{\pi\pr{p_1 + p_2}}{2m}\right) x_0(q_1)\dphib\pr{p_1,q_2}\dphib\pr{p_2,q_3}.
\end{split}    
\end{equation}
The $q_1^2$ term vanishes as before because it is anti-symmetric with respect to the change of integration variables $q_2\leftrightarrow q_3,\ p_1\leftrightarrow p_2$, due to the additional $p_1^2-p_2^2$ factor.

\paragraph{\vbbb:}
\begin{equation}
\begin{split}
    \vbbb \propto  & \intop dx\tanh(mx)\sech(mx)\dphib^3 
    \\
    = & \intop\frac{dp_1}{2\pi}\frac{dp_2}{2\pi}\frac{dp_3}{2\pi} \dphib\pr{p_1}\dphib\pr{p_2}\dphib\pr{p_3} \cdot
    \\
    & \intop dxe^{-ix\pr{p_1+p_2+p_3}} \tanh(mx)\sech(mx)\cdot
    \\
    & \left(\frac{ip_1+m\tanh(mx)}{ip_1+m}\right)\left(\frac{ip_2+m\tanh(mx)}{ip_2+m}\right)\left(\frac{ip_3+m\tanh(mx)}{ip_3+m}\right)
    \\
    = & \frac{\pi}{8m}\sech\pr{\frac{\pi(p_1+p_2+p_3)}{2m}}\cdot
    \\
    &\frac{3m^4 + (-p_1+p_2+p_3)(p_1-p_2+p_3)(p_1+p_2-p_3)(p_1+p_2+p_3) + 2m^2(p_1^2+p_2^2+p_3^2)}{(ip_1+m)(ip_2+m)(ip_3+m)}.
\end{split}
\end{equation}
Hence:
\begin{equation}
\begin{split}
    \vbbb = & \frac{2m^2\bb}{3!}\intop d^2\sg \intop \frac{dp_1}{2\pi}\frac{dp_2}{2\pi}\frac{dp_3}{2\pi} \cdot
    \\
    & \frac{3m^4 + (-p_1+p_2+p_3)(p_1-p_2+p_3)(p_1+p_2-p_3)(p_1+p_2+p_3) + 2m^2(p_1^2+p_2^2+p_3^2)}{(ip_1+m)(ip_2+m)(ip_3+m)}\cdot
    \\
    & \frac{\pi}{8m}\sech\pr{\frac{\pi(p_1+p_2+p_3)}{2m}} \dphib(p_1,\sg)\dphib(p_2,\sg)\dphib(p_3,\sg) \Rightarrow
    \\
    \vbbb = & \frac{\pi m\bb}{24} \int \pr{\prod_{i=1}^3 \frac{d^2q_i}{(2\pi)^2}}(2\pi)^2\de(q_1+q_2+q_3) \intop \frac{dp_1}{2\pi}\frac{dp_2}{2\pi}\frac{dp_3}{2\pi} \cdot
    \\
    & \frac{3m^4 + (-p_1+p_2+p_3)(p_1-p_2+p_3)(p_1+p_2-p_3)(p_1+p_2+p_3) + 2m^2(p_1^2+p_2^2+p_3^2)}{(ip_1+m)(ip_2+m)(ip_3+m)}\cdot
    \\
    & \sech\pr{\frac{\pi(p_1+p_2+p_3)}{2m}} \dphib(p_1,q_1)\dphib(p_2,q_2)\dphib(p_3,q_3).
\end{split}
\end{equation}

%% file: appendices/jacobian.tex
In this appendix, we compute the Jacobian of the change of variables $\phi$ to $X_0$ and $\dphib$. It will be convenient to use the spectral decomposition coefficients $\dphib(p,\sg)$ as variables instead of $\dphib(x,\sg)$ since the latter are constrained by orthogonality to $\sech(m(x-x_0(\sg)))$. So we need:
\begin{equation}
\begin{split}
    1 = & \int \DD X_0(\sg)\DD\dphib(p,\sg) J[X_0,\dphib]\de(\phi(x,\sg)-\phi[X_0(\sg),\dphib(p,\sg)]),
    \\
    J = & \int \DD\psi_0(\sg_1)\DD\psib(p,\sg_1)\DD\chi(x,\sg_2) e^{iS_g} \equiv e^{iS_J},
    \\
    S_g \equiv & -i\int dxd^2\sg_2d^2\sg_1\chi(x,\sg_2)\pr{\psi_0(\sg_1)\frac{\dph(x,\sg_2)}{\de X_0(\sg_1)} + \int \frac{dp}{2\pi}\psib(p,\sg_1)\frac{\de\phi(x,\sg_2)}{\de(\dphib(p,\sg_1))}}.
\end{split}
\end{equation}
After we insert the Faddeev-Popov identity into the PI, the $\DD\phi(x,\sg)$ integration removes the $\de(\phi-\phi[X_0,\dphib])$, the ghost will disappear, and we remain with $\DD X_0(\sg)\DD\dphib(p,\sg)$ integration. Alternatively, we could simply use $\DD\phi(x,\sg) = J \DD X_0(\sg)\DD\dphib(p,\sg)$. However, the ghosts are utilized to present the operators in the determinant in a cleaner way.\footnote{The exponent in $J$ in a discretized notation, using a lattice with $N^3$ sites, is $\chi_i A_{ij} \psi_{0,j} + \chi_i A_{ik} \psi_{\rm bulk,k}$, where the second index splits into two: $1 \leq i \leq N^3,\ 1\leq j \leq N^2,\ 1\leq k \leq (N-1)N^2$. This separation amounts to the separation of the NGB and the BP.  Using the ghosts, we will bring it to the form $\chi_i A_{ij} \psi_j$, and then $J = \det(A)$.}

Using:
\begin{equation}
\begin{split}
    \frac{\de\phi(x,\sg_2)}{\de(\dphib(p,\sg_1))} &= \phi_p^*(x)\de(\sg_1 - \sg_2),
    \\
    \frac{\de\phi(x,\sg_2)}{\de X_0(\sg_1)} &= -\frac{\bb}{\sqrt{8m}}\pd_x\phi(x-x_0(\sg_2),\sg_2)\de(\sg_2 - \sg_1), 
\end{split}
\end{equation}
the ghost variation is:
\begin{equation}
\begin{split}
    & \int d^2\sg_1\pr{\psi_0(\sg_1)\frac{\de\phi(x,\sg_2)}{\de X_0(\sg_1)} + \int \frac{dp}{2\pi}\psib(p,\sg_1)\frac{\de\phi(x,\sg_2)}{\de(\dphib(p,\sg_1))}}
    \\
    = & -\frac{\bb}{\sqrt{8m}}\pd_x\phi(x,\sg_2)\psi_0(\sg_2) + \frac{dp}{2\pi}\phi_p^*(x-x_0(\sg_2))\psib(p,\sg_2)
    \\
    = & \pr{-\frac{2m}{\bb}\sech(m(x-x_0(\sg_2))) -\pd_x\dphib(x-x_0(\sg_2),\sg_2)}\frac{\bb}{\sqrt{8m}}\psi_0(\sg_2) \\
    &+ \int\frac{dp}{2\pi}\phi_p^*(x-x_0(\sg_2))\psib(p,\sg_2).
\end{split}
\end{equation}
So the ghost action can be written compactly as:
\begin{equation}
    S_g = -i\int dxd^2\sg \chi(x+x_0(\sg),\sg)\pr{- \frac{\bb}{\sqrt{8m}}\psi_0(\sg)(\pd_x\dphib(x,\sg) + \frac{2m}{\beta} \sech(mx))
    + \int\frac{dp}{2\pi}\phi_p^*(x)\psib(p,\sg_2)
    }.
\end{equation}
We combine the ghosts $\psi_0(\sg),\psib(p,\sg)$ into a single ghost $\psi(x,\sg)$:
\begin{equation}
\begin{split}
    \psi(x,\sg) & = \int\frac{dp}{2\pi}\phi_p^*(x)\psib(p,\sg) - \sqrt{\frac{m}{2}}\sech(mx)\psi_0(\sg)
    \\
    \Rightarrow \psi_0(\sg) & = - \sqrt{\frac{m}{2}}\int dx'\sech(mx')\psi(x',\sg).
\end{split}
\end{equation}
This transformation has trivial Jacobian due to the same orthogonality considerations as in the direct formalism. Then, the ghost action reads:
\begin{equation}
\begin{split}
    S_g &= -i\int dxd^2\sg dx'd^2\sg' \chi(x+x_0(\sg),\sg)\AAA(x,x';\sg,\sg')\psi(x',\sg'),
    \\
    \AAA(x,x';\sg,\sg') & \equiv \pr{\de(x-x') + \frac{\bb}{4}\pd_x\dphib(x,\sg)\sech(mx')}\de(\sg-\sg')
    \\
    & \equiv I_{3d} + \mathcal{B}.
\end{split}
\end{equation}
One more step is needed, to remove $x_0(\sg)$ from the argument of $\chi$:
\begin{equation}
\begin{split}
    \chi(x+x_0(\sg),\sg) & = e^{x_0(\sg)\pd_x}\chi(x,\sg) 
    \\
    & = \int dx'd\sg' \de(x-x')\de(\sg-\sg')e^{x_0(\sg')\pd_{x'}}\chi(x',\sg') 
    \\
    & \equiv \int dx'd^2\sg'\EE(x,\sg;x',\sg')\chi(x',\sg').
\end{split}
\end{equation}
The total Jacobian factor is:
\begin{equation}
\begin{split}
    J = & \int \DD\psi(x'',\sg'')\DD\chi(x,\sg)\exp(\int dxd^2\sg dx'd^2\sg'dx''d^2\sg''\chi(x,\sg) \EE^T(x,\sg;x',\sg')\AAA(x',\sg';x'',\sg'')\psi(x'',\sg''))
    \\
    = & \det(\EE)\det(\AAA),
\end{split}
\end{equation}
Since the eigenvalues of $\pd_x$ come in pairs $\pm p$, we have $\det(\EE) = 1$. This will be true in any regularization that has $x$-axis translation and reflection symmetries. This triviality is ultimately because a $\sigma$-dependent $x$-translation is also an orthogonal transformation.
Using $\det(\AAA) = e^{\log\det(\AAA)} = e^{\Tr\log(\AAA)}$, the other determinant gives:
\begin{equation}
\begin{split}
    \Tr[\log(\AAA)] = & \Tr[\log(I_{3d} + \BB)]
    \\
    = & -\sum_{n=1}^{\infty}\frac{1}{n}\Tr(-\BB)^n
    \\
    = & -\sum_{n=1}^{\infty}\frac{1}{n}\pr{-\frac{\bb}{4}}^n\de^2(0)\int d^2\sg\prod_{i=1}^n dx_{i}\pd_x\dphib(x_{i},\sg)\sech(mx_{i+1\mod n})
    \\
    = & -\sum_{n=1}^{\infty}\frac{1}{n}\pr{-\frac{\bb}{4}}^n\de^2(0)\int d^2\sg\pr{\int dx\sech(mx)\pd_x\dphib(x,\sg)}^n
    \\
    = & \de^2(0)\int d^2\sg\log\pr{1 + \frac{\bb}{4}\int dx\sech(mx)\pd_x\dphib(x,\sg)}.
\end{split}
\end{equation}
These are a series of counterterms, due to the UV divergence in $\de^2(0)$, for $\dphib$. Finally, we find:
\begin{equation}
\begin{split}
    S_J = & -i\log (J)
    \\
    = & i\de^2(0)\sum_{n=1}^{\infty}\frac{1}{n}\pr{-\frac{\bb}{4}}^n\int d^2\sg\pr{\int dx\sech(mx)\pd_x\dphib(x,\sg)}^n.
\end{split}    
\end{equation}

%% file: appendices/scattering.tex
In this appendix, we calculate the two NGBs to one BP and two NGBs to two NGBs scattering amplitudes in the direct formalism. The results obtained are the same as those calculated in section \ref{sec: scattering} using the stringy formalism. The two NGBs to one BP amplitude involves one diagram, as simple as the stringy formalism. The two NGBs to NGBs amplitude involves four diagrams rather than three in the stringy formalism, and the integral inside is more cumbersome. To be cautious, offshell correlators will be calculated first, and the scattering amplitudes will be deduced from them. While onshell scattering amplitudes match, offshell correlators differ. This is expected since the fields themselves differ. This offshell difference explains how the onshell quantities should match. From the examples calculated in this appendix, this observation can be seen in two facts:
\begin{enumerate}
    \item Although the NGB does not only have derivative interactions in the direct formalism, as a NGB should, they appear in the scattering amplitude when putting the momenta onshell.

    \item The two NGBs to two NGBs amplitude in the stringy formalism got contributions from $s,t,u$-channels with an intermediate BP (and because of the kinematics, the $t$-channel vanished). In the direct formalism, these three diagrams contribute (and now the $t$-channel does not vanish), in addition to a quartic NGB diagram.\footnote{In principle, this formalism allows another three diagrams with an intermediate NGB, but $\vggg$ vanishes due to $x$-reflection symmetry.} 
\end{enumerate}
This appendix is also a nice manifestation of the fact that scattering amplitudes are physical, i.e., independent of the fields used to describe the particles and their interactions. We will use the vertices for the direct formalism calculated in appendix \ref{app_vertices}.

\subsection{2 NGB to 1 BP}

The correlator $\langle \pp_0(k_1)\pp_0(k_2)\pp(p,k_3)\rangle$ has a tree-level contribution from one diagram containing the vertex $\vggb$:
\begin{equation}
\begin{split}
    & \langle \pp_0(-k_1)\pp_0(-k_2)\pp(p,k_3)\rangle_{\rm tree} =
    \vcenter{\hbox{
    \begin{tikzpicture}[baseline=(c.base)]
    \begin{feynhand}
    \vertex (a) at (-1.4,-1.4) {}; 
    \vertex (b) at (1.4,-1.4) {}; 
    \vertex (c) [dot,label=right:{$\vggb$}] at (0,0) {};
    \vertex (d) at (0,1.8) {};
    \propag [scalar, mom={$k_1$}] (a) to (c);
    \propag [scalar, mom={$k_2$}] (b) to (c); 
    \propag [plain, mom={$(p,k_3)$}] (c) to (d);
    \end{feynhand}
    \end{tikzpicture}
    }}
    \\
    &= \frac{i}{k_1^2}\frac{i}{k_2^2}\frac{i}{k_3^2 - p^2 - m^2}(2\pi)^2\de(-k_1-k_2+k_3)\pr{-\frac{i\pi\bb}{8m}}(m+ip)(m^2+p^2)\sech\pr{\frac{\pi p}{2m}}.
\end{split}
\end{equation}
The correlation function is not the same as the correlator in \eqref{eq_sctr_2NGB1BP_cor}. The scattering amplitude is the pole, and onshell, $s = (k_1+k_2)^2 = p^2+m^2$:
\begin{equation}
    iT(\vec k_1,\vec k_2\to (p,\vec k_3)) = -\frac{i\pi\bb s}{8m}(m\pm i\sqrt{s-m^2}){\rm sech}\left(\frac{\pi\sqrt{s-m^2}}{2m}\right)(2\pi)^2\de(- k_1 - k_2 + k_3).
\end{equation}
This is the same as \eqref{eq_sctr_2NGB1BP_sctr} up to a minus sign. This minus appears because of the relative sign between the fields representing the NGB, $\pp_0 = -X_0 + O(\bb)$, see \eqref{eq_change_NGB}.

\subsection{2 NGB to 2 NGB}\label{subsec_scattering_2NGB}

For two NGBs to two NGB scattering, there are four diagrams, one of four NGBs with insertion of $\vgggg$, and three with intermediate BP in $s,t,u$ channels with insertions of two $\vggb$'s (accompanied with a factor of $1/2$). The first one contributes:
\begin{equation}
\begin{split}
    \langle \pp_0(-k_1)\pp_0(-k_2)\pp_0(k_3)\pp_0(k_4) \rangle_{\rm tree,1} & = \vcenter{\hbox{
    \begin{tikzpicture}[baseline=(c.base)]
    \begin{feynhand}
    \vertex (a) at (-1.4,-1.4) {}; 
    \vertex (b) at (1.4,-1.4) {}; 
    \vertex (c) [dot,label=right:{$\vgggg$}] at (0,0) {};
    \vertex (d) at (-1.4,1.4) {};
    \vertex (e) at (1.4,1.4) {};
    \propag [scalar, mom={$k_1$}] (a) to (c);
    \propag [scalar, mom={$k_2$}] (b) to (c); 
    \propag [scalar, mom={$k_3$}] (c) to (d);
    \propag [scalar, mom={$k_4$}] (c) to (e); 
    \end{feynhand}
    \end{tikzpicture}
    }}
    \\
    & = \frac{i}{k_1^2}\frac{i}{k_2^2}\frac{i}{k_3^2}\frac{i}{k_4^2}(2\pi)^2\de(-k_1-k_2+k_3+k_4)\pr{-\frac{im^3\bb^2}{5}}.
\end{split}
\end{equation}
The other three contribute:
\begin{equation}
\begin{split}
    & \langle \pp_0(-k_1)\pp_0(-k_2)\pp_0(k_3)\pp_0(k_4) \rangle_{\rm tree,2} = 
    \\
    &
    \vcenter{\hbox{
    \begin{tikzpicture}[baseline=(c.base)]
    \begin{feynhand}
    \vertex (a) at (-1.4,-1.4) {}; 
    \vertex (b) at (1.4,-1.4) {}; 
    \vertex (c) [dot,label=right:{$\vggb$}] at (0,0) {};
    \vertex (d) [dot,label=right:{$\vggb$}] at (0,1.8) {};
    \vertex (e) at (-1.4,3.2) {};
    \vertex (f) at (1.4,3.2) {};
    \propag [scalar, mom={$k_1$}] (a) to (c);
    \propag [scalar, mom={$k_2$}] (b) to (c); 
    \propag [plain, mom={$(p,k_1+k_2)$}] (c) to (d);
    \propag [scalar, mom={$k_3$}] (d) to (e);
    \propag [scalar, mom={$k_4$}] (d) to (f); 
    \end{feynhand}
    \end{tikzpicture}
    }}
    + 
    \vcenter{\hbox{
    \begin{tikzpicture}[baseline=(c.base)]
    \begin{feynhand}
    \vertex (a) at (-1.4,-1.4) {}; 
    \vertex (b) at (-1.4,1.4) {}; 
    \vertex (c) [dot,label=left:{$\vggb$}] at (0,0) {};
    \vertex (d) [dot,label=right:{$\vggb$}] at (1.8,0) {};
    \vertex (e) at (3.2,-1.4) {};
    \vertex (f) at (3.2,1.4) {};
    \propag [scalar, mom'={$k_1$}] (a) to (c);
    \propag [scalar, mom'={$k_3$}] (c) to (b); 
    \propag [plain, mom={$(p,k_1-k_3)$}] (c) to (d);
    \propag [scalar, mom={$k_2$}] (e) to (d);
    \propag [scalar, mom={$k_4$}] (d) to (f); 
    \end{feynhand}
    \end{tikzpicture}
    }}
    + 
    \vcenter{\hbox{
    \begin{tikzpicture}[baseline=(c.base)]
    \begin{feynhand}
    \vertex (a) at (-1.4,-1.4) {}; 
    \vertex (b) at (-1.4,1.4) {}; 
    \vertex (c) [dot,label=left:{$\vggb$}] at (0,0) {};
    \vertex (d) [dot,label=right:{$\vggb$}] at (1.8,0) {};
    \vertex (e) at (3.2,-1.4) {};
    \vertex (f) at (3.2,1.4) {};
    \propag [scalar, mom'={$k_1$}] (a) to (c);
    \propag [scalar, mom'={$k_4$}] (c) to (b); 
    \propag [plain, mom={$(p,k_1-k_4)$}] (c) to (d);
    \propag [scalar, mom={$k_2$}] (e) to (d);
    \propag [scalar, mom={$k_3$}] (d) to (f); 
    \end{feynhand}
    \end{tikzpicture}
    }}
    \\
    & = \frac{i}{k_1^2}\frac{i}{k_2^2}\frac{i}{k_3^2}\frac{i}{k_4^2}(2\pi)^2\de(-k_1-k_2+k_3+k_4) \pr{-\frac{i\pi^2\bb^2}{64m^2}}\int \frac{dp}{2\pi}\frac{(m^2+p^2)^3}{(k_1+k_2)^2 - p^2 - m^2 + i\ep}\sech^2\pr{\frac{\pi p}{2m}} 
    \\
    & +  (k_1\leftrightarrow -k_3) + (k_1\leftrightarrow -k_3, k_1\leftrightarrow k_2).
\end{split}
\end{equation}
This integral is evaluated in appendix \ref{app_integrals}, and the result is:
\begin{equation}
    \int \frac{dp}{2\pi}\frac{(m^2+p^2)^3}{s - p^2 - m^2 + i\ep}\sech^2\pr{\frac{\pi p}{2m}} = -\frac{2m^5}{15\pi^2}\pr{32 + \frac{20s}{m^2} + \frac{15s^2}{m^4}} - \frac{s^3}{\pi^2}\frac{\PolyG\pr{1,\frac{1}{2} + \frac{\sqrt{-s+m^2}}{2m}}}{\sqrt{-s+m^2}}.
\end{equation}
Adding the three channels onshell, the constant term $\pr{-\frac{i\pi^2\bb^2}{64m^2}}3\pr{-\frac{64m^5}{15\pi^2}} = \frac{im^3\bb^2}{5}$ cancels the quartic vertex contribution. The linear term in $s$ cancels between the $s$ and the $u$ channels. All the rest give the same answer as \eqref{eq_sctr_2GB2GB_sctr}.

%% file: appendices/IR.tex
In this appendix, we give more details for the argument presented in subsection \ref{subsec_sctr_1NGB1BP}, regarding the cancellation of the IR divergences from emitted soft NGBs. 

The process that gives rise to an IR divergence is the insertion of $\vthreeA$ on a BP's external leg. When the NGB becomes soft, the BP's internal propagator becomes onshell and cancels the momentum factor coming from the derivative interactions of the NGB. Then, in the total cross section, an IR divergence appears due to the phase space factor of the NGB, $\frac{d\vec k}{2|\vec k|}$. We claim that this divergence is not related to the dressing of the BP by NGBs but to a recoil of the string. This suggestion relies on two observations discussed in the following two subsections.

\subsection{One-Loop Amplitudes}

The first observation is that the one-loop correction to the BP propagator does not have an IR divergence. There are two diagrams involving a NGB's loop:
\begin{equation}
\begin{split}
    & \langle \dphib(p_1,k)\dphib(-p_2,-k) \rangle_{\rm 1-loop} = 
    \\
    &
    \vcenter{\hbox{
    \begin{tikzpicture}[baseline=(c.base)]
    \begin{feynhand}
    \vertex (a) at (0,0) {}; 
    \vertex (b) [dot,label=below:{$\vthree$}] at (2,0) {};
    \vertex (c) [dot,label=below:{$\vthree$}] at (4,0) {};
    \vertex (d) at (6,0) {}; 
    \propag [plain, mom={$(p_1,k)$}] (a) to (b);
    \propag [plain, mom={$(p,k-q)$}] (b) to (c); 
    \propag [plain, mom={$(p_2,k)$}] (c) to (d);
    \propag[scalar, mom={$q$}] (b) to [in=90, out=90, looseness=2] (c);
    \end{feynhand}
    \end{tikzpicture}
    }}
    +
    \vcenter{\hbox{
    \begin{tikzpicture}
    \begin{feynman}
    \vertex (a);
    \node [right = 2 cm of a] (b) [dot,label=below:{$\vtwo$}];
    \vertex [right = 2 cm of b] (c);
    \vertex [above = 2 cm of b] (d);
    \diagram* {
      (a) -- [plain, momentum'={$(p_1,k)$}] (b) -- [plain, momentum'={$(p_2,k)$}] (c),
      (b) -- [scalar, half left, momentum'=\(q\)] (d) -- [scalar, half left] (b);
    };
    \end{feynman}
    \end{tikzpicture}
    }}.
\end{split}
\end{equation}
One can easily check that the momentum-conserving part $p=p_1=p_2$ does not contain any IR divergences: both corrections
\begin{align}
    \pr{\frac{i\bb}{\sqrt{8m}}}^2(ip)^2\int\frac{d^2q}{(2\pi)^2}\frac{i(-q\cdot (2k-q))}{q^2+i\ep}\frac{iq\cdot (2k-q)}{(k-q)^2-p^2-m^2 + i\ep}+
    \frac{i\bb^2}{8m}p^2\int\frac{d^2q}{(2\pi)^2}\frac{iq^2}{q^2+i\ep}
\end{align}
do not diverge as $q\to 0$. The non-conserving part also does not contain any IR divergences. For $\vtwoB$, the loop integral is the same. For insertion of one $\vthreeA$ and one $\vthreeB$ or two $\vthreeB$'s, the factor $(q\cdot (2k-q))^2$ remains in the numerator and makes the limit $q\to 0$ finite.\footnote{Even when the momentum 
$(p,k)$ is onshell and one considers the momentum-conserving part, the result is IR finite; in particular, it vanishes (using some UV regularization maintaining reflection symmetry in $\sg$ directions). This is expected, as should any other momentum-conserving correction, which is present everywhere in the bulk; see comment in subsection \ref{subsubsec_pre_psu_mom}. The two-$\vthreeA$ diagram gives $\frac{\bb^2p^2}{8m}\int\frac{d^2q}{(2\pi)^2}$ and the $\vtwoA$ diagram gives the opposite value.}

In addition, if the BP should be dressed with the NGB, there should be an IR divergence in the $1$-loop correction to the process of two NGBs to one BP, that cancels the IR divergence for the tree-level process of two NGBs to one BP and one NGB (discussed in subsection \ref{subsec_sctr_1NGB1BP}), when the latter NGB is soft with energy below the detection accuracy. However, it seems that no IR divergences appear in the loop correction; the loop integral is finite near the $q\to 0$ region, where $q$ is the momentum of the NGB running in the loop. Showing this requires a lengthy calculation, which will not be performed here. For a handwaving argument, note that usually (as in QED for example) one can cut the soft particle propagator in the IR-divergent $1$-loop diagram, omit one part of the broken propagator, and obtain the tree-level IR-divergent diagram with the second part as a soft particle emitted. In our case, it is easy to see that none of the loop diagrams can produce the tree-level diagram in this way. There are six $1$-loop diagrams involving a NGB:
\begin{equation}
\begin{split}
    \vcenter{\hbox{
    \begin{tikzpicture}[baseline=(c.base)]
    \begin{feynhand}
    \vertex (a) at (-2.8,-1.4) {}; 
    \vertex (b) at (2.8,-1.4) {};
    \vertex (c) [dot,label=left:{$\vone$}] at (-1.4,0) {};
    \vertex (d) [dot,label=right:{$\vthree$}] at (1.4,0) {}; 
    \vertex (e) [dot,label=right:{$\vthree$}] at (0,1.4) {}; 
    \vertex (f) at (0,2.8) {}; 
    \propag [scalar, mom'={$k_1$}] (a) to (c);
    \propag [scalar, mom={$k_2$}] (b) to (d); 
    \propag [plain] (c) to (d);
    \propag[scalar] (c) to (e);
    \propag[plain] (d) to (e);
    \propag[plain, mom={$(p,k_3)$}] (e) to (f);
    \end{feynhand}
    \end{tikzpicture}
    }}
    &
    \vcenter{\hbox{
    \begin{tikzpicture}[baseline=(c.base)]
    \begin{feynhand}
    \vertex (a) at (-2.8,-1.4) {}; 
    \vertex (b) at (2.8,-1.4) {};
    \vertex (c) [dot,label=left:{$\vthree$}] at (-1.4,0) {};
    \vertex (d) [dot,label=right:{$\vone$}] at (1.4,0) {}; 
    \vertex (e) [dot,label=right:{$\vthree$}] at (0,1.4) {}; 
    \vertex (f) at (0,2.8) {}; 
    \propag [scalar, mom'={$k_1$}] (a) to (c);
    \propag [scalar, mom={$k_2$}] (b) to (d); 
    \propag [plain] (c) to (d);
    \propag[plain] (c) to (e);
    \propag[scalar] (d) to (e);
    \propag[plain, mom={$(p,k_3)$}] (e) to (f);
    \end{feynhand}
    \end{tikzpicture}
    }}
    \\
    \vcenter{\hbox{
    \begin{tikzpicture}[baseline=(c.base)]
    \begin{feynhand}
    \vertex (a) at (-2.8,-1.4) {}; 
    \vertex (b) at (2.8,-1.4) {};
    \vertex (c) [dot,label=left:{$\vone$}] at (-1.4,0) {};
    \vertex (d) [dot,label=right:{$\vtwo$}] at (1.4,0) {}; 
    \vertex (e) at (2.8,1.4) {}; 
    \propag [scalar, mom'={$k_1$}] (a) to (c);
    \propag [scalar, mom={$k_2$}] (b) to (d); 
    \propag [scalar] (c) to [out=45, in=135, looseness=1] (d);
    \propag[plain] (c) to [out=-45, in=-135, looseness=1] (d);
    \propag[plain, mom={$(p,k_3)$}] (d) to (e);
    \end{feynhand}
    \end{tikzpicture}
    }}
    &
    \vcenter{\hbox{
    \begin{tikzpicture}[baseline=(c.base)]
    \begin{feynhand}
    \vertex (a) at (-2.8,-1.4) {}; 
    \vertex (b) at (2.8,-1.4) {};
    \vertex (c) [dot,label=left:{$\vtwo$}] at (-1.4,0) {};
    \vertex (d) [dot,label=right:{$\vone$}] at (1.4,0) {}; 
    \vertex (e) at (-2.8,1.4) {}; 
    \propag [scalar, mom'={$k_1$}] (a) to (c);
    \propag [scalar, mom={$k_2$}] (b) to (d); 
    \propag [scalar] (c) to [out=45, in=135, looseness=1] (d);
    \propag[plain] (c) to [out=-45, in=-135, looseness=1] (d);
    \propag[plain, mom'={$(p,k_3)$}] (c) to (e);
    \end{feynhand}
    \end{tikzpicture}
    }}
    \\
    \vcenter{\hbox{
    \begin{tikzpicture}[baseline=(c.base)]
    \begin{feynhand}
    \vertex (a) at (-2.8,-1.4) {}; 
    \vertex (b) at (2.8,-1.4) {};
    \vertex (c) [dot,label=left:{$\vthree$}] at (-1.4,0) {};
    \vertex (d) [dot,label=right:{$\vthree$}] at (1.4,0) {}; 
    \vertex (e) [dot,label=right:{$\vthree$}] at (0,1.4) {}; 
    \vertex (f) at (0,2.8) {}; 
    \propag [scalar, mom'={$k_1$}] (a) to (c);
    \propag [scalar, mom={$k_2$}] (b) to (d); 
    \propag [plain] (c) to (d);
    \propag[scalar] (c) to (e);
    \propag[scalar] (d) to (e);
    \propag[plain, mom={$(p,k_3)$}] (e) to (f);
    \end{feynhand}
    \end{tikzpicture}
    }}
    &
    \vcenter{\hbox{
    \begin{tikzpicture}[baseline=(c.base)]
    \begin{feynhand}
    \vertex (a) at (-2.8,-1.4) {}; 
    \vertex (b) at (2.8,-1.4) {};
    \vertex (c) [dot,label=left:{$\vone$}] at (-1.4,0) {};
    \vertex (d) [dot,label=right:{$\vone$}] at (1.4,0) {}; 
    \vertex (e) [dot,label=right:{$\vbbb$}] at (0,1.4) {}; 
    \vertex (f) at (0,2.8) {}; 
    \propag [scalar, mom'={$k_1$}] (a) to (c);
    \propag [scalar, mom={$k_2$}] (b) to (d); 
    \propag [scalar] (c) to (d);
    \propag[plain] (c) to (e);
    \propag[plain] (d) to (e);
    \propag[plain, mom={$(p,k_3)$}] (e) to (f);
    \end{feynhand}
    \end{tikzpicture}
    }}.
\end{split}    
\end{equation}
Out of them, cutting the NGB such that it connects to a BP propagator is possible in the first four. They give the scattering process of two NGBs to one BP and one NPB, but in the $t,u$ channels and not in the $s$-channel.

\subsection{Recoil of the String}

The second observation is that the IR divergence vanishes when the $x$-momentum of the BPs is conserved, and the string does not recoil. For the scattering of two NGBs to one BP and one NGB discussed in subsection \ref{subsec_sctr_1NGB1BP}, this can be seen from the $p$ factor appearing in the first term of \eqref{eq_sctr_amp_1NP1NGB}. This part contains the $s$-channel from the $\vthreeA$ vertex, and it is finite when $\vec k_3 \to 0$ (and $p \to \pm\sqrt{s-m^2}$), and hence causes the IR divergence in the total cross section. When $p=0$ it vanishes. 

Consider now a general process involving incoming and outgoing BPs. Adding one soft NGB will cause an IR divergence only via a $\vthreeA$ insertion in one of the BP's external legs. If the BP is incoming with momentum $(p,k_1)$ and the NGB has momentum $k_2$, the diagram is modified as follows:
\begin{equation}
    \vcenter{\hbox{
    \begin{tikzpicture}[baseline=(c.base)]
    \begin{feynhand}
    \vertex (a) at (0,-1.4) {}; 
    \vertex (b) [dot,label=left:{$\vthreeA$}] at (0,0) {};
    \vertex (c) [NWblob] at (0,1.4) {};
    \vertex (d) at (1.4,0) {}; 
    \propag [plain, mom={$(p,k_1)$}] (a) to (b);
    \propag [plain, mom={$(p,k_1 - k_2)$}] (b) to (c); 
    \propag [scalar, mom={$k_2$}] (b) to (d);
    \end{feynhand}
    \end{tikzpicture}
    }} \to 
    \frac{i\bb}{\sqrt{8m}}(-k_2\cdot (k_1+(k_1-k_2))ip)\frac{i}{(k_1-k_2)^2-p^2-m^2} = -\frac{i\bb p}{\sqrt{8m}}.
\end{equation}
If the BP is outgoing with momentum $(p,k_1)$ and the NGB has momentum $k_2$, the diagram is modified as follows:
\begin{equation}
    \vcenter{\hbox{
    \begin{tikzpicture}[baseline=(c.base)]
    \begin{feynhand}
    \vertex (a) [NWblob] at (0,-1.4) {}; 
    \vertex (b) [dot,label=left:{$\vthreeA$}] at (0,0) {};
    \vertex (c) at (0,1.4) {};
    \vertex (d) at (1.4,0) {}; 
    \propag [plain, mom={$(p,k_1+k_2)$}] (a) to (b);
    \propag [plain, mom={$(p,k_1)$}] (b) to (c); 
    \propag [scalar, mom={$k_2$}] (b) to (d);
    \end{feynhand}
    \end{tikzpicture}
    }} \to 
    \frac{i\bb}{\sqrt{8m}}(-k_2\cdot ((k_1+k_2)+k_1)ip)\frac{i}{(k_1+k_2)^2-p^2-m^2} = \frac{i\bb p}{\sqrt{8m}}.
\end{equation}
Hence, for each external BP with $x$-momentum $p$, the diagram is multiplied by $f(p) \equiv \frac{i\bb}{\sqrt{8m}}p$ to get the IR divergence, and all in all the diagram is multiplied by the factor: $\sum_i f(p_i) = \frac{i\bb}{\sqrt{8m}}\sum_i p_i$, where incoming momenta are counted with a minus sign. Thus, the IR divergence vanishes if the $x$-momentum is conserved $\sum_i p_i = 0$.

When two or more soft NGBs are added, this factor exponentiates, maintaining the elimination of the IR divergence when the $x$-momentum is conserved. We will show this for two additional NGBs and the generalization to three or more is simple. Adding the two NGBs together with $\vtwoA$ does not give a finite contribution when the NGBs become soft. Taking an outgoing BP for example gives
\begin{equation}
\begin{split}
    &
    \vcenter{\hbox{
    \begin{tikzpicture}[baseline=(c.base)]
    \begin{feynhand}
    \vertex (a) [NWblob] at (0,-1.4) {}; 
    \vertex (b) [dot,label=left:{$\vtwoA$}] at (0,0) {};
    \vertex (c) at (0,1.4) {};
    \vertex (d) at (1.4,-1.2) {}; 
    \vertex (e) at (1.4,1.2) {}; 
    \propag [plain, mom={$(p,k_1+k_2+k_3)$}] (a) to (b);
    \propag [plain, mom={$(p,k_1)$}] (b) to (c); 
    \propag [scalar, mom={$k_2$}] (b) to (d);
    \propag [scalar, mom={$k_3$}] (b) to (e);
    \end{feynhand}
    \end{tikzpicture}
    }}
    \to 
    \\
    & \frac{i\bb^2}{\sqrt{4m}}(-(k_2\cdot k_3)p^2)\frac{i}{(k_1+k_2+k_3)^2-p^2-m^2} 
    = \frac{\bb^2 p^2}{8m}\frac{k_2\cdot k_3}{k_1\cdot k_2 + k_1\cdot k_3 + k_2\cdot k_3}.
\end{split}
\end{equation}
This factor vanishes when $k_2$ or $k_3$ goes to zero ($k_1$ is finite). So only NGBs added via $\vthreeA$ lead to IR divergences. They can be on two different legs or on the same leg. In the first case, the diagram is modified by $2f(p_i)f(p_j)$; the factor of $2$ comes from the two possibilities of inserting the two NGBs (they have different momenta). In the second case, there are also two possibilities regarding the order of the two NGBs. Taking an outgoing BP for example:
\begin{equation}
\begin{split}
    & 
    \vcenter{\hbox{
    \begin{tikzpicture}[baseline=(c.base)]
    \begin{feynhand}
    \vertex (a) [NWblob] at (0,-1.4) {}; 
    \vertex (b) [dot,label=left:{$\vthreeA$}] at (0,0) {};
    \vertex (c) [dot,label=left:{$\vthreeA$}] at (0,1.4) {};
    \vertex (d) at (0,2.8) {};
    \vertex (e) at (1.4,0) {}; 
    \vertex (f) at (1.4,1.4) {}; 
    \propag [plain, mom={$(p,k_1+k_2+k_3)$}] (a) to (b);
    \propag [plain, mom={$(p,k_1 + k_2)$}] (b) to (c); 
    \propag [plain, mom={$(p,k_1)$}] (c) to (d); 
    \propag [scalar, mom={$k_3$}] (b) to (e);
    \propag [scalar, mom={$k_2$}] (c) to (f);
    \end{feynhand}
    \end{tikzpicture}
    }} + (k_2 \leftrightarrow k_3) 
    \to 
    \\
    &
    \pr{\frac{i\bb}{\sqrt{8m}}ip}^2\frac{i(-k_3\cdot (2k_1+2k_2+k_3))}{(k_1+k_2+k_3)^2-p^2-m^2}\frac{i(-k_2\cdot (2k_1+k_2))}{(k_1+k_2)^2-p^2-m^2} + (k_2 \leftrightarrow k_3) 
    \\
    = & -\frac{\bb^2p^2}{8m}\frac{k_3\cdot (k_1+k_2)}{k_1\cdot k_2 + k_1\cdot k_3 + k_2\cdot k_3} + (k_2 \leftrightarrow k_3) = -\frac{\bb^2}{8m}\pr{1 + \frac{k_2\cdot k_3}{k_1\cdot k_2 + k_1\cdot k_3 + k_2\cdot k_3}}.
\end{split}    
\end{equation}
The first term is $f(p)^2$, and the second term is the same as encountered with the $\vtwoA$ vertex, vanishing in the soft limit (they also cancel each other, so the total contribution is exactly $f(p)^2$). Therefore, the total contribution is:
\begin{equation}
    \sum_{i\neq j}2f(p_i)f(p_j) + \sum_i f(p_i)^2 = \pr{\sum_i f(p_i)}^2,
\end{equation}
as we claimed.

%% file: appendices/integrals.tex
In this appendix, we derive some of the identities used in the paper.

\subsection{Flux-Tube Profile}\label{app: flux tube profile}

We demonstrate the cancellation of the divergence \eqref{eq: normal ordered vertex op times divergence}, at first nontrivial order ($O\pr{\bb}$). There, we have:
\bea
\left\langle \pd_x\phi\right\rangle _{\bb} = & \underbrace{\left\langle :e^{-ipx_0\pr{y,t}}:_M\right\rangle _{\bb^{0}}\frac{2\pi}{\bb}\sech\pr{\frac{\pi p}{2m}}\frac{1}{2}\frac{\bb^2}{8m}\frac{1}{2\pi\epsilon_M}p^2}_{E_1}\\
+ & \underbrace{{\left\langle :e^{-ipx_0\pr{y,t}}:_M\right\rangle _{\bb^2}}\frac{2\pi}{\bb}\sech\pr{\frac{\pi p}{2m}}}_{E_2}\\
+ & \underbrace{\left\langle :e^{-ipx_0\pr{y,t}}:_M\right\rangle _{{\rm \bb^{0}}}\intop dxe^{-ipx}{\left\langle \pd_x\dphib\pr{x,y,t}\right\rangle _{\bb}}}_{E_3}\\
+ & \underbrace{\intop dxe^{-ipx}{\left\langle :e^{-ipx_0\pr{y,t}}:_M\pd_x\dphib\pr{x,y,t}\right\rangle _{\bb;{\rm con}}}}_{E_4}.
\eea
$E_2$ vanishes, because $\left\langle :e^{-ipx_0\pr{y,t}}:_M\right\rangle _{\bb^2}$ vanishes. The only viable diagram comes from $\vthree$, but this vanishes by symmetry considerations in the $\dphib$ loop. $E_4$ vanishes because there are no viable diagrams. We have:
\bea
E_1 =\left\langle :e^{-ipx_0\pr{y,t}}:_M\right\rangle _{\bb^{0}}\bb A_1, \ A_1 \equiv\frac{\pi}{8m}\sech\pr{\frac{\pi p}{2m}}p^2\frac{1}{2\pi\epsilon_M}.
\eea

We need to examine $E_3$ also. Consider:
\bea\label{eq: A_2 and E_3}
E_3 = - \left\langle :e^{-ipx_0\pr{y,t}}:_M\right\rangle _{\bb^{0}} A_2, \ A_2 \equiv - \intop dxe^{-ipx}\left\langle \pd_x\dphib\pr{x,y,t}\right\rangle _{\bb} / \bb.
\eea
This gets a contribution from the vertex $\vbbb$, from the cubic vertex $\vone$, from the tadpole counterterms:
\beq
-\bb\delta m_2^2\left(2\tanh(mx)\sech(mx)\right)\dphib,
\eeq
and from the Jacobian:
\beq
\frac{\bb}{4}\delta^2\pr{0}\sech(mx)\pd_x\dphib.
\eeq
The contribution from the $\vone$
vertex cancels formally with the Jacobian counterterm because:
\beq
-\frac{\bb}{4}\left\langle \pr{\pd x_0}^2\right\rangle_{\bb^0} =-\frac{\bb}{4}\intop\frac{d^2p}{\left(2\pi\right)^2}\frac{iq_a\pr{-iq^a}}{p^2}=-\frac{\bb}{4}\delta^2\pr{0},
\eeq
which is just right to be canceled by the counterterm. Notice that
the cancellation is complete, even when we compactify the $y$ direction,
since the counterterm changes too, as it arises from the Jacobian. This is in contrast with ordinary counterterms in the Lagrangian, whose coefficients cannot depend on non-local information such as $L_y$. The remaining contributions give:
\bea
-\left\langle \pd_x\dphib\pr{x,\sg}\right\rangle _{\bb} / \bb & =  \left\langle \pd_x\dphib\pr{x,y,t} \pr{\frac{2m^2}{3!}\tanh(mx)\sech(mx)\pr{\dphib}^3+ {\rm c.t.}}  \right \rangle_{\bb^0}  \\
& = \intop d{x'}\left(m^2G_{{\rm bulk}}\pr{{x'},0;{x'},0}-2\de m_2^2\right)\sinh(m{x'})\sech^2(m{x'})\pd_xG_{{\rm bulk};\perp}\pr{x;{x'}},
\eea
with:
\bea
G_{{\rm bulk};\perp}\pr{x;{x'}} \equiv \intop\frac{dp}{2\pi}\frac{1}{p^2+m^2}e^{ip\pr{x-{x'}}}\frac{\left(ip+m\tanh\pr{m{x'}}\right)\left(-ip+m\tanh(mx)\right)}{p^2+m^2}.
\eea
$G_{{\rm bulk};\perp}(x,x')$ appears after integrating the propagator $G_{{\rm bulk}}\pr{x,\sg;x',\sg'}$ over the vertex's position in the parallel directions $\sg'$. The BP loop $G_{{\rm bulk}}\pr{{x'},0;{x'},0}$ is placed at $\sg=0$ by translation symmetry. It is natural to separate the loop into a ``bulk part'' and a ``string part'':
\bea
 & \left(m^2G_{{\rm bulk}}\pr{{x'},0;{x'},0}-2\de m_2^2\right)
 \\
= & m^2\underbrace{\left(\frac{\mu^{-\ep}}{L_y}\sum_{n_y}\intop\frac{d^{2+\ep}p}{\left(2\pi\right)^{2+\ep}}\frac{1}{p^2+\pr{\frac{2\pi n_y}{L_y}}^2+m^2}-2\frac{\de m_2^2}{m^2}\right)}_{\equiv I_1}\\
+ & \underbrace{\left(-\frac{\mu^{-\ep}}{L_y}\sum_{n_y}\intop\frac{d^{1+\ep}q}{\left(2\pi\right)^{1+\ep}}\frac{dp}{2\pi}\frac{1}{q^2+\pr{\frac{2\pi n_y}{L_y}}^2+p^2+m^2}\frac{2m}{p^2+m^2}\right)}_{\equiv I_2}\frac{1}{2}m^3{\sech}^2\pr{m{x'}}.
\eea
Note that in $I_1$, $p$ denotes the momenta in all $2+\epsilon$ noncompact directions (including $x)$. Thus:
\bea
A_2= & I_1B_1+I_2B_2,
\eea
\bea
B_1 & =m^2\intop dxe^{-ipx}\intop d{x'}\sinh(m{x'})\sech^2(m{x'})\pd_xG_{{\rm bulk};\perp}\pr{x;{x'}}\\
 & =m\intop dxe^{-ipx}\intop d{x'}\sech(m{x'})\pd_x\pd_{{x'}}G_{{\rm bulk};\perp}\pr{x;{x'}},
\eea
\bea
B_2 & =\frac{1}{2}m^3\intop dxe^{-ipx}\intop d{x'}\sinh(m{x'})\sech^4(m{x'})\pd_xG_{{\rm bulk};\perp}\pr{x;{x'}}\\
 & =\frac{1}{6}m^2\intop dxe^{-ipx}\intop d{x'}\sech^3(m{x'})\pd_x\pd_{{x'}}G_{{\rm bulk};\perp}\pr{x;{x'}}.
\eea
The $B$'s contain the $p$ dependence, while the $I$'s act as
coefficients. The term $I_1B_1$ represents the ``pure bulk''
contribution, and is finite, thanks to the counterterm. Our main interest
is in $I_2B_2$, which is divergent, and needs to cancel the divergence
in $A_1$. The next subsections are devoted to the calculation of these four quantities. The results, \eqref{eq: I1 coefficient of B1}, \eqref{eq: I2 coefficient of B2}, \eqref{eq: B1 px dependence} and \eqref{eq: B2 px dependence} are:
\begin{equation}
\begin{split}
    B_1 & =\frac{\pi}{4m}\sech\pr{\frac{\pi p}{2m}}\pi p\tanh\pr{\frac{\pi p}{2m}},
    \\
    I_1 & =-\frac{\log\pr{1-e^{-mL_y}}}{2\pi L_y},
    \\
    B_2 & =\frac{\pi}{8m}\sech\pr{\frac{\pi p}{2m}}p^2,
    \\
    I_2 & =\frac{1}{2\pi\ep_M}+\frac{1}{2\pi}\log\pr{\frac{m}{2M}}+O\pr{L_y^{-1}}.
\end{split}
\end{equation}
We see that $B_2$ has the right $p$ dependence to cancel the
divergence in $A_1$, and $I_2$ has the right coefficient:
\bea\label{eq: NLO e-field profile}
\left\langle \pd_x\phi\right\rangle _{\bb} & =\left\langle :e^{-ipx_0\pr{y,t}}:_M\right\rangle _{\bb^{0}}\frac{\pi}{8m}\sech\pr{\frac{\pi p}{2m}}p^2\left(-\frac{1}{2\pi\ep_M}+I_2\right)+\cdots\\
 & =\left\langle :e^{-ipx_0\pr{y,t}}:_M\right\rangle _{\bb^{0}}\frac{\pi}{8m}\sech\pr{\frac{\pi p}{2m}}p^2\pr{\frac{1}{2\pi}\log\pr{\frac{m}{2M}}+O\pr{L_y^{-1}}}+\cdots .
\eea

\subsubsection{\texorpdfstring{$I_1$}{I1}}

\bea
I_1=\frac{\mu^{-\ep}}{L_y}\sum_{n_y}\intop\frac{d^{2+\ep}p}{\left(2\pi\right)^{2+\ep}}\frac{1}{p^2+\pr{\frac{2\pi n_y}{L_y}}^2+m^2}-2\frac{\de m_2^2}{m^2}.
\eea
Conveniently, the counterterm just gives the non-compact limit of
the integral:
\bea
2\frac{\de m_2^2}{m^2}=\mu^{-\ep}\sum_{n_y}\intop\frac{d^{3+\ep}p}{\left(2\pi\right)^{3+\ep}}\frac{1}{p^2+m^2},
\eea
so after summing over $n_y$ (and integrating over $p_y$ for the counterterm)
we can combine the two terms under the same integral sign and remove
the regulator. Then we change to $E=\sqrt{p^2+m^2}$:
\bea\label{eq: I1 coefficient of B1}
I_1 & =\mu^{-\ep}\intop\frac{d^{2+\ep}p}{\left(2\pi\right)^{2+\ep}}\frac{\coth\pr{\frac{1}{2}L_y\sqrt{m^2+p^2}}}{2\sqrt{m^2+p^2}}-2\frac{\de m_2^2}{m^2}\\
 & =\mu^{-\ep}\intop\frac{d^{2+\ep}p}{\left(2\pi\right)^{2+\ep}}\frac{\coth\pr{\frac{1}{2}L_y\sqrt{m^2+p^2}}-1}{2\sqrt{m^2+p^2}}\\
 & =\intop\frac{d^2p}{\left(2\pi\right)^2}\frac{\coth\pr{\frac{1}{2}L_y\sqrt{m^2+p^2}}-1}{2\sqrt{m^2+p^2}}+O\pr{\ep}\\
 & =\frac{1}{4\pi}\intop_{m}^{\infty}dE\pr{\coth\pr{\frac{1}{2}L_yE}-1}\\
 & =\frac{m}{4\pi}-\frac{\log\left(2\sinh\pr{\frac{mL_y}{2}}\right)}{2\pi L_y}\\
 & =-\frac{\log\pr{1-e^{-mL_y}}}{2\pi L_y}.
\eea

\subsubsection{\texorpdfstring{$I_2$}{I2}}

\bea
I_2= & -\intop\frac{d^2qdp}{\left(2\pi\right)^3}\frac{1}{q^2+p^2+m^2}\frac{2m}{p^2+m^2}\\
\to & -\mu^{-\ep}\frac{1}{L_y}\sum_{n_y}\intop\frac{d^{1+\ep}q}{\left(2\pi\right)^{1+\ep}}\frac{dp}{2\pi}\frac{1}{q^2+\pr{\frac{2\pi}{L_y}n_y}^2+p^2+m^2}\frac{2m}{p^2+m^2}.
\eea
The non-compact case:
\bea
I_{2;\text{non-compact}}= & -\mu^{-\ep}\intop\frac{d^{2+\ep}q}{\left(2\pi\right)^{2+\ep}}\frac{dp}{2\pi}\frac{1}{q^2+p^2+m^2}\frac{2m}{p^2+m^2}\\
= & \mu^{-\ep}\intop\frac{dp}{2\pi}\frac{m2^{-\ep-1}\pi^{-\frac{\ep}{2}}\csc\pr{\frac{\pi\ep}{2}}\pr{k^2+m^2}^{\frac{\ep}{2}-1}}{\Gamma\pr{\frac{\ep+2}{2}}}\\
= & \frac{1}{\ep}2^{-\ep-1}\mu^{-\ep}\pi^{-\frac{\ep}{2}-\frac{3}{2}}m^{\ep}\Gamma\pr{\frac{1}{2}-\frac{\ep}{2}}\\
= & \frac{1}{2\pi\ep}+\frac{-2\log(\frac{\mu}{m})+\gamma-\log(\pi)}{4\pi}.
\eea
To compute the compact case, we add and subtract the non-compact:
\bea
I_2=I_{2;\text{non-compact}}+\left(I_2-I_{2;\text{non-compact}}\right).
\eea
To evaluate the difference, we can carry out first the $n_y$ sum (and $q_y$ integral for the non-compact subtraction) to combine the two under the same integral sign, and then remove the regulator. It is then simple to change to polar coordinates in the $q_{t}-p$ plane and integrate over the angle $\theta$. The final integral over $r=\sqrt{p^2+q_{t}^2}$ cannot be evaluated in closed form, but can be cleaned up a bit by changing variables to $E=\sqrt{r^2+m^2}$ and then $x=E/m-1\in\left(0,\infty\right)$:
\bea
 & I_2-I_{2;\text{non-compact}}\\
= & -\left(\mu^{-\ep}\frac{1}{L_y}\sum_{n_y}\intop\frac{d^{1+\ep}q}{\left(2\pi\right)^{1+\ep}}\frac{dp}{2\pi}\frac{1}{q^2+\pr{\frac{2\pi}{L_y}n_y}^2+p^2+m^2}\frac{2m}{p^2+m^2}-{\rm reg}\right)\\
= & -\left(\mu^{-\ep}\intop\frac{d^{1+\ep}q}{\left(2\pi\right)^{1+\ep}}\frac{dp}{2\pi}\frac{m\coth\pr{\frac{1}{2}L_y\sqrt{p^2+m^2+q^2}}}{\pr{p^2+m^2}\sqrt{p^2+m^2+q^2}}-{\rm reg}\right)\\
= & -\pr{\intop\frac{dq}{2\pi}\frac{dp}{2\pi}\frac{m\pr{\coth\pr{\frac{1}{2}L_y\sqrt{p^2+m^2+q^2}}-1}}{\pr{p^2+m^2}\sqrt{p^2+m^2+q^2}}+O\pr{\ep}}\\
= & -\intop_{0}^{\infty}dx\frac{1}{\pi\pr{e^{mL_y+x}-1}\pr{mL_y+x}}+O\pr{\ep}\\
= & \frac{1}{\pi}\sum_{n=1}^{\infty}\text{Ei}\pr{-mnL_y}+O\pr{\ep}\\
= & \frac{1}{\pi}\sum_{n=1}^{\infty}e^{-mnL_y}\left(-\frac{1}{mnL_y}+O\pr{L_y^{-2}}\right)+O\pr{\ep}.
\eea
In summary:
\bea\label{eq: I2 coefficient of B2}
I_2= & \frac{1}{2\pi\ep}+\frac{-2\log(\frac{\mu}{m})+\gamma-\log(\pi)}{4\pi}-\intop_{0}^{\infty}dx\frac{1}{\pi\pr{e^{mL_y+x}-1}\pr{mL_y+x}}+O\pr{\ep}\\
= & \frac{1}{2\pi\ep}+\frac{-2\log(\frac{\mu}{m})+\gamma-\log(\pi)}{4\pi}+\frac{1}{\pi}\sum_{n=1}^{\infty}e^{-mnL_y}\left(-\frac{1}{mnL_y}+O\pr{L_y^{-2}}\right)+O\pr{\ep}.
\eea

\subsubsection{\texorpdfstring{$B_1$}{B1}}

Recall that:
\bea
G_{{\rm bulk};\perp}\pr{x;x_2}=\intop\frac{dp'}{2\pi}\frac{1}{p'^2+m^2}e^{ip'\pr{x-x_2}}\frac{\left(ip'+m\tanh\pr{mx_2}\right)\left(-ip'+m\tanh(mx)\right)}{p'^2+m^2}.
\eea

Let us compute $B_1$:
\bea
B_1 & =m\intop dxe^{-ipx}\intop dx_2\sech(mx_2)\pd_x\pd_{x_2}G_{{\rm bulk};\perp}\pr{x;x_2}.
\eea
First note:
\bea
 \, & \intop dx_2\sech\pr{mx_2}\pd_{x_2}\left(e^{p'x_2}\left(ip'+m \tanh\pr{mx_2}\right)\right)\\
= & \intop dx_2e^{-ip'x_2}\sech\pr{mx_2}\left(\cancel{-ip'\left(ip'+m\tanh\pr{mx_2}\right)}+m^2\sech^2\pr{mx_2}\right)\\
= & m^2\intop dx_2e^{-ip'x_2}\sech^3\pr{mx_2}\,\text{(by orthogonality)}\\
= & \frac{\pi}{2m}\pr{m^2+p'^2}\sech\pr{\frac{\pi p'}{2m}}.
\eea
We will need another side-integral
:
\bea
 \intop dx\intop\frac{dp'}{2\pi}\sech\pr{\frac{\pi p'}{2m}}e^{i\pr{p'-p}x}\frac{\tanh(mx)}{p'^2+m^2}= \frac{1}{\pi m}\frac{-i\pr{\pi\pr{m^2+p^2}\tanh\pr{\frac{\pi p}{2m}}-2mp}\pi\sech\pr{\frac{\pi p}{2m}}}{2m\pr{p^2+m^2}}.
\eea

Putting it all together:
\bea
 \, & \intop dxdx_2e^{-ipx}\sech\pr{mx_2}\pd_x\pd_{x_2}G_{{\rm \perp}}\pr{x,x_2}\\
= & \intop dxdx_2e^{-ipx}\sech\pr{mx_2}\pd_x\pd_{x_2}\intop\frac{dp'}{2\pi}\left(e^{ip'\pr{x-x_2}}\frac{\left(ip'+m\tanh\pr{mx_2}\right)\left(-ip'+m\tanh(mx)\right)}{\pr{p'^2+m^2}^2}\right)\\
= & \frac{\pi}{2m}ip\intop dx\intop\frac{dp'}{2\pi}\sech\pr{\frac{\pi p'}{2m}}e^{i\pr{p'-p}x}\frac{\left(-ip'+m\tanh(mx)\right)}{p'^2+m^2}\\
= & \frac{\pi}{2m}\sech\pr{\frac{\pi p}{2m}}\frac{p^2}{p^2+m^2}+\frac{\pi}{2}ip\intop dx\intop\frac{dp'}{2\pi}\sech\pr{\frac{\pi p'}{2m}}e^{i\pr{p'-p}x}\frac{\tanh(mx)}{p'^2+m^2}\\
= & \frac{\pi}{2m}\sech\pr{\frac{\pi p}{2m}}\frac{p^2}{p^2+m^2}+\frac{p}{2m}\frac{\pr{\pi\pr{m^2+p^2}\tanh\pr{\frac{\pi p}{2m}}-2mp}\pi\sech\pr{\frac{\pi p}{2m}}}{2m\pr{p^2+m^2}}\\
= & \pr{\frac{\pi}{2m}}^2\sech\pr{\frac{\pi p}{2m}}p\tanh\pr{\frac{\pi p}{2m}}.
\eea
So we find:
\bea\label{eq: B1 px dependence}
B_1 & =\frac{\pi^2}{4m}\sech\pr{\frac{\pi p}{2m}}p\tanh\pr{\frac{\pi p}{2m}}.
\eea

\subsubsection{\texorpdfstring{$B_2$}{B2}}
\bea
B_2=\frac{1}{6}m^2\intop dxe^{-ipx}\intop dx_2\sech^3(mx_2)\pd_x\pd_{x_2}G_{{\rm bulk};\perp}\pr{x;x_2}.
\eea
First, note that:
\bea
 \, & \intop dx_2\sech^3\pr{mx_2}\pd_{x_2}\left(e^{-ip'x_2}\left(ip'+m\tanh\pr{mx_2}\right)\right)\\
= & \intop dx_2e^{-ip'x_2}\sech^3\pr{mx_2}\left(-ip'\left(ip'+m\tanh\pr{mx_2}\right)+m^2\sech^2\pr{mx_2}\right)\\
= & \intop dx_2e^{-ip'x_2}\left(p'^2\sech^3\pr{mx_2}-ip'm\sech^4\pr{mx_2}\sinh\pr{mx_2}+m^2\sech^{5}\pr{mx_2}\right)\\
= & \intop dx_2e^{-ip'x_2}\pr{\frac{2}{3}p'^2\sech^3\pr{mx_2}+m^2\sech^{5}\pr{mx_2}}\\
= & \frac{3\pi\pr{p'^2+m^2}{}^2\sech\pr{\frac{\pi p'}{2m}}}{8m^3}.
\eea
The factor of $\pr{p'^2+m^2}^2$ is just right to cancel the denominator in $G_{{\rm bulk};\perp}$. We use this to carry out the $x_2$ integral in:
\bea
 \, & \intop dxdx_2e^{-ipx}\sech^3\pr{mx_2}\pd_x\pd_{x_2}G_{{\rm \perp}}\pr{x,x_2}\\
= & \intop dxdx_2\frac{dp'}{2\pi}\sech^3\pr{mx_2}e^{-ipx}\pd_x\pd_{x_2}\left(e^{ip'\pr{x-x_2}}\frac{\left(ip'+m\tanh\pr{mx_2}\right)\left(-ip'+m\tanh(mx)\right)}{\pr{p'^2+m^2}^2}\right)\\
= & \frac{3\pi}{8m^3}\intop dx\frac{dp'}{2\pi}\sech\pr{\frac{\pi p'}{2m}}ipe^{i\pr{p'-p}x}\left(-ip-ip'+\left(ip+m\tanh(mx)\right)\right)\\
= & \frac{3\pi}{4m^3}\sech\pr{\frac{\pi p}{2m}}p^2+\frac{3\pi}{8m^3}ip\intop dx\frac{dp'}{2\pi}\sech\pr{\frac{\pi p'}{2m}}e^{i\pr{p'-p}x}\left(ip+m\tanh(mx)\right)\\
= & \frac{3\pi}{4m^3}\sech\pr{\frac{\pi p}{2m}}p^2+\frac{3}{8m^2}ip\cancel{\intop dx\sech(mx)e^{-ipx}\left(ip+m\tanh(mx)\right)}\\
= & \frac{3\pi}{4m^3}\sech\pr{\frac{\pi p}{2m}}p^2\,
\eea
where in the before-last line we used orthogonality of the eigenfunctions.
In summary, we find:
\bea\label{eq: B2 px dependence}
B_2=\frac{\pi}{8m}\sech\pr{\frac{\pi p}{2m}}p^2.
\eea

\subsection{Ground-State Energy}

\subsubsection{PI Framework}

For $E_{{\rm string}}$ we evaluate the sum-integral:
\beq
\frac{1}{2}\sum_{n_y}\mu^{-\ep}\intop\frac{d^{1+\ep}q}{\left(2\pi\right)^{1+\ep}}\left(\log\left(q^2+\pr{\frac{2\pi n_y}{L_y}}^2\right)-\intop\frac{dp}{2\pi}\log\left(q^2+\pr{\frac{2\pi n_y}{L_y}}^2+p^2+m^2\right)\frac{2m}{p^2+m^2}\right).
\eeq
Before regularization and compactification, this is \eqref{eq_ground_Egs}:
\beq
\frac{1}{2}L_y\intop\frac{d^2q}{\left(2\pi\right)^2}\pr{\log\pr{q^2}-\intop\frac{dp}{2\pi}\log\pr{q^2+p^2+m^2}\frac{2m}{p^2+m^2}}.
\eeq
We can carry out the integral over the pseudo-momentum
$p$, this commutes with dim-regularization and $y$-compactification:
\bea
\dots= & -L_y\intop\frac{d^2q}{\left(2\pi\right)^2}{\rm arctanh}\pr{\frac{m}{\sqrt{m^2+q^2}}}.
\eea
Now, we regulate, compactify, and spherically integrate in the noncompact directions:
\bea
 \, & -L_y\intop\frac{d^2q}{\left(2\pi\right)^2}{\rm arctanh}\pr{\frac{m}{\sqrt{m^2+q^2}}}\\
\to & -\sum_{n_y}\mu^{-\ep}\intop\frac{d^{1+\ep}q}{\left(2\pi\right)^{1+\ep}}{\rm arctanh}\pr{\frac{m}{\sqrt{m^2+q^2+\pr{\frac{2\pi n_y}{L_y}}^2}}}\\
= & -\frac{2\mu^{-\ep}\pi^{\frac{1+\ep}{2}}}{\left(2\pi\right)^{1+\ep}\Gamma\pr{\frac{1+\ep}{2}}}\sum_{n_y}\intop dqq^{\ep}{\rm arctanh}\pr{\frac{m}{\sqrt{m^2+q^2+\pr{\frac{2\pi n_y}{L_y}}^2}}}.
\eea
Integration by parts gives:
\bea\label{eq: E string calc checkpoint 1}
\dots= & -m\frac{2\mu^{-\ep}\pi^{\frac{1+\ep}{2}}}{\left(2\pi\right)^{1+\ep}\Gamma\pr{\frac{1+\ep}{2}}\left(1+\ep\right)}\sum_{n_y}\intop dqq^{2+\ep}\frac{1}{\pr{\pr{\frac{2\pi n_y}{L_y}}^2+q^2}\sqrt{\pr{\frac{2\pi n_y}{L_y}}^2+m^2+q^2}}.
\eea
Note that the prefactor outside the sum-integral is finite at $\ep\to0$.
By strategic addition and subtraction of a helper term, we separate
the expression into a finite-but-complicated part (for which we remove
the regulator) and a divergent-but-simple part, and evaluate the $q$-integral:
\bea\label{eq: E string calc checkpoint 2}
\, & \sum_{n_y}\intop dqq^{2+\ep}\frac{1}{\pr{\pr{\frac{2\pi n_y}{L_y}}^2+q^2}\sqrt{\pr{\frac{2\pi n_y}{L_y}}^2+m^2+q^2}}\\
= & \sum_{n_y}\intop dqq^2\frac{1}{\sqrt{\pr{\frac{2\pi n_y}{L_y}}^2+m^2+q^2}}\pr{\frac{1}{\pr{\frac{2\pi n_y}{L_y}}^2+q^2}-\frac{1}{\pr{\frac{2\pi n_y}{L_y}}^2+m^2+q^2}}\\
+ & \sum_{n_y}\intop dqq^{2+\ep}\frac{1}{\pr{\pr{\frac{2\pi n_y}{L_y}}^2+m^2+q^2}^{3/2}}\\
= & \sum_{n_y}\left(1-\frac{2\pi\left|n_y\right|}{mL_y}\arctan\pr{\left|\frac{mL_y}{2\pi n_y}\right|}\right)
+ \frac{\Gamma\pr{\frac{\ep+3}{2}}m^{\ep}}{\sqrt{\pi}}\Gamma\left(-\frac{\ep}{2}\right)\sum_{n_y}\pr{\pr{\frac{2\pi n_y}{mL_y}}^2+1}^{\ep/2}.
\eea
First, let us evaluate the divergent sum. Writing $a\equiv\frac{mL_y}{2\pi}$, we can factor $\pr{\frac{n_y}{a}}^2+1=a^{-2}\pr{n_y+ia}\pr{n_y-ia}$ and combine the two linear polynomials using the Feynman trick:
\bea
 \, & \Gamma\left(-\frac{\ep}{2}\right)\sum_{n_y}\left(1+\pr{\frac{n_y}{a}}^2\right)^{\ep/2}\\
= & \Gamma\left(-\frac{\ep}{2}\right)+2\frac{\Gamma\left(-\ep\right)}{\Gamma\left(-\frac{\ep}{2}\right)}a^{-\ep}\intop_{0}^{1}dxx^{-1-\ep/2}\pr{1-x}^{-1-\ep/2}\sum_{n_y>0}\left(ia\pr{2x-1}+n_y\right)^{\ep}\\
= & \Gamma\left(-\frac{\ep}{2}\right)+2\frac{\Gamma\left(-\ep\right)}{\Gamma\left(-\frac{\ep}{2}\right)}a^{-\ep}\intop_{0}^{1}dxx^{-1-\ep/2}\pr{1-x}^{-1-\ep/2}\zeta(-\ep,1+ia\pr{2x-1})\\
= & \Gamma\left(-\frac{\ep}{2}\right)+\frac{\Gamma\left(-\ep\right)}{\Gamma\left(-\frac{\ep}{2}\right)}a^{-\ep}\intop_{0}^{1}dxx^{-1-\ep/2}\pr{1-x}^{-1-\ep/2}\left(\zeta(-\ep,1+ia\pr{2x-1})+{\rm c.c.}\right),
\eea
where $\zeta\pr{z,w}$ is the Hurwitz zeta function, and on
the last we line used the reflection $x\to1-x$ to make the integrand
manifestly real. We again separate into finite-but-complicated and
divergent-but-simple, this time by adding and subtracting the values
where $\zeta$ is evaluated at the endpoint $x=1$, and remove
the regulator for the finite part:
\bea
\dots= & \Gamma\left(-\frac{\ep}{2}\right)+a^{-\ep}\Gamma\left(-\frac{\ep}{2}\right)\left(\zeta(-\ep,1+ia)+{\rm c.c.}\right)\\
+ & 2\intop_{0}^{1}dxx^{-1}\pr{1-x}^{-1}\left(\zeta(0,1+ia\pr{2x-1})-\zeta(0,1+ia)+{\rm c.c.}\right).
\eea
It turns out that at $\ep=0$, the real part of $\zeta(0,1+ia\pr{2x-1})$
is $x$-independent, and so the entire finite part cancels exactly.
We also find that the divergences in the putative divergent part cancel,
leaving:
\bea
\dots= & \Gamma\left(-\frac{\ep}{2}\right)+\left(1-\ep\log\pr{a}+O\pr{\ep^2}\right)\Gamma\left(-\frac{\ep}{2}\right)\left(-\frac{1}{2}-\frac{1}{2}+\ep\left(\pd_{\ep}\zeta(-\ep,1+ia)\Big|_{\ep=0}+{\rm c.c.}\right)+O\pr{\ep^2}\right)\\
+ & 2\intop_{0}^{1}dxx^{-1}\pr{1-x}^{-1}\left(-1-\pr{-1}\right)\\
= & 2\left(-\log\pr{a}-\left(\pd_{\ep}\zeta(-\ep,1+ia)\Big|_{\ep=0}+{\rm c.c.}\right)+O\pr{\ep^{1}}\right)\\
= & 2\left(-\log\pr{a}+\left(\log\pr{\Gamma\pr{1+ia}\Gamma\pr{1-ia}}-\log\left(2\pi\right)\right)+O\pr{\ep^{1}}\right)\\
= & 2\left(-\log\pr{a}+\log\pr{\pi a{\rm csch}\pr{\pi a}}+O\pr{\ep^{1}}\right)\\
= & 2\pr{\log\pr{\frac{1}{2}}-\log\pr{\sinh\pr{\pi a}}+O\pr{\ep^{1}}}.
\eea
Thus:
\beq
\frac{\Gamma\pr{\frac{\ep+3}{2}}m^\ep}{\sqrt{\pi}}\Gamma\left(-\frac{\ep}{2}\right)\sum_{n_y}\pr{\pr{\frac{2\pi n_y}{mL_y}}^2+1}^{\ep/2}=2\pr{\log\pr{\frac{1}{2}}-\log\pr{\sinh\pr{\frac{mL_y}{2}}}}.
\eeq
For the finite part in \eqref{eq: E string calc checkpoint 2}, we evaluate it by taking an $a$-derivative and integrating back:
\bea
\dots & \sum_{n_y-\infty}^{\infty}\left(1-\frac{2\pi\left|n_y\right|}{mL_y}\arctan\pr{\frac{mL_y}{2\pi\left|n_y\right|}}\right)\\
= & \frac{2\pi}{mL_y}\sum_{n_y-\infty}^{\infty}\intop_{0}^{\frac{mL_y}{2\pi}}da\pd_{a}\left(a-\left|n_y\right|\arctan\pr{\frac{a}{\left|n_y\right|}}\right)\\
= & \frac{2\pi}{mL_y}\sum_{n_y-\infty}^{\infty}\intop_{0}^{\frac{mL_y}{2\pi}}da\frac{a^2}{a^2+n_y^2}\\
= & \frac{2\pi}{mL_y}\intop_{0}^{\frac{mL_y}{2\pi}}da\pi a\coth\pr{\pi a}\\
= & \frac{1}{mL_y}\frac{1}{6}\pr{\frac{3}{2}m^2L_y^2-6\text{Li}_2\pr{e^{-mL_y}}+6mL_y\log\pr{1-e^{-mL_y}}+\pi^2}.
\eea
Putting together the finite and (putative-)divergent parts back into \eqref{eq: E string calc checkpoint 2}:
\bea
\dots= & \frac{1}{mL_y}\frac{1}{6}\pr{\frac{3}{2}m^2L_y^2-6\text{Li}_2\pr{e^{-mL_y}}+6mL_y\log\pr{1-e^{-mL_y}}+\pi^2}\\
+ & \frac{\Gamma\pr{\frac{3}{2}}}{\sqrt{\pi}}2\pr{\log\pr{\frac{1}{2}}-\log\pr{\sinh\pr{\pi\frac{mL_y}{2\pi}}}}+O\pr{\ep^{1}}\\
= & -\frac{mL_y}{4}+\frac{\pi^2}{6mL_y}-\frac{1}{mL_y}\text{Li}_2\pr{e^{-mL_y}}+O\pr{\ep}.
\eea
Remembering the prefactor from \eqref{eq: E string calc checkpoint 1}, we get:
\bea
\, & -m\frac{2\mu^{-\ep}\pi^{\frac{1+\ep}{2}}}{\left(2\pi\right)^{1+\ep}\Gamma\pr{\frac{1+\ep}{2}}\left(1+\ep\right)}\left(-\frac{mL_y}{4}+\frac{\pi^2}{6mL_y}-\frac{1}{mL_y}\text{Li}_2\pr{e^{-mL_y}}+O\pr{\ep^{1}}\right)\\
= & \frac{m^2L_y}{4\pi}-\frac{\pi}{6L_y}+\frac{1}{\pi L_y}\text{Li}_2\pr{e^{-mL_y}}+O\pr{\ep^{1}},
\eea
which is the final answer. In summary (suppressing regularization and compactification):
\bea \label{eq: E string calc checkpoint 3}
 \, & \frac{1}{2}L_y\intop\frac{d^2q}{\left(2\pi\right)^2}\pr{\log\pr{q^2}-\intop\frac{dp}{2\pi}\log\pr{q^2+p^2+m^2}\frac{2m}{p^2+m^2}}\\
= & \frac{m^2L_y}{4\pi}-\frac{\pi}{6L_y}+\frac{1}{\pi L_y}\text{Li}_2\pr{e^{-mL_y}}.
\eea

There is one further contribution to $\de E_{\rm string}$ at this order, stemming from the on-shell evaluation of the cosine-potential counterterm:
\beq
\frac{4\de m_2^2}{m} L_y.
\eeq
The value of $\de m_2^2$ can be obtained from the renormalization condition, setting the true mass of the photon to $m$. It is tuned to cancel the single 1-loop contribution to the photon propagator (in the vacuum, not the 1-string background). We find:
\bea
\de m_2^2 =\frac{m^2}{2}\mu^{-\ep}\intop\frac{d^{3+\ep}q}{\left(2\pi\right)^{3+\ep}}\frac{1}{q^2+m^2}
 =-\frac{m^{3}}{8\pi}.
\eea
Combining this with \eqref{eq: E string calc checkpoint 3}, we find:
\bea\label{eq: appendix E string final result}
\de E_{\rm string} = & -\frac{m^2L_y}{4\pi}-\frac{\pi}{6L_y}+\frac{1}{\pi L_y}\text{Li}_2\pr{e^{-mL_y}}.
\eea

\subsubsection{Canonical Framework}

We need to evaluate the sum \eqref{eq_ground_deE_dis}: 
\begin{equation}\label{eq_integrals_can_1}
\begin{split}
    \sum_{l = -\infty}^\infty e^{-s|q_l|} &\Bigg( \frac{q_l\arctan\pr{\frac{m^2-q_l^2}{2mq_l}} - q_l\arctan\pr{\frac{m}{q_l}}}{\pi} + \frac{m}{2\pi}\log \pr{\frac{m^2+q_l^2}{4\La^2}} 
    \\
    &+ \frac{\pi\sqrt{m^2+q_l^2}-2m}{2\pi} + \frac{1}{2}\pr{|q_l| - \sqrt{m^2+q_l^2}} \Bigg) =
    \\
    \sum_{l = -\infty}^\infty e^{-s|q_l|} &\pr{ \frac{2q_l\arctan\pr{\frac{m^2-q_l^2}{2mq_l}} - 2q_l\arctan\pr{\frac{m}{q_l}} + \pi|q_l| - 2m}{2\pi} + \frac{m}{2\pi}\log\pr{\frac{m^2+q_l^2}{4\La^2}}}.
\end{split}    
\end{equation}
The first term in the parenthesis is finite; it goes as $\frac{1}{l^2}$ as $l\to\infty$. Therefore, we can evaluate it with $s=0$. In order to evaluate the sum, we take a derivative with respect to $m$, evaluate the sum, and then integrate it again. The integration constant can be fixed by requiring the sum to vanish for $m=0$, for which the numerator vanishes $2q_l\pr{-\frac{\pi}{2\text{sign}(q_l)}} - 0 + \pi|q_l| - 0 =0$. Hence:
\begin{equation}
\begin{split}
    &\sum_{l = -\infty}^\infty \pr{ \frac{2q_l\arctan\pr{\frac{m^2-q_l^2}{2mq_l}} - 2q_l\arctan\pr{\frac{m}{q_l}} + \pi|q_l| - 2m}{2\pi} } = 
    \\
    -\frac{m}{\pi} + 2&\sum_{l=1}^\infty \pr{ \frac{2q_l\arctan\pr{\frac{m^2-q_l^2}{2mq_l}} - 2q_l\arctan\pr{\frac{m}{q_l}} + \pi|q_l| - 2m}{2\pi} } =
    \\
    -\frac{m}{\pi} + 2&\int_0^m dm' \sum_{l=1}^\infty \pr{-\frac{1}{\pi}\frac{m'^2}{m'^2+q_l^2} } = 
    -\frac{m}{\pi} + 2\int_0^m dm' \sum_{l=1}^\infty \pr{\frac{2 - L_ym'\coth\pr{\frac{L_ym'}{2}}}{4\pi} } =
    \\
    &-\frac{m^2}{4\pi}L_y - \frac{\pi}{6L_y} -\frac{m\log\pr{1-e^{-mL_y}}}{\pi } + \frac{\text{Li}_2\pr{e^{-mL_y}}}{\pi L_y}. 
\end{split}    
\end{equation}
The second term in \eqref{eq_integrals_can_1} diverges and will be canceled with the counterterm divergence. We need:
\begin{equation}
\begin{split}
    \sum_{l = -\infty}^\infty e^{-s|q_l|} \frac{m}{2\pi}\log\pr{\frac{m^2+q_l^2}{4\La^2}} 
    = \frac{m}{\pi}\log\pr{\frac{m}{2\La}} + 2\sum_{l=1}^\infty e^{-s|q_l|} \frac{m}{2\pi}\log\pr{\frac{m^2+q_l^2}{4\La^2}}. 
\end{split}    
\end{equation}
For $l\neq 0$, it is convenient to split it into a convergent part and a divergent part:
\begin{equation}
    \frac{m}{2\pi}\log\pr{\frac{m^2+q_l^2}{4\La^2}} = \frac{m}{\pi}\log\pr{\frac{\pi }{\La L_y}} + \frac{m}{\pi}\log(|l|) + \frac{m}{2\pi}\log\left(1+\pr{\frac{mL_y}{2\pi l}}^2\right).
\end{equation}
The sum over the first two terms is divergent and requires the regulator $s$, while the third one does not. The sum of the first term is trivial:
\begin{equation}\label{eq_integrals_log_1}
    2\sum_{l=1}^\infty\frac{m}{\pi}\log\pr{\frac{\pi }{\La L_y}}e^{-\frac{2\pi sl}{L_y}} = \frac{2m}{\pi}\log\pr{\frac{\pi }{\La L_y}}\pr{-\frac{1}{2} + \frac{L_y}{2\pi s} + O(s)}.
\end{equation}
For the second term we use $\log(l) = \frac{d l^\ep}{d\ep}\big|_{\ep=0}$, and then:
\begin{equation}\label{eq_integrals_log_2}
\begin{split}
    2\sum_{l=1}^\infty\frac{m}{\pi}l^\ep e^{-\frac{2\pi sl}{L_y}} &= \frac{2m}{\pi}\text{Li}_{-\ep}\pr{e^{-\frac{2\pi s}{L_y}}} \Rightarrow
    \\
    2\sum_{l=1}^\infty\frac{m}{\pi}\log(l)e^{-\frac{2\pi sl}{L_y}} &= \frac{2m\log(2\pi)}{L_y} - \frac{2m}{\pi s} \pr{\log\pr{\frac{2\pi s}{L_y}} + \gamma } + O(s).
\end{split}    
\end{equation}
For the third term, we set $s=0$, differentiate by $m$, sum, and integrate. It should vanish for $m=0$. Hence:
\begin{equation}\label{eq_integrals_log_3}
\begin{split}
    2\sum_{l=1}^\infty \frac{m}{2\pi}\log\left(1+\pr{\frac{mL_y}{2\pi l}}^2\right)
    = \frac{m}{\pi}\int_0^m dm'\sum_{l=1}^\infty \frac{2L_y^2m'}{L_y^2m'^2 + 4\pi^2 l^2} = 
    \\ 
    \frac{m}{\pi}\int_0^m dm'\frac{-2 + m'L_y \coth\pr{\frac{m'L_y}{2}}}{2m} = \frac{m}{\pi} \pr{\log\pr{\sinh\pr{\frac{mL_y}{2}}} - \log\pr{\frac{mL_y}{2}}}.
\end{split}    
\end{equation}
Adding \eqref{eq_integrals_log_1}, \eqref{eq_integrals_log_2}, \eqref{eq_integrals_log_3} together, one finds:
\begin{equation}\label{eq_integrals_log_tot}
\begin{split}
    \sum_{l = -\infty}^\infty e^{-s|q_l|} \frac{m}{2\pi}\log\pr{\frac{m^2+q_l^2}{4\La^2}} = \frac{m}{\pi}\log\left(2\sinh\pr{\frac{mL_y}{2}}\right) - \frac{mL_y}{\pi ^2 s}(\log(2\La s) + \g) + O(s).
\end{split}
\end{equation}
With the counterterm \eqref{eq_ground_cntr_can}, the correction \eqref{eq_ground_deE_dis} is:
\begin{equation}
\begin{split}
    \de E_{\rm string} = 
    & -\frac{m^2}{4\pi}L_y - \frac{\pi}{6L_y} -\frac{m\log\pr{1-e^{-mL_y}}}{\pi } + \frac{\text{Li}_2\pr{e^{-mL_y}}}{\pi L_y}
    \\
    & + \frac{m}{\pi}\log\left(2\sinh\pr{\frac{mL_y}{2}}\right) - \frac{mL_y}{\pi ^2 s}(\log(2\La s) + \g) + 
    \\
    & +\frac{m^2\bb^2}{4}\pr{\frac{1}{\pi^2 s}(\g + \log(2\La s)) - \frac{m}{2\pi}}\frac{4L_y}{m\bb^2}, 
\end{split}
\end{equation}
which is \eqref{eq_ground_final}, using:
\begin{equation}
    -\frac{m\log\pr{1-e^{-mL_y}}}{\pi} + \frac{m}{\pi}\log\left(2\sinh\pr{\frac{mL_y}{2}}\right) - \frac{m^2L_y}{2\pi} = 0.
\end{equation}

\subsection{Scattering}

Below, the integrals used for calculating scattering cross sections are evaluated. All are Wick-rotated to Euclidean variable $s_L \to -s_E$, where the propagator is changed:
\begin{equation}
    \frac{1}{s_L-p^2-m^2+i\ep} \to -\frac{1}{s_E+p^2+m^2},
\end{equation}
which will be our starting point. The answer is rotated back to Lorentzian variable by $s_E\to -s_L$. The factor $\sqrt{s_E + m^2}$ appearing in the Euclidean result is analytically-continued to $(-i)^{\theta\pr{s_L-m^2}}\sqrt{|s_L-m^2|}$. The sign of the square root for $s_L>m^2$ can be dictated by the imaginary part \eqref{eq_sctr_opt_3} being positive. It can also be seen by the usual relation $E_E = -iE_L$. Going from Lorentzian energy to Euclidean energy amounts to the trajectory $E_L = Ee^{i\pp}$ for $0\leq \pp\leq \frac{\pi}{2}$ in the complex Lorentzian energy. For the complex Euclidean energy it corresponds to $E_E = -iEe^{i\pp}$ or $s_E = -s e^{2i\pp}$. The factor $\sqrt{s_E + m^2}$ has a branch cut on the negative real $s_E$ axis, starting from $s_E = -m^2$. For $s_L = -s_E$ between $-m^2$ and $0$, there is no ambiguity, and the positive square root remains. The region $s_L>m^2$ is on the branch cut. Going back along $s_E = s e^{-2i\pp}$ for $0\leq \pp\leq \frac{\pi}{2}$ amounts to taking the square root with negative imaginary part, which also dictates that $\sqrt{s_E + m^2} = -i\sqrt{s_L-m^2}$ in this region.

\subsubsection{$I_{\rm 2NGB}$}

This integral is evaluated by Fourier-transforming to position space:
\begin{equation}
    -\int\frac{dp}{2\pi}\frac{m^2 + p^2}{s + p^2 + m^2}\sech\pr{\frac{\pi p}{2m}}^2 = -\frac{m}{2\pi^3}\int dpdxdy  \frac{p^2+1}{\frac{s}{m^2}+p^2+1}\sech(x)\sech(y)e^{ip(x+y)},
\end{equation}
where we used the Fourier transform of $\sech$:
\begin{equation}
    \sech\left(\frac{\pi p}{2}\right) =  \frac{1}{\pi}\int dx e^{ipx} \sech(x).
\end{equation}
The integral over $p$ is now a simple Fourier transform:
\begin{align}
    \int dp \frac{e^{ipx}}{a^2+p^2} = \frac{\pi e^{-a|x|}}{a}.
\end{align}
Changing variables $x=z+w,y=z-w$, the integrals of $w$ and $z$ are:
\begin{align}
    \int dw\sech(w+z)\sech(z-w) &= 2z\csch(z) \sech(z),
    \\
    \int_0^{\infty}dz\frac{z e^{-az}}{\sinh(2z)} &= \frac{1}{8}\psi _1\left(\frac{a+2}{4}\right).
\end{align}

The evaluation of the integral in subsection \ref{subsec_scattering_2NGB} follows similar lines to the evaluation of $I_{\rm 2NGB}$. One needs a few more Fourier transforms:
\begin{align}
    \int dp\frac{p^2e^{ipx}}{a^2+p^2} &= - \pi ae^{-a|x|} + 2\pi\delta(x),
    \\
    \int dp\frac{p^4 e^{ipx}}{a^2+p^2} &= \pi a^3e^{-a|x|} -2\pi a^2 \delta(x) - 2\pi\delta''(x),
    \\
    \int dp\frac{p^6 e^{ipx}}{a^2+p^2} &= -\pi a^5e^{-a|x|} + 2\pi a^4\delta(x) + 2\pi a^2 \delta''(x) + 2\pi\delta^{(4)}(x).
\end{align}
The integral over $w$ gives the same answer, and the integral over $z$ contains other integrals over $\de$-function and its derivative, which are easy to evaluate.

\subsubsection{$I_{\rm 1NGB1BP}$}

First, we manipulate the fraction:
\begin{equation}
\begin{split}
    & -\int\frac{dp'}{2\pi}{\csch}\pr{\frac{\pi(p+p')}{2m}}{\sech}\pr{\frac{\pi p'}{2m}}\frac{p'^2-p^2}{s+p'^2+m^2} = 
    \\
    & -\frac{m}{\pi^2}\int dp \frac{1}{\cosh(p)\sinh(a+p)} + \frac{m(a^2+b^2)}{\pi^2}\int dp \frac{1}{\cosh(p)\sinh(a+p)}\frac{1}{b^2+p^2},
\end{split}    
\end{equation}
where $a \equiv \frac{\pi p}{2m},\ b \equiv \frac{\pi}{2}\sqrt{\frac{s}{m^2}+1}$. The first integral is solved by a change of variables $x=e^{2p}$:
\begin{equation}
\begin{split}
    \int dp \frac{1}{\cosh(p)\sinh(a+p)} = 2a\sech(a).
\end{split}    
\end{equation}
The second integral is more involved. First, we manipulate it:
\begin{equation}
    \int dp\frac{1}{b^2+p^2}\frac{1}{\cosh(p)\sinh(a+p)} = 4\sech(a)e^{2a}\int dy\frac{1}{e^{2a}-e^{-y}}\frac{1}{4b^2+y^2} - 4\sech(a)\int dy\frac{1}{1+ e^{-y}}\frac{1}{4b^2+y^2},  
\end{equation}
with $y=2p$. The last integral is solved by using $1 = \frac{1}{1+e^y} + \frac{e^y}{1+e^y}$ and changing $y\to -y$ inside the integral. The result is:
\begin{equation}\label{eq_integrals_I5}
    \int dy\frac{1}{1+ e^{-y}}\frac{1}{4b^2+y^2} = \frac{\pi}{4b}.
\end{equation}
The integral remaining is transformed into a series upon the residue theorem. On the upper half of the complex plane, it has poles at $2ib, \{2\pi ik, k\in \mathbb{Z}^{>0}\}$. Then:
\begin{equation}\label{eq_integrals_I6}
    \int dy\frac{1}{e^{2a}-e^{-y}}\frac{1}{4b^2+y^2} = -2\pi\left(\sum _{k=1}^\infty \frac{\pi  a e^{-2a}k}{2\left(a^2+(b-\pi  k)^2\right) \left(a^2+(b+\pi  k)^2\right)}+\frac{\cos(2b)-e^{2a}}{4b \left(-2e^{2a} \cos(2b)+e^{4 a}+1\right)}\right).
\end{equation}
The series can be summed up as follows:
\begin{equation}
    \sum_{k=1}^\infty \frac{k}{\left(a^2+(b-\pi k)^2\right)\left(a^2+(b+\pi  k)^2\right)} = \frac{i}{8\pi^2ab}\left(-H_{\frac{ia-b}{\pi}}-H_{\frac{b-ia}{\pi }}+H_{\frac{b+ia}{\pi}} + H_{-\frac{b+ia}{\pi}}\right).
\end{equation}
This result is simplified by using the identity:
\begin{equation}
    H_{-z} = H_z - \frac{1}{z} + \pi\cot(\pi z).
\end{equation}

\subsubsection{$I^1_{\rm 2BP}$}

For solving the integral:
\begin{equation}
\begin{split}
    -\int \frac{dp}{2\pi} &\frac{\sech\pr{\frac{\pi(p_1+p_2+p)}{2m}}\sech\pr{\frac{\pi p}{2m}}}{s + p^2 + m^2}\cdot
    \\
    & \pr{3m^4 + (-p_1+p_2+p)(p_1-p_2+p)(p_1+p_2-p)(p_1+p_2+p) + 2m^2(p_1^2+p_2^2+p^2)},    
\end{split}
\end{equation}
we write the fraction as $\frac{a_4p^4 + a_2p^2 + a_0}{p^2 + b^2} = a_2 - a_4b^2 + a_4p^2 + \frac{a_0 - a_2b^2 + a_4b^4}{b^2 + p^2}$, with $b$ as before. The integral of the polynomial part times the two sech's is known:
\begin{align}
    \int dp \sech(p)\sech(p+a) &= 2a\csch(a),
    \\
    \int dp \sech(p)\sech(p+a)p^2 &= \frac{a(4a^2+\pi^2)}{6}\csch(a).
\end{align}
The remaining integral is:
\begin{equation}
    \int dp \frac{\sech(p)\sech(p+a)}{b^2+p^2} = -4\csch(a)\int dy\frac{1}{4 b^2+y^2}\left(\frac{1}{1+e^{-y}} -\frac{e^{2a}}{e^{2a}+e^{-y}}\right).
\end{equation}
The first integral was solved in \eqref{eq_integrals_I5}. The second one is similar to \eqref{eq_integrals_I6}, but its poles are shifted by $i\pi$, and its sum could not be solved directly by Mathematica. However, it can be decomposed as follows:
\begin{equation}
    \frac{1}{e^{2a}+e^{-y}}=\frac{2e^{2a}}{e^{4a}-e^{-2y}}-\frac{1}{e^{2a}-e^{-y}}.
\end{equation}
These two integrals are the same as in \eqref{eq_integrals_I6}.

\subsubsection{$I^2_{\rm 2BP}$}

This integral can be solved using \eqref{eq_integrals_I6}. First, we decompose
\begin{equation}
\begin{split}
     -\int\frac{dp}{2\pi}&\frac{1}{t + p^2 + m^2}\frac{(p_1^2-p^2)(p_2^2-p^2)}{p^2 + m^2}\csch\pr{\frac{\pi(p_1-p)}{2m}}\csch\pr{\frac{\pi(p_2+p)}{2m}} = 
     \\
     &\frac{m}{\pi^2}\int dp \csch(p-a_1) \csch(a_2+p)  
    \\
    + &\frac{m}{\pi^2}\frac{1}{b^2-c^2}\int dp\csch(p-a_1) \csch(a_2+p) \left(\frac{\left(a_1^2+c^2\right) \left(a_2^2+c^2\right)}{c^2+p^2}-\frac{\left(a_1^2+b^2\right) \left(a_2^2+b^2\right)}{b^2+p^2}\right),
\end{split}    
\end{equation}
where $a_{1,2} \equiv \frac{\pi p_{1,2}}{2m},\ c \equiv \frac{\pi}{2}$ and $b$ as before. The first integral gives:
\begin{equation}
    \int dp \csch(p-a_1)\csch(a_2+p)=\int dp \csch(p)\csch(a_1+a_2+p) = -2(a_1+a_2)\csch(a_1+a_2).
\end{equation}
The second integral can be decomposed further:
\begin{equation}
\begin{split}
    & \int dp \frac{1}{p^2+c^2}\csch(p-a_1)\csch(p+a_2) = 
    \\
    & \frac{8e^{a_1-a_2}}{e^{2a_1}-e^{-2a_2}}\pr{e^{-2a_1}\int dy \frac{1}{y^2 +4c^2}\frac{1}{e^{-2a_1}-e^{-y}} - e^{2a_2}\int dy \frac{1}{y^2 +4c^2}\frac{1}{e^{2a_2}-e^{-y}}}.
\end{split}    
\end{equation}
These integrals are again the same as in \eqref{eq_integrals_I6}.